\documentclass[%
aps, 
prd,
12pt,
onecolumn,
tightenlines,
showpacs,
nofootinbib,
floatfix,
amssymb,
amsmath
]{revtex4}
\usepackage{bm}
\usepackage[dvips]{color,graphicx}
\usepackage{epsfig}
\usepackage[figuresright]{rotating}
\allowdisplaybreaks[4]

\def    \ie             {i.e.}
\def    \etal             {{\it et al.\/}}
\def    \eg             {e.g.}
\newcommand{\ud}     {\mathrm{d}}
\newcommand{\gev}    {\:\mathrm{GeV}}

\newcommand{\gevsq}  {\:\mathrm{GeV}^2}

\renewcommand{\Im}{\mathop{\mathrm{Im}}}
\renewcommand{\Re}{\mathop{\mathrm{Re}}}

\newcommand{\ceps}{\varepsilon}

\newcommand{\bra}[1]{\langle{#1}|}
\newcommand{\ket}[1]{|{#1}\rangle}
\newcommand{\average}[1]{\left\langle{#1}\right\rangle}

\newcommand{\eq}[1]{Eq.(\ref{#1})}
\newcommand{\Eqs}[2]{Eqs.~(\ref{#1}) and (\ref{#2})}
\newcommand{\eqs}[1]{Eqs.(\ref{#1})}

\newcommand{\sigmabar}{\bar{\sigma}}

\newcommand{\as}{\alpha_{\scriptscriptstyle S}}
\newcommand{\sgls}{S_{\mathrm{GLS}}}
\newcommand{\sadler}{S_{\mathrm{A}}}

\newcommand{\FLpcac}{ F_L^{\textsc{pcac}} }
\newcommand{\FLac}{ F_L^{\textsc{ac}} }

\newcommand{\FLvc}{ F_L^{\textsc{vc}} }
\newcommand{\FTvc}{ F_T^{\textsc{vc}} }
\newcommand{\shrinkvspace}{\vspace{-3ex}}
\begin{document}
%
\title{%
Neutrino inclusive inelastic scattering off nuclei
}

\author{S.~A.~Kulagin}
\affiliation{Institute for Nuclear Research, 117312 Moscow, Russia}
\author{R.~Petti}
\affiliation{Department of Physics and Astronomy, University of South Carolina, Columbia SC 29208, USA }


\begin{abstract}

\noindent
We present a detailed description of high-energy neutrino and antineutrino
inelastic inclusive scattering off nuclei in terms of nuclear structure
functions. In our approach we take into account a QCD description of the
nucleon structure functions as well as a number of basic nuclear effects
including nuclear shadowing, Fermi motion and binding, nuclear pion excess
and off-shell correction to bound nucleon structure functions. These
effects prove to be important in the studies of charged-lepton deep-inelastic scattering.
We discuss similarities and dissimilarities in the
calculation of nuclear effects for charged-lepton and neutrino scattering
caused by nonconserved axial current in neutrino scattering.
We examine the Adler and the
Gross-Llewellyn-Smith sum rules for nuclear structure functions
and find a remarkable cancellation between nuclear shadowing and off-shell
corrections in these sum rules.
We present calculations of differential cross
sections for inclusive neutrino and antineutrino scattering in comparison
with recent data on different target materials.

\end{abstract}


\maketitle

\section{Introduction}
\label{sec:intro}

The interest in the study of neutrino interactions with nuclei at intermediate
and high energy has grown considerably in recent years (see, e.g., \cite{nuint}).
New experimental results on neutrino cross-sections are also obtained
for different nuclear targets \cite{nutev-xsec,nomad-xsec,chorus-xsec}.
The presence of an axial-vector component in the weak
current and the quark flavor selection distinguish neutrino interactions
from charged-lepton (CL) and hadron collisions. This feature makes the
neutrino ($\nu$) and antineutrino ($\bar{\nu}$) data a unique source of information
on the nucleon and nuclear structure. The understanding of neutrino propagation in matter
is also important for astrophysics, cosmology and even geology applications.

The use of heavy nuclear targets in neutrino experiments is motivated
by the need to collect a statistically significant number of interactions.
For this reason, the understanding of nuclear effects is of primary
importance for a correct interpretation of experimental results and for
the evaluation of the corresponding uncertainties. An example of such synergy
is provided by the precision electroweak measurements from neutrino interactions.
The role of nuclear corrections to neutrino structure functions has recently been
emphasized~\cite{nutev-nucl} in the context of the anomalous measurement
of the weak mixing angle ($\sin^2 \theta_{W}$) reported by the NuTeV collaboration~\cite{nutev-sin2w}.

An accurate account of nuclear effects is not only
important in the determination of electroweak parameters in neutrino
scattering experiments, but also for the understanding of neutrino masses
and mixing. The next generation experiments would imply precision
measurements to disentangle small oscillation signals from neutrino and antineutrino
interactions on nuclei. This in turn would require us to improve our knowledge
of $\nu(\bar{\nu})$-nucleus cross-sections in order to reduce systematic uncertainties.

The availability of high-intensity neutrino beams from the recent
NuMI~\cite{numi} and JPARC~\cite{JPARC} facilities offers new opportunities
for detailed studies of cross sections and nuclear effects in neutrino interactions.
One such example is MINER$\nu$A~\cite{minerva}, a dedicated experiment at Fermilab which will
collect data in few years. On a longer time scale, the construction of a neutrino
factory~\cite{nufact} would then allow a further step forward, finally reaching the ultimate
precision of the neutrino probe.

In this work we present a calculation of inelastic neutrino-nucleus
structure functions (SF) and differential cross sections in a wide
kinematical range of the Bjorken variable $x$ and momentum transfer
squared $Q^2$.
Our goal is to develop a quantitative model incorporating existing data,
which would be useful in interpreting  $\nu(\bar{\nu})$ experiments.

Scattering experiments with charged leptons show that even in the deep-inelastic
scattering (DIS) region, for which the energy and momentum transfer are larger
than the nucleon mass $M$, nuclear SF differ from a simple sum over bound nucleons and
a significant nuclear effect is present even in the scaling regime (see, e.g., \cite{Arneodo:1994wf}).
As evidenced by numerous studies (see \cite{p-w-rev,thomas-rev} and also \cite{KP04}),
the lepton-nucleus DIS is characterized by different mechanisms in different kinematical
regions of the Bjorken variable $x$.
In a recent paper~\cite{KP04} we studied the charged-lepton DIS off
nuclear targets and developed a quantitative model for nuclear structure
functions, taking into account major nuclear effects including nuclear
shadowing, Fermi motion and binding, nuclear pion excess and off-shell
correction to bound nucleon structure functions. This approach showed a
very good agreement with data on the EMC effect for light and heavy nuclei
for both the $x$ and $Q^2$ dependence.
In the present paper we extend this approach to neutrino-nucleus inelastic
scattering.

Although nuclear scattering mechanisms for high-energy charged leptons and
neutrinos are similar in many respects
(especially in the impulse approximation),
there are important differences due to the presence of the axial-vector
current in neutrino interactions. The interference between the vector and
the axial-vector currents introduces $C$-odd terms in neutrino cross
sections, which are described by SF $F_3$.
In the calculation of nuclear corrections, we separate
the contributions to different SF according to their $C$-parity.
We examine in detail the dependence of nuclear effects on $C$-parity and
show that it is particulary important in the nuclear shadowing region.
Results are then used to predict nuclear structure functions for
neutrino and antineutrino interactions, respectively.

In contrast to the electromagnetic current, the axial current is not conserved, and for this reason the
neutrino structure functions in the region of low $Q^2$ and small $x$ are
rather different from those of electroproduction. In particular, the
neutrino longitudinal SF $F_L$ (as well as $F_2$) does not vanish at low
$Q^2$. Its value is determined by the divergence of the axial-vector current
which is linked to the virtual pion cross-section (Adler relation \cite{Adler:1964yx}).
We discuss the derivation of this relationship for structure functions and
examine corrections to that. We also address the phenomenological
implications of this relation in the description of low-$Q^2$ and low-$x$
neutrino differential cross sections. The nuclear shadowing effect on
the virtual pion cross section results in a significant suppression of
the corresponding nuclear structure functions.

The paper is organized as follows. In Sec.~\ref{sec:basic} we introduce
the general formalism of neutrino inclusive inelastic scattering, define
the structure functions and present the differential cross sections for
charged-current (CC) and neutral-current (NC) neutrino scattering. In
Sec.~\ref{sec:sf:high-q} we summarize our model of the nucleon SF in the
region of high $Q^2$; in Sec.~\ref{sec:sf:low-q} we deal with the
low-$Q^2$ region and discuss the separation of the axial current
contribution from the vector component. In Sec.~\ref{sec:nuke} we discuss
the nuclear effects in neutrino DIS. Particular attention is paid to the
calculation of neutrino $C$-even and $C$-odd structure functions. In
Sec.~\ref{sec:ASR} and \ref{sec:GLS} we examine the Adler and the
Gross--Llewellyn-Smith sum rules for nuclear targets. In
Sec.~\ref{sec:res} we present the results of numerical analysis of SF and
differential cross sections for a number of nuclear targets and compare
our predictions with the recent data
\cite{nomad-xsec,nutev-xsec,chorus-xsec}.

\section{General formalism of neutrino inclusive scattering}
\label{sec:basic}

To leading order in the weak coupling constant, neutrino scattering is
described by the standard one-boson exchange process.
Neutrino interactions can be mediated by the charged $W^\pm$ boson
(\emph{charged curent} or CC) or by the neutral $Z$ boson (\emph{neutral
curent} or NC).
In the Standard Model (SM) the leptonic and hadronic charged currents can
be written as
\begin{subequations}
\label{cc}
\begin{align}
\label{cc:l}
j_\lambda^- &= \bar e O_\lambda^L\nu_e + \bar \mu
O_\lambda^L\nu_\mu, \\
J_\lambda^- &= \bar d' O_\lambda^L u + \bar s' O_\lambda^L c,
\label{cc:h}
\end{align}
\end{subequations}
where $O_\lambda^L=(1-\gamma_5)\gamma_\lambda$, and
$d'=d\,\cos\theta_C+s\,\sin\theta_C$,
$s'=-d\,\sin\theta_C+s\,\cos\theta_C$ 
are the superpositions of $d$- and $s$-quark
states and $\theta_C$ is a Cabibbo angle ($\sin^2 \theta_C \approx 0.05$).
For simplicity, we neglect contributions from the third generation
of quarks and leptons. The superscript $L$ in Eqs.(\ref{cc}) indicates that
only left doublets participate in the CC weak interaction.

The neutral current can be written as
\begin{subequations}\label{NC}
\begin{align}
\label{NC:l}
j_\lambda^0 &= \sum_{\ell} \bar\ell(g_V^l-g_A^l\gamma_5)\gamma_\lambda
\ell, \\
\label{NC:h}
J_\lambda^0 &= \sum_{q} \bar q(g_V^q-g_A^q\gamma_5)\gamma_\lambda q,
\end{align}
\end{subequations}
where the sum is taken over all types of leptons ($\ell=e,\nu_e,\mu,\nu_\mu$)
and quarks ($q=u,d,c,s$). The vector and axial charges of quarks and
leptons in the SM in terms of the weak mixing angle $\theta_W$ can be found in,
e.g., \cite{IoKhLi84}.

\subsection{Charged-curent neutrino scattering}
\label{sec:cc}

We first consider the CC (anti)neutrino inelastic inclusive scattering.
In inclusive scattering, the final hadronic state is not
detected and the differential cross section can be written as
(see, e.g., \cite{IoKhLi84})
\begin{equation}
\ud^3 \sigma_{\text{CC}} =
\frac{{2\pi G_F^2}/{(4\pi)^3}}{(1+Q^2/M_W^2)^2}
L_{\mu\lambda} W_{\mu\lambda}
\frac{\ud^3k'}{(p\cdot k)E'} ,
\label{CC:xsec}
\end{equation}
where $G_F$ is the Fermi weak coupling constant; $M_W$ is the $W$-boson mass;
$p$ is the target four-momentum; $k=(E,\bm{k})$ and $k'=(E',\bm{k'})$ are four-momenta of the
incoming and outgoing lepton, respectively; $q=k-k'$ is
four-momentum transfer and $Q^2=-q^2$.
The tensors $L_{\mu\lambda}$ and $W_{\mu\lambda}$
describe the interaction of the $W$ boson with leptons and hadrons, respectively.
If the polarization of
the outgoing lepton is not detected, then the leptonic tensors for
neutrino and antineutrino scattering are
%
\begin{align}
\label{CC:L}
L_{\mu\lambda}^{(\nu,\bar \nu)} &=
8\left(
k_\mu k'_\lambda + k_\lambda k'_\mu - k\cdot k' g_{\mu\lambda }
\pm i\,\varepsilon_{\mu\lambda}(k,k') \right),
\end{align}
%
where the sign $+(-)$ corresponds to the neutrino (antineutrino).
We use the contracted notation
$\ceps_{\mu\lambda}(a,b)=\ceps_{\mu\lambda\alpha\beta}a^\alpha b^\beta$.
The hadronic tensor $W_{\mu\lambda}$ is the sum of CC matrix
elements over all possible final hadronic states.
For the neutrino we have
\begin{align}
W_{\mu\lambda}^{(\nu)}(p,q) &= \frac{1}{4\pi}\sum_{f}
(2\pi)^4\delta(p+q-p_f)
\bra{p}{J_\mu^+}^\dagger(0)\ket{f}\bra{f}J_\lambda^+(0)\ket{p},
\label{CC:W}
\end{align}
where the current in the SM is given by \eq{cc:h}.
The antineutrino tensor corresponds to the exchange of the $W^-$ boson and is given
by a similar equation with the current $J_\lambda^+$ replaced by $J_\lambda^-$.

For the scattering from an unpolarized target, there are five independent
structure functions in the hadronic tensor (\ref{CC:W}) for either
the neutrino or antineutrino (see, e.g., \cite{IoKhLi84})
\begin{equation}
\begin{split}
W_{\mu\lambda }(p,q) &=
\left( \frac{q_\mu q_\lambda}{q^2} - g_{\mu\lambda}\right) F_1
+
     \left(p_\mu - q_\mu\frac{p\cdot q}{q^2}\right)
   \left(p_\lambda - q_\lambda\frac{p\cdot q}{q^2}\right)
\frac{F_2}{p\cdot q}
\\
&
+ i\,\varepsilon_{\mu\lambda}(p,q)\frac{F_3}{2p\cdot q}
+\frac{q_\mu q_\lambda}{Q^2}F_4
+\frac{q_\mu p_\lambda + q_\lambda p_\mu}{p\cdot q}F_5 .
\end{split}
\label{CC:SF}
\end{equation}
We use the normalization of states
$\langle p|p'\rangle = 2E_{\bm{p}}(2\pi)^3 \delta(\bm{p}-\bm{p}')$
for both the bosons and fermions.
In this normalization the hadronic tensor (\ref{CC:W})
and the structure functions are dimensionless.
The terms with $F_1$ and $F_2$ are similar to those in charged-lepton
scattering. They originate from vector-vector and
axial-axial correlations in \eq{CC:W}. The antisymmetric term ($F_3$)
describes parity-violating vector-axial and axial-vector transitions. 
The terms $F_4$ and $F_5$ are present because of nonconservation of the axial current.

The neutrino and antineutrino structure functions are apparently
different. We will also consider the sum and the difference of neutrino and
antineutrino structure functions, which have definite $C$-parity
%
\begin{align}\label{SF:C}
F_i^{\nu \pm \bar\nu} &= F_i^\nu \pm F_i^{\bar\nu},
\end{align}
%

Contracting the leptonic and hadronic tensors, we obtain the explicit form of
the differential cross section in terms of the structure functions.
Using the usual variables $x=Q^2/(2p\cdot q)$
and $y=p\cdot q/p\cdot k$ we have
\begin{align}
\label{CC:xsec:xy}
\frac{\ud^2\sigma^{(\nu,\bar\nu)}_{\mathrm{CC}}}{\ud x\ud y} &=
\frac{G_F^2 (p\cdot k)/\pi}{(1+Q^2/M_W^2)^2}
\sum_{i=1}^5 Y_i F_i^{(\nu,\bar\nu)}.
\end{align}
The kinematical factors $Y_i$ read as follows
\begin{subequations}\label{Y}
\begin{eqnarray}
Y_1 &=& y^2x \frac{{Q'}^2}{Q^2} \left(1-\frac{{m'}^2}{2Q^2}\right),\\
Y_2 &=& \left(1-\frac{y{Q'}^2}{2Q^2}\right)^2 -
        \frac{y^2{Q'}^2}{4Q^2}\left(1+\frac{p^2Q^2}{(p\cdot q)^2}\right),\\
Y_3 &=& \pm\, xy \left(1-\frac{y{Q'}^2}{2Q^2}\right),\\
Y_4 &=& \frac{y{Q'}^2}{4Q^2}\frac{{m'}^2}{p\cdot k}, \\
Y_5 &=& -\frac{{m'}^2}{p\cdot k},
\end{eqnarray}
\end{subequations}
where the sign $+(-)$ refers to neutrino
(antineutrino) scattering, $m'$ is the mass of the outgoing charged
lepton and ${Q'}^2=Q^2+{m'}^2$. We keep the lepton mass terms for
completeness. Although these terms are tiny in $\nu_e$ and $\nu_\mu$
scattering, they are not negligible in $\nu_\tau$ scattering. The
contributions from the structure functions $F_4$ and $F_5$ to the neutrino
production cross section are suppessed at high energy by a small ratio
${m'}^2/p\cdot k$.

The structure functions can be related to the virtual
boson helicity cross sections by projecting \eq{CC:SF} onto the states with
definite polarizations of an intermediate vector boson
[for simplicity, we suppress explicit notation for (anti)neutrino hadronic
tensor]
\begin{equation}
W_h = {e^\mu_h}^* W_{\mu\nu}{e^\nu_h},
\label{SF:hel}
\end{equation}
where $e^\mu_h$ is the polarization vector of the virtual boson in the helicity
state $h$ ($h=\pm 1$ corresponds to two transverse polarizations, and
$h=0$ to the longitudinal polarization).
The relation between the helicity structure functions $W_h$ and
$F_{1,2,3}$ can be derived from \eq{CC:SF} using an
explicit form of polarization vectors
and the orthogonality and normalization conditions for polarization
vectors (see, e.g., \cite{IoKhLi84,KP04}). We have
\begin{subequations}\label{hel:123}
\begin{align}
W_{\pm 1} &= F_1 \pm \gamma F_3,
\\
W_0 &= \gamma^2 F_2/(2x)-F_1.
\end{align}
\end{subequations}
One concludes from
\eqs{hel:123} that $F_1$ is the transverse structure function averaged
over transverse polarizations and $F_3$ is determined by the right-left
asymmetry in helicity structure functions $W_{+1} - W_{-1}$.
Below, we also use $F_T=2xF_1$, $F_L=2xW_0$ and $R=F_L/F_T$.

\subsection{Neutral-current neutrino scattering}
\label{sec:nc}

The NC hadronic tensor is given by \eq{CC:W}, in which the charged
current must be replaced by the neutral current.  The leptonic tensor
in this case is given by \eq{CC:L} with the overall factor $1/2$, reflecting the
fact that the neutrino can only be left-polarized (and the antineutrino,
right-polarized). As a result, the NC cross section has the overall
factor $1/2$ compared to \eq{CC:xsec:xy},
\begin{eqnarray}
\label{NC:xsec:xy}
\frac{\ud^2\sigma^{(\nu,\bar\nu)}_{\text{CC}}}{\ud x\ud y} &=&
\frac{G_F^2 (p\cdot k)/(2\pi)}{(1+Q^2/M_Z^2)^2}
\sum_{i=1}^5 Y_i F_i^Z,
\end{eqnarray}
where $M_Z$ is the mass of the $Z$ boson and
the kinematical factors $Y_i$ are obtained from
Eqs.(\ref{Y}) by taking the limit $m'\to0$. Therefore, the structure
functions $F_{4}$ and $F_{5}$ do not contribute to the NC cross
section.
We also comment that neutrino and antineutrino NC structure functions of the type $i$
are identical, because the neutrino and antineutrino in NC scattering couple
to the same hadronic current.

\subsection{Neutrino structure functions at high $Q^2$}
\label{sec:sf:high-q}

At high $Q^2$, above the
resonance region, the structure functions are usually described in QCD as
a series in $Q^{-2}$ on the basis of the operator product expansion of
the correlator of the weak currents (twist expansion) \cite{OPE}. The
leading twist (LT) terms correspond to quasifree scattering off quarks
corrected for gluon radiation effects. In this approximation the structure
functions are given in terms of parton distributions or PDFs
(see, e.g., \cite{handbook,IoKhLi84}). The PDFs
are universal high-momentum transfer characteristics of the target. The
PDFs have a nonperturbative origin and their dependence on the Bjorken
variable $x$ can not be calculated in perturbative QCD. However, the
dependence of PDFs on momentum transfer is governed by the perturbation
theory in strong coupling constant $\as$ at the scale $Q^2$ (DGLAP
evolution equations \cite{DGLAP}). Currently, the coefficient and the
splitting functions, which determine the QCD evolution, are known up to
two-loop approximation (NNLO approximation) \cite{C-NNLO,P-NNLO}.

Along with the LT terms, the higher
order $Q^{-2}$ terms have to be taken into account (higher twists or HT).
The HT terms can be of two different sources: (i) the kinematical target mass
terms which can be approximately resummed and absorbed into the LT terms
according to \cite{GeoPol76,Nacht} and (ii) the dynamical HT terms which
are related to quark-gluon correlation effects.
In summary, the structure function of type $i$ can be written as a power
series
\begin{equation}
F_i(x,Q^2) = F_i^{\text{TMC}}(x,Q^2)
        + H_i^{(4)}/Q^{2}
        + \cdots,
\end{equation}
where $F_i^{\text{TMC}}$ is the LT structure function
corrected for the target mass
effects and $H_i^{(t)}$ are the functions of $x$ and $Q^2$ describing the
strength of the HT terms of twist $t$. In this paper the target mass
corrections are computed using the approach of Ref.\cite{GeoPol76}. In
order to keep the correct threshold behaviour of structure functions for $x
\to 1$, we expand $F_i^{\text{TMC}}$ in a power series in
$Q^{-2}$ and keep the terms to the order $Q^{-2}$~\cite{KP04}.
 
In the numerical analysis the LT structure
functions are computed using both the PDFs and the coefficient
functions to the NNLO approximation.
We use the PDFs and HT of Ref.\cite{a02}, obtained from
a new fits optimized at low $Q^2$ and including additional data.
It should be noted that unlike the PDFs the
HT terms are not universal and generally depend on the type of the
structure function $i$ as well as on the type of interaction.
In order to evaluate HT in neutrino DIS, we assume that the ratio $H_i/F_i^{\rm LT}$
is similar for CL and neutrino structure functions as a working hypothesis.
That allows us to use the phenomenological HT terms extracted from the fits
to CL DIS data. We also assume no HT terms for $F_3$ in the calculations of
nuclear structure functions and differential cross sections presented in
Sec.~\ref{sec:res}.

In the calculation of neutrino cross sections for the structure function $F_5$
we use the Albright--Jarlskog relation $2xF_5=F_2$  \cite{AJ75}. Recently,
it was argued that this relation survives QCD higher order corrections and
the target mass corrections \cite{KR02}. For $F_4$ we will use the
relation $F_4=F_2/(2x)-F_1$ which replaces the Albright--Jarlskog relation
$F_4=0$ for massless quarks if the target mass corrections are taken into
account \cite{KR02}.

\subsection{Neutrino structure functions at low $Q^2$}
\label{sec:sf:low-q}

We note that even for high-energy scattering in the fixed-target
experiments the events with small Bjorken $x$ typically have the
values of $Q^2$ below $1\gevsq$. In this Section we discuss the
vector- and the axial-current contributions to neutrino structure
functions at low $Q^2$ and outline the model which will be used in our
studies of neutrino differential cross sections.
In this context it is useful to review the requirements which follow from
conservation of the vector current and partial conservation of the axial current.
It follows from \Eqs{SF:hel}{CC:W} that the structure functions
$F_T$ and $F_L$ can be written as
\begin{equation}
\label{SF:TL}
F_{T,L} = \frac{\gamma}{\pi} Q^2 \sigma_{T,L},
\end{equation}
where $\sigma_{T(L)}$ refers to the total cross section
induced by the transverse (longitudinal) component of
the weak current $J_\lambda$,
\begin{equation}
\label{sigma:TL}
\sigma_{T,L} =
\frac{1}{4 j}\sum_{f}(2\pi)^4\delta(p+q-p_f)
|\bra{p}e_{T,L}\cdot J(0)\ket{f}|^2,
\end{equation}
where $j=\left((p\cdot q)^2-p^2 q^2\right)^{1/2}$ is the invarint flux of
the virtual boson with four-momentum $q$ on the target with four-momentum $p$
(for the nucleon at rest $j=M|\bm{q}|$, with $M$ the nucleon mass) and the
sum is taken over all possible final states.
The transverse cross section does
not vanish in the limit $Q^2\to 0$; therefore $F_T$ vanishes as
$Q^2$ in this limit. This holds for both the vector- and the axial-current
contributions.

In the longitudinal channel the low-$Q$ behavior of the vector- and
axial-current parts are different. We first consider the contribution from
the \emph{vector current}.
The conservation of the vector current (CVC) suggests $q_\mu W_{\mu\nu}=0$ for the vector-current
part of the hadronic tensor. From this condition we conclude that $\FLvc$
vanishes faster than $\FTvc$ at low $Q^2$ and $\FLvc/\FTvc\sim Q^2/q_0^2$,
similar to the charged-lepton case.

In contrast to the vector current, the axial current is not
conserved and, according to the hypothesis of partial conservation of the axial current (PCAC),
its divergence is proportional
to the pion field \cite{Adler:1964yx},
\begin{equation}
\label{pcac}
\partial A^\pm = f_\pi m_\pi^2 \varphi^\pm,
\end{equation}
where $m_\pi$ is the pion mass and $f_\pi=0.93 m_\pi$ is the pion decay
constant and $\varphi^\pm$ is the pion field in the corresponding charge
state. The PCAC relation can be used to compute the axial current
contribution to the longitudinal structure function
$\FLac$ in the region of low momentum transfer.
In this context it is useful to explicitly separate
the pion contribution to the axial current,
\begin{equation}
\label{A:prim}
A_\mu = -f_\pi\partial_\mu\varphi + A_\mu',
\end{equation}
where the operator $A'$ describes the contributions from heavy hadron states, i.e. the
contributions from axial-vector meson states, $\rho\pi$ continuum, etc.
It is important to notice that the pion derivative term
does not contribute to the structure functions \cite{Kopeliovich:1992ym}. This
is because the matrix elements of $\partial_\mu\varphi$ are proportional
to the momentum transfer $q_\mu$. These terms give a vanishing contribution
when contracted with the boson polarization vectors because of the
orthogonality condition $e_h\cdot q=0$ for any helicity state $h$.
Therefore, the axial-vector
operator $A_\mu$ can be replaced by $A_\mu'$ everywhere in \eqs{CC:W}.
The PCAC relation for $A'$ follows from \eq{pcac}:
\begin{equation}
\label{pcac:prim}
\partial {A'}^\pm = f_\pi j_\pi^\pm,
\end{equation}
where $j_\pi^\pm=(\partial^2+m_\pi^2)\varphi^\pm$ is the pion current.
In order to calculate $F_L$ we separate the contribution of
the axial current to the hadronic tensor (\ref{CC:W}), contract it
with the polarization vector $e_0$ and replace $A$ with $A'$.%
\footnote{To this end a covariant form of longitudinal vector
$e_0^\mu = \left(q^\mu+p^\mu Q^2/p\cdot q\right)/Q$
is useful.}
Using \eq{pcac:prim} we have
\begin{equation}
\label{FL:ac}
\FLac = \gamma^{3} \FLpcac
+ 2\gamma^3\frac{Q^2 f_\pi \sigma'_\pi}{\pi |\bm{q}|}
+ \widetilde{F}_L^{\textsc{ac}},
\end{equation}
where $\FLpcac=f_\pi^2\sigma_\pi/\pi$ and $\sigma_\pi=\sigma_\pi(s,Q^2)$
is the total cross section for scattering of a virtual pion with
four-momentum $q$ and the center-of-mass energy squared $s=(p+q)^2$
\begin{equation}
\label{xsec:pi}
\sigma _\pi = \frac{1}{4 j}\sum_{f}(2\pi)^4\delta(p+q-p_f)
|\bra{p} j_\pi(0)\ket{f}|^2.
\end{equation}
The flux $j$ is defined in \eq{sigma:TL}. Note also that $\pi^+$ cross
section corresponds to the neutrino, and $\pi^-$ to the antineutrino. The quantity
$\sigma'_\pi$ describes the interference between the divergence of the
axial current and the operator $A'_z$ in the hadronic tensor
\begin{equation}
\label{sig:prim}
\sigma'_\pi = \frac{1}{4 j}\sum_{f'}(2\pi)^4\delta(p+q-p_{f'})
\Im\left( \bra{p} j_\pi(0)\ket{f'}\bra{f'} {A'_z(0)}^\dagger \ket{p}\right).
\end{equation}
The last term in \eq{FL:ac} is similar to $\FLvc$ and vanishes as $Q^4$.

It follows from the relation $F_2 = (F_L+F_T)/\gamma^2$ and (\ref{FL:ac})
that the structure function $F_2$ at low $Q^2$ is dominated by the
$\FLpcac$ term.
The PCAC contribution to the neutrino cross sections at small values of $Q^2$
was experimentally tested in \cite{PCAC:BEBC}.
On the other hand, the PCAC term should vanish at high $Q^2$. 
In order to interpolate between low and high $Q^2$, we introduce a
form factor $f_{\textsc{pcac}}(Q^2)$
\begin{equation}
\label{FL:PCAC:2}
\FLpcac = \frac{f_\pi^2\sigma _\pi}{\pi}f_{\textsc{pcac}}(Q^2).
\end{equation}
In the numerical analysis we assume a
dipole form $f_{\textsc{pcac}}(Q^2)=(1+Q^2/M_{\textsc{pcac}}^2)^{-2}$
with $M_{\textsc{pcac}}$ the mass scale controlling
the PCAC mechanism. Since the pion pole does not contribute
to the structure functions, the scale $M_{\textsc{pcac}}$ is not
determined by the pion mass, but rather it is related to higher mass states like
$a_1$ meson, $\rho \pi$ continuum etc.
In the present paper we fix the numerical value of $M_{\textsc{pcac}}$ phenomenologically
using low-$Q^2$ cross section data from the CHORUS experiment~\cite{chorus-xsec}
(see Sec.~\ref{sec:res:xsec}).

\section{Nuclear aspects of neutrino scattering}
\label{sec:nuke}

In this Section we summarize the theoretical framework used in this
paper to calculate the (anti)neutrino structure functions and
cross sections for nuclear targets.
A similar approach was recently applied in the studies of charged-lepton
nuclear DIS~\cite{KP04}.
In Sec.~\ref{sec:nuke:incoh} we review calculations of nuclear structure
functions in the impulse approximation with off-shell corrections, in
Sec.~\ref{sec:PI} we discuss nuclear pion correction, in
Sec.~\ref{sec:nuke:coh} we deal with nuclear effects from
coherent nuclear interactions of hadronic component of an intermediate boson,
which are relevant at small $x$.

\subsection{Incoherent scattering from bound nucleons}
\label{sec:nuke:incoh}

In the region of large Bjorken $x$ the nuclear DIS is usually treated in
impulse approximation, \ie as an incoherent sum over bound protons and
neutrons.
In this approximation the nuclear structure functions can be written in terms
of the convolution of the nuclear spectral function and off-shell nucleon
structure functions.
In the target rest frame we have (for a derivation and
more details see \cite{KP04,Ku89,KPW94,Ku98})
\begin{subequations}
\label{FA}
\begin{align}
F_T^A(x,Q^2) &= \sum_{\tau=p,n}\int [\ud p]
        {\mathcal P}^\tau(\varepsilon,\bm{p})\,
                \left(1+\frac{\gamma p_z}{M}\right)
\left(
F_T^\tau + \frac{2{x'}^2\bm{p}_\perp^2}{Q^2}F_2^\tau
\right),
\label{FA-T}
 \\
F_L^A(x,Q^2) &= \sum_{\tau=p,n}\int [\ud p]
        {\mathcal P}^\tau(\varepsilon,\bm{p})\,
                \left(1+\frac{\gamma p_z}{M}\right)
\left(
F_L^\tau + \frac{4{x'}^2\bm{p}_\perp^2}{Q^2}F_2^\tau
\right),
\label{FA-L}
\\
\gamma^2 F_2^A(x,Q^2) &=
 \sum_{\tau=p,n}\int [\ud p]
        {\mathcal P}^\tau(\varepsilon,\bm{p})\,
                \left(1+\frac{\gamma p_z}{M}\right)
\left({\gamma'}^2 +\frac{6{x'}^2 \bm{p}_\perp^2}{Q^2} \right)
F_2^\tau. 
\label{FA-2}
\\
xF_3^A(x,Q^2) &=
 \sum_{\tau=p,n}\int [\ud p]
        {\mathcal P}^\tau(\varepsilon,\bm{p})\,
                \left(1+\frac{p_z}{\gamma M}\right)
        x'F_3^\tau. 
\label{FA-3}
\end{align}
\end{subequations}
where the integration is taken over the four-momentum of the bound nucleon
$p=(M+\ceps,\,\bm{p})$ and $[\ud p]=\ud\ceps\,\ud^3\bm{p}/(2\pi)^4$.
The axis $z$ is chosen such that
$q=(q_0,\bm{0}_\perp,-|\bm{q}|)$,
$\bm{p}_\perp$ is the transverse nucleon momentum, and
$\gamma=|\bm{q}|/q_0$.
The energy and momentum distribution of bound nucleons is
described by the proton (neutron) nuclear spectral function
$\mathcal{P}^{p(n)}(\varepsilon,\bm{p})$ which is normalized to the
proton (neutron) number in a nucleus.
In the integrand $F_i^{p(n)}$ are the structure
functions of bound proton (neutron), which
depend on the Bjorken variable $x'=Q^2/(2p\cdot q)$, momentum transfer
square $Q^2$ and generally on the nucleon invariant mass squared
$p^2=(M+\ceps)^2-\bm{p}^2$.
The relation between the $F_T$, $F_L$ and $F_2$ of the off-shell nucleon
is similar to (\ref{hel:123}), \ie ${\gamma'}^2 F_2=F_T+F_L$ with
${\gamma'}^2=1+4{x'}^2p^2/Q^2$. Note that $\gamma'=\gamma$ for the
vanishing momentum of the bound nucleon. It should be also noted that the
transverse motion of the bound nucleon in the target causes the mixture of
different structure functions in \Eqs{FA-T}{FA-L} to order
$\bm{p}_\perp^2/(M|\bm{q}|)$.

\subsubsection{Isoscalar and isovector contributions}
\label{sec:I01}

Since complex nuclei typically have different numbers of protons and
neutrons, the nuclear structure functions generally have both the
\emph{isoscalar} and the \emph{isovector} parts.
In order to separate the isoscalar and isovector contributions in
Eqs.(\ref{FA}), we write the isoscalar
($\mathcal{P}^{p+n}=\mathcal{P}^p+\mathcal{P}^n$)
and isovector ($\mathcal{P}^{p-n}=\mathcal{P}^p-\mathcal{P}^n$)
nuclear spectral functions as
\begin{subequations}
\label{spfn:01}
\begin{eqnarray}
\label{spfn:0}
\mathcal{P}^{p+n} &=& A \mathcal{P}_0,\\
\mathcal{P}^{p-n} &=& (Z-N)\mathcal{P}_1,
\label{spfn:1}
\end{eqnarray}
\end{subequations}
where the reduced spectral functions $\mathcal{P}_{0}$ and
$\mathcal{P}_{1}$ are normalized to unity.
Using \eq{spfn:01} we explicitly write Eqs.(\ref{FA}) in
terms of isoscalar and isovector contributions
\begin{equation}
\label{nuke:FA}
F_i^A/A = \left\langle F_i^N \right\rangle_0 +
	\frac{\beta}{2}\left\langle F_i^{p-n} \right\rangle_1,
\end{equation}
where $F_i^N=\tfrac12(F_i^p+F_i^n)$, $F_i^{p-n}=F_i^p-F_i^n$ for the
structure function of type $i$ and
the parameter
$\beta=(Z-N)/A$ describes the excess of protons over neutrons in a nucleus.
The quantities $\left\langle F \right\rangle_{0}$ and
$\left\langle F \right\rangle_{1}$ are the
contracted notations of the integration in
Eqs.(\ref{FA}) with reduced spectral
functions $\mathcal{P}_0$ and $\mathcal{P}_1$, respectively.
The model of $\mathcal{P}_0$ and $\mathcal{P}_1$, which is used in
numerical applications in this paper, is discussed in more
detail in Ref.\cite{KP04}.


Equation (\ref{nuke:FA}) is generic and can be applied to any structure function.
We now discuss in more detail the $\nu \pm \bar\nu$
combinations of neutrino structure functions with definite $C$-parity (see \eq{SF:C}).
Let us first consider symmetric $\nu + \bar\nu$ combination. From \eq{nuke:FA} we have
\begin{equation}
\label{nuke:C-even}
F_i^{(\nu+\bar\nu)A}/A = \left\langle F_i^{(\nu+\bar\nu)N} \right\rangle_0
+ ({\beta}/{2})\left\langle F_i^{(\nu+\bar\nu)(p-n)} \right\rangle_1
\end{equation}
for any type $i$ of the structure function.

In the absence of heavy quark contributions we have
$F_i^{(\nu+\bar\nu)p}=F_i^{(\nu+\bar\nu)n}$ because of the isospin
symmetry. For this reason the isovector term in
\eq{nuke:C-even} should vanish.
However, we should remark that the isospin relations for structure
functions for neutrino CC scattering are violated by the mixing of different
quark generations and the $c$-quark mass effect, even in the presence of
exact isospin symmetry on the PDF level. This effect results in the
nonzero difference $F_i^{(\nu+\bar\nu)(p-n)} \propto \sin^2\theta_C$ with
$\theta_C$  the Cabbibo mixing angle. Since the parameter $\beta$ is
small, this effect is suppressed in \eq{nuke:C-even} and, therefore, it is
a good approximation to keep only the isoscalar term in \eq{nuke:C-even}.

\begin{figure}[htb]
\begin{center}
\shrinkvspace
\epsfig{file=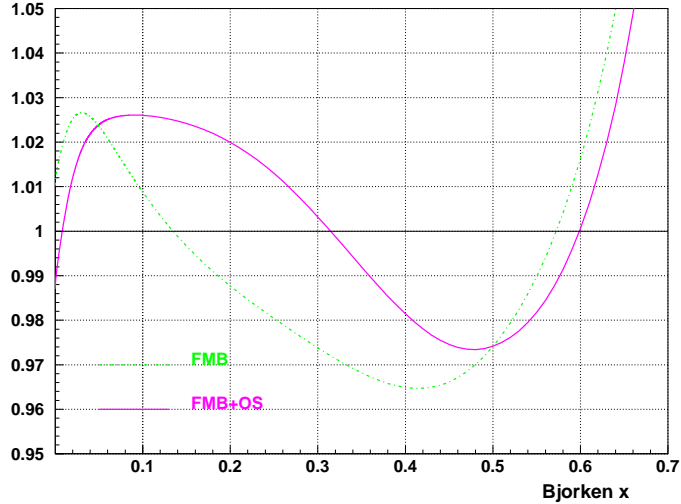,width=10.0cm}
\caption{%
The isovector nuclear effect
$\langle F_2^{(\nu-\bar\nu)p}\rangle_1/F_2^{(\nu-\bar\nu)p}$ calculated for
${}^{207}$Pb at $Q^2=5\gevsq$ with and without off-shell correction using
the isovector nuclear spectral function of Ref.\cite{KP04}. The labels on
the curves correspond to effects due to Fermi motion and nuclear binding
(FMB) and off-shell correction (OS), which was calculated using
\eq{delf:fit}.
}
\label{fig:non0}
\shrinkvspace
\shrinkvspace
\end{center}
\end{figure}

Let us now discuss the $\nu{-}\bar\nu$ asymmetry in the nuclear structure
functions. From \eq{nuke:FA} we have for the nuclear structure function
of the type $i$
\begin{equation}
\label{nuke:C-odd}
F_i^{(\nu-\bar\nu)A}/A =
\left\langle F_i^{(\nu-\bar\nu)N} \right\rangle_0 +
({\beta}/{2})\left\langle F_i^{(\nu-\bar\nu)(p-n)} \right\rangle_1.
\end{equation}
The application of this equation to $F_2$ and $xF_3$ requires somewhat
more attention. We first consider $F_2^{\nu-\bar\nu}$ (a similar discussion also
applies to $F_T$ and $F_L$). This structure function is $C$-odd and
dominated by the isovector quark distributions. In the absence of the Cabbibo
mixing effect we have
$F_2^{(\nu-\bar\nu)p}=-F_2^{(\nu-\bar\nu)n}$. It follows from this
relation that the first term in the right side of \eq{nuke:C-odd} vanishes
and the nuclear structure function is determined by the second (isovector)
term. Nuclear effects in this case are illustrated in Fig.~\ref{fig:non0}.

However, the isospin relations for structure functions
are violated by the heavy quark effect and
the Cabbibo mixing angle, as was discussed above. This effect generates a
nonzero value of  $F_2^{(\nu{-}\bar\nu)N}\propto\sin^2\theta_C$.
Surprisingly, this effect should not be
neglected in the analysis of nuclear corrections for $\nu-\bar \nu$
asymmeries even in the first approximation. This is because the isovector
contribution in \eq{nuke:C-odd} is suppressed by the factor $\beta$ and
thus the relative effect of $F_2^{(\nu{-}\bar\nu)N}$ is enhanced.
This effect is further discussed in Sec.~\ref{sec:res}, where we present
the results of numerical calculation of nuclear structure functions.

Let us discuss \eq{nuke:C-odd} in application to $xF_3$.
Note that $xF_3^{\nu-\bar\nu}$ is $C$ even
and includes the contribution from the
light quarks, which determines the isovector nuclear correction in
\eq{nuke:C-odd}, and the $s$-quark contribution to
the isoscalar part in \eq{nuke:C-odd}. In fact, the difference
$xF_3^{(\nu{-}\bar\nu)N}$ is driven by the $s$-quark distribution [recall
that in the parton model $xF_3^{(\nu{-}\bar\nu)N}=x(s+\bar s)$]. Because
$\beta$ is a small parameter,  the relative contribution from the isoscalar
term in \eq{nuke:C-odd} is enhanced and we conclude that the asymmetry
$xF_3^{\nu-\bar\nu}$ in nuclei is dominated by the strange quark
contribution.

\subsubsection{Off-shell effects}
\label{sec:OS}

The structure functions of the bound proton and neutron can differ from
those of the free proton and neutron. In the approach discussed, this
effect is related to analytical continuation of the structure functions to
the off-shell region and the dependence on the nucleon invariant mass $p^2$.
One can separate the two sources of $p^2$ dependence: 
(i) the target mass correction leading to the terms of the order $p^2/Q^2$
and 
(ii)
off-shell dependence of LT structure functions (or parton distributions).
The off-shell dependence of the target mass effect is evaluated by the
replacement $M^2\to p^2$ in the expressions of Ref.\cite{GeoPol76}.
In order to facilitate the discussion of the off-shell dependence of the LT
structure functions, we note that characteristic momenta and energies of
the bound nucleon are small compared to the nucleon mass, and the nucleon
virtuality $v=(p^2-M^2)/M^2$ is a small parameter.
We then expand the LT structure functions in
this parameter and keep the terms to the first order in $v$,
\begin{align}
\label{SF:OS}
F_2^{\mathrm{LT}}(x,Q^2,p^2) &=
F_2^{\mathrm{LT}}(x,Q^2)\left( 1+\delta f_2(x,Q^2)\,v \right),\\
\label{delf}
\delta f_2 &= \partial\ln F_2^{\mathrm{LT}}/\partial\ln p^2,
\end{align}
where the first term on the right in \eq{SF:OS} is the structure
function of the on-mass-shell nucleon and the derivative is evaluated
at $p^2=M^2$. Similar expressions can be written for other
structure functions.

The off-shell correction $\delta f_2$ was studied phenomenologically
in \cite{KP04} by analyzing the data on the ratios of structure
functions of different nuclei (EMC effect). As a result it was found
that $\delta f_2$ is independent of $Q^2$ with a good accuracy and can
be parametrized as
\begin{equation}
\label{delf:fit}
\delta f_2 = C_N (x-x_1)(x-x_0)(1+x_0 - x)
\end{equation}
with the parameters extracted from the fit
$C_N=8.10\pm0.30(\mathrm{stat.})\pm0.53(\mathrm{syst.})$,
$x_0=0.448\pm0.005(\mathrm{stat.})\pm0.007(\mathrm{syst.})$,
and $x_1=0.05$.

Although the off-shell function (\ref{delf}) generally
depends on the structure function type,
it was assumed in Ref.\cite{KP04} that
the relative off-shell effect is common for all types of
structure functions (or partons) and described by the off-shell
function (\ref{delf:fit}). It was tested that this assumption leads to
fairly good cancellation between off-shell and nuclear shadowing effects
in the normalization of the valence quark distribution. In this paper the
off-shell function by \eq{delf:fit} is applied in our calculations of
nuclear (anti)neutrino structure functions. The interplay between off-shell (OS) and
nuclear shadowing (NS) corrections in the DIS sum rules is further studied in
Sec.~\ref{sec:GLS} and \ref{sec:ASR}, where we discuss the
Gross--Llewellyn-Smith and the Adler sum rules for nuclear targets.

\subsection{Nuclear pion correction}
\label{sec:PI}

The lepton scattering off the meson fields in nuclei
gives rise to mesonic contribution to the nuclear structure functions
(for a review see, e.g., \cite{p-w-rev,thomas-rev}).
We note that because of binding the nucleons do not
carry all of the nuclear light-cone momentum and the meson
contribution is important for balancing the missing momentum. At high $Q^2$
the mesonic correction to nuclear structure functions can
approximately be taken into account using the convolution model
and the kinematics of the Bjorken limit neglecting power corrections. Then
the pion correction to $F_2^A$ can be written as
\begin{equation}
\label{SF:picor}
\delta_\pi F_2^{A}(x,Q^2) = \sum_{\pi}\int_x\ud y\,
                            f_{\pi/A}(y)F_2^\pi(x/y,Q^2),
\end{equation}
where $f_{\pi/A}(y)$ is the distribution of nuclear pion excess%
\protect\footnote{The contribution from the nucleon meson cloud should
be subtracted since it is accounted for in the nucleon structure
function.}
and $F_2^\pi$ is the pion structure function and the sum is taken over
different pion states $\pi=\pi^+,\pi^-,\pi^0$. Similar equations can be
written for the pion correction to $F_{T}$ and $F_{L}$ with the same pion
distribution function.
We also note that the pion correction vanishes for
$F_3$. In the numerical analysis we also assume no off-shell effect in the
virtual pion structure function, which would be the second order correction
because (\ref{SF:picor}) is a small correction by itself.

We apply \eq{SF:picor} to $\nu\pm\bar\nu$ combinations of
neutrino structure functions.
It is convenient to separate the isoscalar and isovector contributions in
the sum over pion states in \eq{SF:picor}. We also apply the isospin
relations for the CC neutrino-pion SFs ($F_2^{(\nu+\bar\nu)\pi}$ is
isoscalar and independent of the pion charge state, and
$F_2^{(\nu-\bar\nu)\pi^+}=-F_2^{(\nu-\bar\nu)\pi^-}$).%
\footnote{We neglect here the violation of the isospin relations because of
the Cabbibo mixing angle (see Sec.~\ref{sec:I01}).}
We then have for the nuclear pion correction
\begin{subequations}\label{picor:C}
\begin{align}
\label{picor:C-even}
\delta_\pi F_2^{(\nu+\bar\nu)A}(x,Q^2) &=
        \int_x\ud y\,f_{\pi/A}^0(y)F_2^{(\nu+\bar\nu)\pi^+}(x/y,Q^2),
\\
\label{picor:C-odd}
\delta_\pi F_2^{(\nu-\bar\nu)A}(x,Q^2) &=
        \int_x\ud y\,f_{\pi/A}^1(y)F_2^{(\nu-\bar\nu)\pi^+}(x/y,Q^2),
\end{align}
\end{subequations}
where $f_{\pi/A}^0=f_{\pi^+/A}+f_{\pi^-/A}+f_{\pi^0/A}$
and  $f_{\pi/A}^1=f_{\pi^+/A}-f_{\pi^-/A}$
are the isoscalar and the isovector pion distributions in a nucleus.
In the applications discussed below, we neglect possible isovector
contributions and use a model isoscalar pion distribution function, which
obeys the constraints due to the nuclear light-cone momentum balance equation
and equations of motion of the pion field in the nucleus (for more details
see \cite{KP04,Ku89}).

\subsection{Coherent nuclear effects (shadowing and antishadowing)}
\label{sec:nuke:coh}

At small $x$, at which the DIS longitudinal distance
$L_I=1/(Mx)$ is large compared to typical nuclear sizes, the nuclear DIS can
be viewed as a two-step process in which the intermediate boson first
fluctuates into a quark-antiquark pair (or hadronic configuration) which
then scatters off the nucleus.
As an average lifetime of this fluctuation
is large (of order $L_I$), it can undergo multiple interactions with bound
nucleons (for a recent review see, \eg, \cite{p-w-rev}).
In this section we discuss nuclear effects in
neutrino scattering associated with this mechanism
and derive corrections for different SF types for $\nu$ and $\bar\nu$
interactions.

The structure functions at small $x$ can be presented
as a superposition of contributions from different hadronic states the
virtual boson fluctuates to. For the helicity structure functions $W_0$ and
$W_{\pm}$, as defined in \eq{hel:123}, we have
\begin{equation}
W_h = \sum_i w_h(i) \sigma_h(i,s),
\label{sum-h}
\end{equation}
where $\sigma_h$ is the total cross section of scattering of the
hadronic state $i$ with the given helicity $h=0,\pm 1$ off the target
nucleon (or nucleus) with the center-of-mass energy $s=Q^2(1/x-1)+M^2$,
and $w_h$ describes the weight of the given hadronic state in the wave
function of the intermediate boson.

We first consider the transverse structure functions with $h=\pm 1$
and approximate the sum over hadronic states in \eq{sum-h} by
a factorized form \cite{KP04}
\begin{equation}
W_h(x,Q^2) = w_h(x,Q^2) \sigmabar_h(s),
\label{sum-h-aprx}
\end{equation}
where $\sigmabar_h$ is an \emph{effective} cross section averaged over
hadronic configurations and $w_h$ is the remaining normalization factor for
the given helicity $h$. At low $Q^2$ and small $x$ the effective cross
section can be approximated by average vector meson cross section in the 
vector meson dominance model (VMD) (see, \eg, \cite{Bauer:iq}).
As $Q^2$ increases, the averaging in (\ref{sum-h-aprx}) involves 
the rising number of active hadronic states.

In the longitudinal case ($W_0$) we explicitly separate the PCAC term,
\eq{FL:PCAC:2}, and assume that the rest, $\widetilde F_L$, can be expressed
by an equation similar to \eq{sum-h-aprx} with an effective longitudinal cross
section.

In this paper we are concerned with the
calculation of nuclear corrections to the structure
functions and cross sections.
In the following we will assume that the normalization
factors $w_h$ and the pion
decay constant $f_\pi$ are not affected by nuclear effects. Then for the
transverse structure functions with $h=\pm 1$ the relative nuclear
correction is determined by the corresponding
correction to the effective cross section,
\begin{equation}
\delta\mathcal{R}_h
     = \frac{\delta W_h^A}{Z W_h^p + N W_h^n}
  = \frac{\delta\sigmabar_h^A}{Z \sigmabar_h^p + N \sigmabar_h^n},
\label{relcor}
\end{equation}
where $\delta W_h^A$ is the nuclear structure function
of helicity $h$ with the incoherent term subtracted,
$\delta W_h^A = W_h^A - Z W_h^p - N W_h^n$,
and a similar definition is used for $\delta\sigmabar_h^A$.
For the longitudinal structure function the corresponding relation to
the cross section will be discussed in Sec.\ref{sec:MS:L}.

\subsubsection{Multiple scattering corrections to helicity amplitudes}
\label{sec:MS}

In order to calculate nuclear modifications of effective cross
sections $\delta\sigma$ we apply the multiple scattering theory
\cite{Glauber:1955qq,Gribov:1968gs}.  Let $a^{p(n)}$ be the proton
(neutron) effective scattering amplitude in forward direction. We will
choose the normalization of the scattering amplitude such that the optical
theorem reads $\Im a(s)=\sigma(s)/2$ with $\sigma$ the total cross
section. For the following discussion it is convenient to write the
forward scattering amplitude as
\begin{equation}\label{amp}
a=(i+\alpha)\sigma/2,
\end{equation}
where $\alpha=\Re a/\Im a$.
The nuclear scattering amplitude $a^A$ can be written as
\begin{equation}\label{amp:A}
a^A = Z\,a^p + N\,a^n + \delta a^A,
\end{equation}
where the first two terms on the right side are
the incoherent contributions from bound protons and neutrons.%
\footnote{The incoherent term is discussed in detail in
Sec.\ref{sec:nuke:incoh} together with nuclear binding, Fermi motion and
off-shell effects. Here we focus on multiple scattering corrections.}
The term $\delta a^A$ is the multiple scattering correction (MS), which can
be written as
\begin{align}\label{mult:A}
& \delta a^A = i\int_{z_1<z_2}\hspace{-1.5em}
		\ud^2\bm{b}\ud z_1\ud z_2\,e^{i k_L(z_1-z_2)}
                a\cdot\rho(1)\, a\cdot\rho(2)
\,e^{S(1,2,a)},
\\
\label{mult:phi}
& S(1,2,a) = i\int_{z_1}^{z_2}\ud z'
                    a\cdot\rho(\bm{b},z'), 
\end{align}
where $a\cdot\rho=a^p\rho^p+a^n\rho^n$ and $\rho^{p}(\rho^{n})$ is the proton
(neutron) nuclear density normalized to the proton (neutron) number
and the integration is performed along the collision axis, which is
chosen to be the $z$ axis, and over the transverse positions of nucleons
(impact parameter $\bm{b}$).
We also use a contracted notation of the nucleon position on
the collision axis $\rho(1)=\rho(\bm{b},z_1)$, etc.
The exponential factor  in \eq{mult:A} accounts for multiple
scattering effects (see, \eg, \cite{Bauer:iq}) and
for double scattering approximation $e^S$ should be replaced by 1.

Amplitude (\ref{amp:A}) describes the scattering of virtual hadron
configuration of the intermediate boson $W^*$ off a nucleus. In
intermediate states in the MS series one considers the scattering of
on-mass-shell hadronic states, and for this reason there appears a
nonzero longitudinal momentum transfer $k_L$. The quantity $L_c=1/k_L$
is the longitudinal coherence length of this state.
If $m$ is the mass of intermediate state,
then $k_L=Mx(1+m^2/Q^2)$.
Since the cross section in \eq{sum-h-aprx} reffers to the effective
intermediate hadronic state, we treat $m^2=m_{\rm eff}^2$ as a parameter.

Note that \eq{mult:A} was derived in the optical approximation
assuming that the nuclear wave function factorizes into the product of
the single particle wave functions. In this approximation possible
effects of correlation between bound nucleons are neglected. We
comment in this context that the correlations are relevant only if the
coherence length $L_c$ is comparable to the short-range
repulsive part of the nucleon-nucleon force, which is about 0.5\:Fm.
However, this region is limited to relatively large $x$ at
which shadowing is a small correction (see discussion in Ref.\cite{p-w-rev}).
The transverse momentum dependence of elastic scattering amplitudes
was also neglected, that is justified by a small transverse size of
the meson-nucleon amplitude compared to the nuclear radius.

We now apply \eq{mult:A} to describe nuclear effects generated by
scattering of charged intermediate
$W^+$ and $W^-$ bosons. Let us first consider the contributions from
the light ($u$ and $d$) quarks. In this case the scattering amplitude
obeys the requirements of the isospin symmetry and we can write the relations
$a(W^\pm p)=a(W^\mp n)=a^0\pm\tfrac12a^1$,
where $a^0$ and $a^1$ are the scattering amplitudes corresponding to the
isoscalar and isovector nucleon configuration. Using these relations we have
\begin{equation}\label{a.rho}
a(W^\pm)\cdot \rho = \left(a^0 \pm \tfrac12{\beta} a^1\right)\rho_A,
\end{equation}
where $\rho_A=\rho_p + \rho_n$ is the (isoscalar) nucleon density and
$\beta=(\rho_p - \rho_n)/\rho_A$ is the relative proton--neutron density
asymmetry. In order to facilitate the discussion of isovector effects,
we assume similar shapes of the proton and neutrion densities.
Then $\beta=(Z-N)/A$ is independent of the nucleon coordinate for
nuclear interior. We also benefit from the fact that $\beta$ is a small
parameter for heavy nuclei, and expand the multiple scattering series
(\ref{mult:A}) in $\beta$ and keep the terms to first order in $\beta$.
Then for the scattering of $W^+$ and $W^-$ bosons we obtain
\begin{equation}\label{mult:01}
\delta a^A(W^\pm) = \mathcal{T}^A(a^0)\pm
                    \tfrac12{\beta} a^1\mathcal{T}^A_1(a^0),
\end{equation}
where
\begin{subequations}\label{mult:T}
\begin{align}
\mathcal{T}^A(a) &= ia^2\mathcal{C}_2^A(a),\\
\mathcal{T}^A_1(a) &=\tfrac{\partial}{\partial a} \mathcal{T}^A(a)
= 2i a\,\mathcal{C}_2^A(a) - a^2\mathcal{C}_3^A(a),
\end{align}
\end{subequations}
and the quantities $\mathcal{C}_{2,3}^A$ incorporate the multiple
scattering effects and read as follows:
\begin{subequations}
\begin{align}
\label{C2A}
\mathcal{C}_2^A(a) &= \int_{z_1<z_2}\hspace{-1.5em}
		\ud^2\bm{b}\ud z_1\ud z_2\, e^{ik_L(z_1-z_2)}
        \rho_A(1)\rho_A(2)
e^{S(1,2,a)},
\\
\label{C3A}
\begin{split}
\mathcal{C}_3^A(a) &= -i \tfrac{\partial}{\partial a}\mathcal{C}_2^A(a) =
\\
& \protect\phantom{=}
\int_{z_1<z_2<z_3}\hspace{-3em}
		\ud^2\bm{b}\ud z_1\ud z_2\ud z_3\, e^{ik_L(z_1-z_3)}
        \rho_A(1)\rho_A(2)\rho_A(3)
\, e^{S(1,3,a)},
\end{split}
\end{align}\end{subequations}
where the $e^S$ factor is given by \eq{mult:phi} with $a\cdot\rho$
replaced by $a\rho_A$.
The rate of multiple scattering interactions is controlled by
the value of the mean free path of hadronic fluctuation in a nucleus
$l_f=(\rho_A\sigma)^{-1}$.
If $l_f$ is large compared with the nuclear radius (i.e.
at low nucleon density or/and small effective cross
section), then the $e^S$ factor can be neglected and the
coefficients $\mathcal{C}_2^A$ and $\mathcal{C}_3^A$ determine the
double ($\sim \rho^2$) and the tripple ($\sim \rho^3$) scattering
terms in the MS series. If $l_f$ is small enough
then the $e^S$ factor should be taken into account.

We find from \eq{mult:01} that nuclear correction to
the sum of $W^+$ and $W^-$ scattering amplitudes
is determined by the isoscalar
amplitude $a^0$, and the nuclear $W^+ - W^-$ asymmetry is
proportional to $\beta a^1$.
Note that this asymmetry also depends on $a^0$ through multiple scattering
effects. The implications to (anti)neutrino structure functions are
discussed in the next section.

\subsubsection{Dependence of nuclear effects on $C$-parity and isospin}

Up to now the helicity dependence of nuclear correction was
implicit. In this section we apply the results of
Sec.\ref{sec:MS} to the combinations of helicity amplitudes which
correspond to the structure functions of interest. Recall that $F_T$
is given by the average $(W_{+1}+W_{-1})/2$, and $F_3$ is
determined by the asymmetry $W_{+1}-W_{-1}$. We will assume that
the helicity is conserved in the MS series so that \eq{mult:A} generalizes to
scattering amplitude with given helicity $a_h$.%
\footnote{However, the helicity-flip processes can occur because of spin
effects, which are not considered here.}

The neutrino and antineutrino cross section correspond to the interaction of
$W^+$ and $W^-$, respectively. It will be convenient to discuss the
$\nu+\bar\nu$ average and $\nu-\bar\nu$ asymmetry since these combinations
have definite $C$-parity. We use the notation
$a_T^I=\tfrac12(a_{+1}^I+a_{-1}^I)$ for the average transverse amplitude
with the isospin $I=0,1$ and $a_\Delta^I=\tfrac12(a_{+1}^I-a_{-1}^I)$ for
the corresponding asymmetry. Note that the amplitudes $a_T^0$ and
$a_\Delta^1$ are $C$-even, while $a_\Delta^0$ and $a_T^1$ are $C$-odd.%
\footnote{The symmetry of the effective amplitude $a_h^I$
becomes apparent if we recall the correspondence with the structure functions:
$F_T^{\nu+\bar\nu} \sim \Im a_T^0$,
$F_T^{\nu-\bar\nu} \sim \Im a_T^1$,
$F_3^{\nu+\bar\nu} \sim \Im a_\Delta^0$,
$F_3^{\nu-\bar\nu} \sim \Im a_\Delta^1$.}

Let us first consider the $\nu+\bar\nu$ sum of amplitudes
and use \eq{mult:01} for helicity $h=+1$ and $h=-1$.
In particular, we consider the helicity average ($T$) and asymmetry ($\Delta$).
Taking into account that $|a_\Delta^0|\ll |a_T^0|$ in the low-$x$ region,
we expand $\mathcal{T}^A(a_{\pm 1}^0)$ about $a_T^0$ to order
$(a_\Delta^0)^2$ and obtain
\begin{subequations}\label{mult:pl}
\begin{align}
\label{mult:pl:T}
\delta a^{(\nu+\bar\nu)A}_T &= 2\,\mathcal{T}^A(a_T^0),\\
\delta a^{(\nu+\bar\nu)A}_\Delta &= 2\,a_\Delta^0\mathcal{T}^A_1(a_T^0).
\label{mult:pl:D}
\end{align}
\end{subequations}
%


Using these equations we compute ratios (\ref{relcor}) for
effective cross sections.
We recall that in
the Born approximation $a^{(\nu+\bar\nu)A}=A\,2a^0$ for either
helicity state $T$ or $\Delta$ and, therefore, the ratio \eq{relcor} is
$\delta\mathcal{R}^{\nu+\bar\nu}=\Im\delta a^{(\nu+\bar\nu)A}/(2A\Im a^0)$.
Using Eqs.(\ref{mult:pl}) and (\ref{amp}) we obtain
\begin{subequations}\label{delR:pl}
\begin{align}
\label{delR:pl:T}
\delta\mathcal{R}_T^{\nu+\bar\nu} &=
\sigma_T^0 \Re \left[(i+\alpha_T^0)^2\mathcal{C}_2^A(a_T^0)\right]/(2A),
\\
\delta\mathcal{R}_\Delta^{\nu+\bar\nu} &=
\Im \left[(i+\alpha_\Delta^0)\mathcal{T}_1^A(a_T^0)\right]/A,
\label{delR:pl:D}
\end{align}
\end{subequations}
where $\sigma_T^0=2\Im a_T^0$ is the effective cross section corresponding to
the transverse isoscalar amplitude and $\alpha_h^I=\Re a_h^I/\Im a_h^I$.
If the real part of the amplitude is small, then the MS
correction is negative because of destructive interference of
forward scattering amplitudes on the upstream nucleons that causes
\emph{shadowing} of virtual hadron interactions. On the other hand, if the
real part is large then the interference in the double scattering term is
constructive, which would lead to an \emph{antishadowing} effect.
It should be remarked that $\delta\mathcal{R}_\Delta^{\nu+\bar\nu}$
(the relative nuclear correction to the structure function $F_3^{\nu+\bar\nu}$)
depends on the $C$-even cross section $\sigma_T^0$. The result is also
affected by the interference of the real parts of the amplitudes in the
$C$-even and $C$-odd channels. If we only keep the double scattering term
in \Eqs{delR:pl:T}{delR:pl:D}, we arrive at the following relation:
\begin{equation}\label{Del:T}
\delta\mathcal{R}_\Delta^{\nu+\bar\nu}/\delta\mathcal{R}_T^{\nu+\bar\nu}
= 2\frac{1-\alpha_\Delta\alpha_T}{1-\alpha_T^2},
\end{equation}
where $\alpha_T$ and $\alpha_\Delta$ refer to $I=0$ amplitudes.
Factor 2 in \eq{Del:T} is a generic enhancement of the double
scattering correction for the cross section asymmetry which has a
combinatorial origin (in the context of $F_3$ it was discussed in
\cite{Kulagin:1998wc,KP04}).
Ratio (\ref{Del:T}) can be futher enhanced if the real
parts of the $C$-odd and $C$-even amplitudes have different sign (this
is indeed a realistic case as will be discussed below).


We now discuss the $\nu-\bar\nu$ asymmetry.
We apply \eq{mult:01} to the amplitudes $a_T$ and $a_\Delta$.
Similar to the $\nu+\bar\nu$ case, 
we expand $\mathcal{T}_1^A(a_{\pm 1}^0)$ about the
average amplitude $a_T^0$ in a series in $a_\Delta^0$ up to the first
order term. We obtain
\begin{subequations}\label{mult:mn}
\begin{align}
\label{mult:mn:T}
\delta a^{(\nu-\bar\nu)A}_T &= \beta\left[
  a_T^1 \mathcal{T}^A_1(a_T^0)+
  \tfrac14 a_\Delta^1 a_\Delta^0\mathcal{T}^A_2(a_T^0)
\right],\\
\delta a^{(\nu-\bar\nu)A}_\Delta &= \beta\left[
  a_\Delta^1 \mathcal{T}^A_1(a_T^0) +
  a_T^1 a_\Delta^0\mathcal{T}^A_2(a_T^0)
\right],
\label{mult:mn:D}
\end{align}
\end{subequations}
where $\mathcal{T}^A_2(a)=\tfrac{\partial}{\partial a}\mathcal{T}^A_1(a)$
[see \eqs{mult:T}].
Note that the first equation in (\ref{mult:mn}) is $C$-odd and the second $C$-even.
The quadratic terms $a^1 a^0$ in (\ref{mult:mn}) produce a
correction to the leading term of order $|a_\Delta^0/a_T^0|$.
Taking into account the correspondence between effective amplitudes and
the structure functions, we find that by order of magnitude
$|a_\Delta^0/a_T^0|$ is the ratio of valence to sea quark distribution
in the proton. This ratio is small in the region of small $x$ ($x\ll
0.1$), where coherent nuclear effects are most important.
For this reason in the following discussion we keep
only the first term in the right side of (\ref{mult:mn}). In this
approximation the structure of \eqs{mult:mn} is similar to that of
\eq{mult:pl:D}.

Using \eqs{mult:mn} we calculate the relative corrections to the
cross sections. For this purpose we recall that in the Born approximation
$a^{(\nu-\bar\nu)A}=(Z-N)a^1$ for either helicity state $T$ or $\Delta$.
Then the ratio
$\delta\mathcal{R}^{\nu-\bar\nu}=\Im\delta a^{(\nu-\bar\nu)A}/[(Z-N)\Im a^1]$
can be written as
\begin{align}
\label{delR:mn}
\delta\mathcal{R}_{T,\Delta}^{\nu-\bar\nu} &=
\Im \left[(i+\alpha_{T,\Delta}^1)\mathcal{T}_1^A(a_T^0)\right]/A,
\end{align}
where $\alpha_{T}^1$ and $\alpha_{\Delta}^1$ are the $\Re/\Im$ ratios for the amplitudes
$a_T^1$ and $a_\Delta^1$, respectively.

\begin{figure}[htb]
\begin{center}
\hspace{-1.5em}
\includegraphics[width=0.52\textwidth]{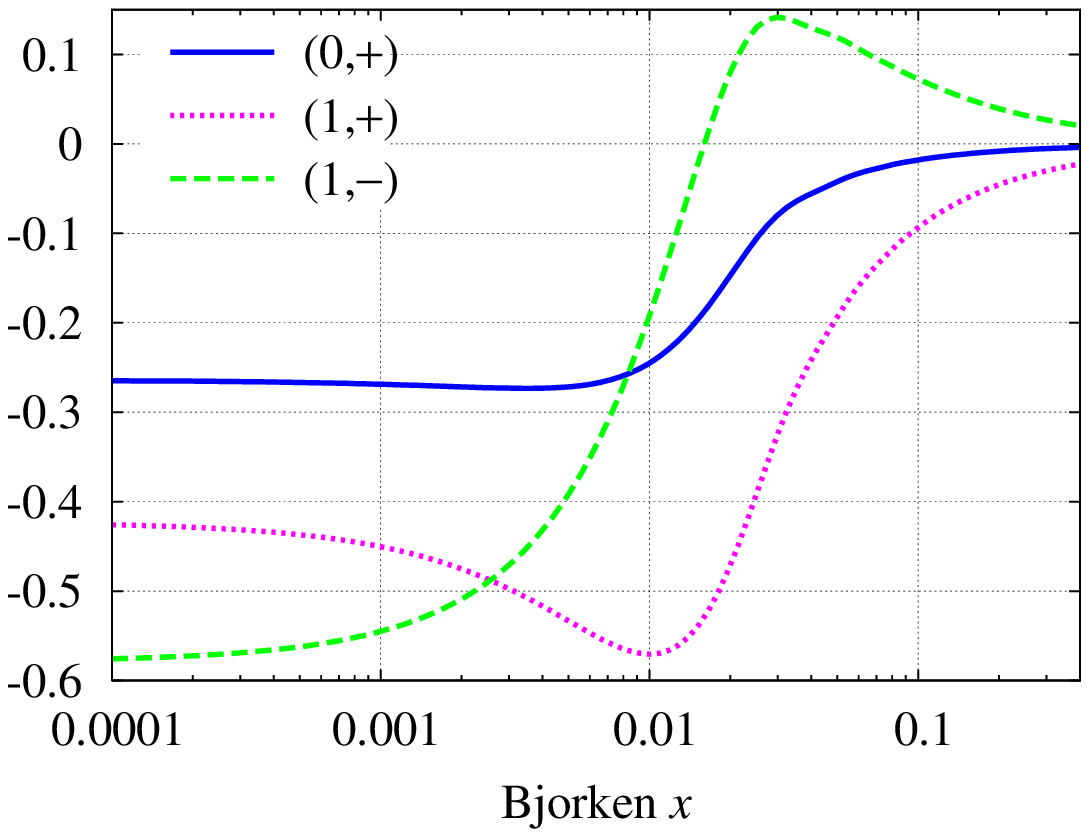}%
\hspace{-1.0em}
\includegraphics[width=0.52\textwidth]{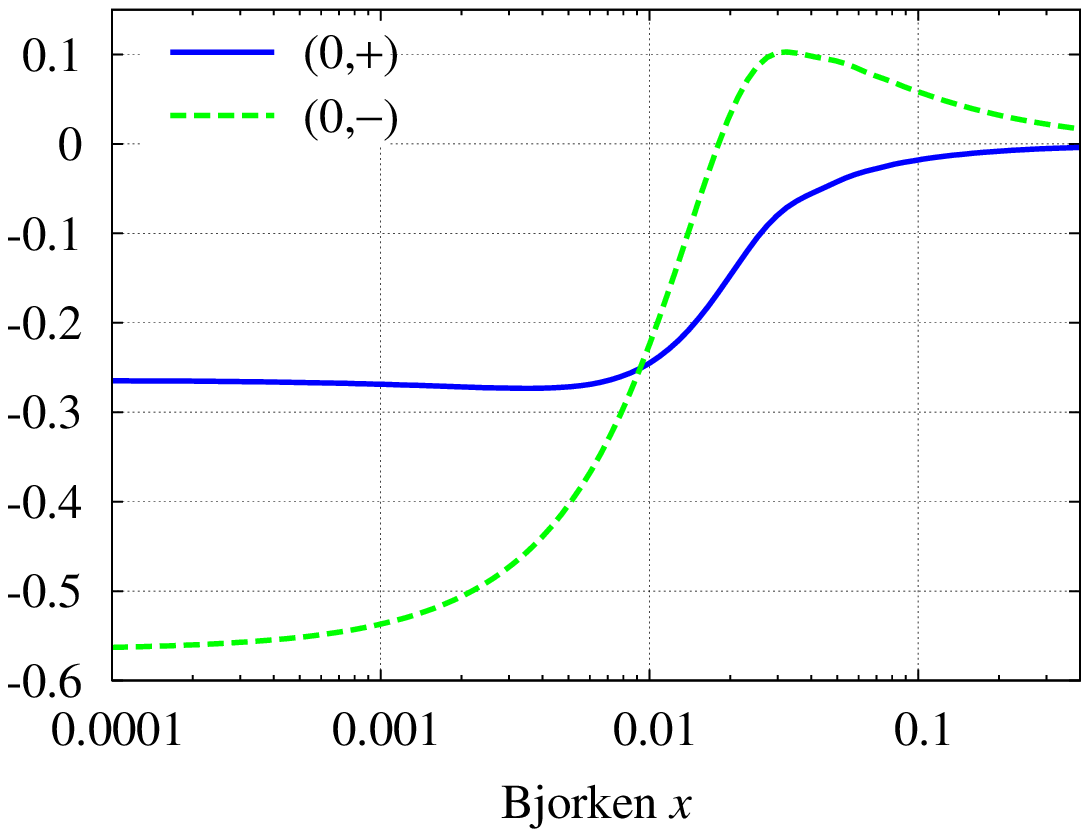}
\caption{%
The ratio $\delta\mathcal{R}^{(I,C)}$ calculated for different isospin and
$C$-parity scattering states for ${}^{207}$Pb at $Q^2=1\gevsq$ using the
parameters of the scattering amplitudes from Sec.~\ref{sec:aeff}.
The labels on the curves mark the values of the isospin $I$ and
$C$-parity, $(I,C)$.
}
\label{fig:deltaRIC}
\shrinkvspace
\end{center}
\end{figure}

Note also that in terms of $(I,C)$ classification
$\delta\mathcal{R}_T^{\nu+\bar\nu} = \delta\mathcal{R}^{(0,+)}$,
$\delta\mathcal{R}_\Delta^{\nu+\bar\nu} = \delta\mathcal{R}^{(0,-)}$,
$\delta\mathcal{R}_{T}^{\nu-\bar\nu}=\delta\mathcal{R}^{(1,-)}$,
and $\delta\mathcal{R}_{\Delta}^{\nu-\bar\nu}=\delta\mathcal{R}^{(1,+)}$.
Figure~\ref{fig:deltaRIC} illustrates the $x$ dependence of the
ratios $\delta\mathcal{R}^{(I,C)}$ for different isospin and $C$-parity
states.


\subsubsection{Coherent nuclear correction to $F_T$ and $F_3$}

We now apply these results to neutrino structure functions. 
We recall that the discussion given above
applies to the $u$- and $d$-quark contribution and make
use of the isospin symmetry relations for the effective amplitude.
In order to take into account the strange quark contribution,
we explicitly separate the ligh-quark ($u$ and $d$) and
the $s$-quark contributions to neutrino structure functions $i=T,3$
\begin{subequations}\label{F:01s}
\begin{align}
\label{F:pl}
F_i^{\nu+\bar\nu} &= F_i^{(\nu+\bar\nu)(0)} + F_i^{(\nu+\bar\nu)(s)},
\\
F_i^{\nu-\bar\nu} &= F_i^{(\nu-\bar\nu)(1)} + F_i^{(\nu-\bar\nu)(s)},
\label{F:mn}
\end{align}
\end{subequations}
where the superscripts indicate the contribution from the ligh quarks
with total isospin 0 or 1, or the $s$-quark contribution. Although this
separation is quite general, we illustrate the terms in \eqs{F:01s}
in the parton model. For the transverse structure function we have
$F_T^{(\nu+\bar\nu)(0)}=x(u+d+\bar u+\bar d)$,
$F_T^{(\nu-\bar\nu)(1)}=x(d-u+\bar u-\bar d)$. The $s$-quark contribution
is relevant for the $C$-even $F_T^{(\nu+\bar\nu)(s)}=x(s+\bar s)$. The
$C$-odd $F_T^{(\nu-\bar\nu)(s)}$ is proportional to $s-\bar s$ and can be
neglected. Similarly, for $xF_3$ we have
$xF_3^{(\nu+\bar\nu)(0)}=x(u+d-\bar u -\bar d)$,
$xF_3^{(\nu-\bar\nu)(1)}=x(d-u-\bar u+\bar d)$, and
$xF_3^{(\nu-\bar\nu)(s)}=x(s+\bar s)$. The $C$-odd
$xF_3^{(\nu+\bar\nu)(s)}$ is similar to $F_T^{(\nu-\bar\nu)(s)}$ and can be
neglected.

In general, the $s$-quark terms should be corrected differently
from those of light quarks. Taking into account \eqs{F:01s} we have for
coherent nuclear correction to the structure functions $F_T$ and $xF_3$ per
one nucleon
\begin{subequations}\label{delFA}
\begin{align}
\label{delFA:0+}
\delta F_T^{(\nu+\bar\nu)A} &= \delta\mathcal{R}^{(0,+)} F_T^{(\nu+\bar\nu)(0)}
+
\delta\mathcal{R}^{(s,+)} F_T^{(\nu+\bar\nu)(s)},\\
\label{delFA:0-}
\delta F_3^{(\nu+\bar\nu)A} &= \delta\mathcal{R}^{(0,-)}F_3^{(\nu+\bar\nu)(0)}, \\
\label{delFA:1-}
\delta F_T^{(\nu-\bar\nu)A} &= \beta\delta\mathcal{R}^{(1,-)}F_T^{(\nu-\bar\nu)(1)}, \\
\label{delFA:1+}
\delta F_3^{(\nu-\bar\nu)A} &= \beta\delta\mathcal{R}^{(1,+)} F_3^{(\nu-\bar\nu)(1)} +
                            \delta\mathcal{R}^{(s,+)} F_3^{(\nu-\bar\nu)(s)},
\end{align}
\end{subequations}
where the structure functions in the right side should be taken for the
free proton and the ratios $\delta\mathcal{R}$ are given by
\Eqs{delR:pl}{delR:mn}.
Note that the $s$-quark contribution in
\eq{delFA:1+} is isoscalar and for this reason it is not suppressed by the factor $\beta$.
{}From \eqs{delFA} we derive coherent nuclear corrections to the
(anti)neutrino structure functions.
We define the ratio
$\delta\mathcal{R}^{\nu}_i (A/N)=\delta F_i^{\nu A}/\delta F_i^{\nu N}$
describing the nuclear correction of the neutrino structure
function of type $i$ in the units of the isoscalar nucleon structure function
$F_i^{\nu N}=(F_i^{\nu p}+F_i^{\nu n})/2$
[and similar definitions for the antineutrino ratio
$\delta\mathcal{R}^{\bar\nu}_i (A/N)$]. We have
\begin{subequations}
\label{delRA}
\begin{align}
\label{delR:nu}
\delta\mathcal{R}^{\nu(\bar\nu)}_T (A/N) &=
   \delta\mathcal{R}^{(0,+)}
\pm
    \beta\delta\mathcal{R}^{(1,-)}
     \frac{F_T^{(\nu-\bar\nu)(1)}}
          {2 F_T^{\nu(\bar\nu)N}}
+
    (\delta\mathcal{R}^{(s,+)}-\delta\mathcal{R}^{(0,+)})
         \frac{F_T^{(\nu+\bar\nu)(s)}}
              {2 F_T^{\nu(\bar\nu)N}},
\\
\label{delR:antinu}
\delta\mathcal{R}^{\nu(\bar\nu)}_3 (A/N) &=
   \delta\mathcal{R}^{(0,-)}
\pm
    \beta\delta\mathcal{R}^{(1,+)}
         \frac{F_3^{(\nu-\bar\nu)(1)}}
              {2\, F_3^{\nu(\bar\nu)N}}
\pm
    (\delta\mathcal{R}^{(s,+)}-\delta\mathcal{R}^{(0,-)})
         \frac{F_3^{(\nu-\bar\nu)(s)}}
              {2\, F_3^{\nu(\bar\nu)N}},
\end{align}
\end{subequations}
where the sign $+(-)$ corresponds to neutrino (antineutrino).
In numerical applications in Sec.~\ref{sec:res}
we will assume similar nuclear effects for $I=0$
light quarks and the $s$-quark,
$\delta\mathcal{R}^{(s,+)}=\delta\mathcal{R}^{(0,+)}$.


\subsubsection{Coherent nuclear corrections to $F_L$ and $F_2$}
\label{sec:MS:L}

We now discuss coherent nuclear corrections to the longitudinal structure
function.
Nuclear modification of the PCAC term is
driven by the corresponding correction to the virtual pion cross
section, and we have $\delta F^{\rm PCAC}_L/F^{\rm PCAC}_L=\delta
\sigma_{\pi A}/(A\sigma_{\pi N})=\delta\mathcal R_\pi(A/N)$.
The nuclear correction to the pion cross section can be analyzed along the
lines discussed in Sec.~\ref{sec:MS}. We consider the isoscalar $\pi
N$ scattering amplitude $a_\pi^0=(a_{\pi^+p}+a_{\pi^-p})/2$ and the
isovector asymmetry $a_\pi^1=a_{\pi^+p}-a_{\pi^-p}$.
Nuclear corrections to these amplitudes are given by equations similar to
(\ref{delR:pl:T}) and (\ref{delR:mn}), and we have
\begin{subequations}\label{delR:pi}
\begin{align}
\delta\mathcal R_\pi^0 &= \sigma_\pi^0 \Re
   \big[
\left(i+\alpha_\pi^0\right)^2 \mathcal C_2^A(a_\pi^0)
   \big]/(2A),
\\
\delta\mathcal R_\pi^1 &= \Im
   \left[
\left(i+\alpha_\pi^1\right)\mathcal T_1^A(a_\pi^0)
   \right]/A,
\end{align}
\end{subequations}
where $\sigma_\pi^0$ is the $\pi N$ cross section in the isoscalar
state and $\alpha_\pi=\Re a_\pi /\Im a_\pi$ for the pion amplitude in the
corresponding isospin state. In numerical applications we calculate
$\delta\mathcal R_\pi$
using $\pi$N scattering amplitude of \cite{piN} (see
Appendix~\ref{sec:piN}).  Since the PCAC relation involves the cross
section of the virtual pion with four-momentum $q$, the longitudinal
momentum transfer in \eq{mult:A} is
$k_L^\pi=Mx(1+m_\pi^2/Q^2)$.

The ratios $\delta\mathcal R_\pi$ for the cross sections of $\pi^+$
and $\pi^-$ mesons can be readily obtained from \eqs{delR:pi}, and we have
\begin{equation}
\delta\mathcal R_{\pi^\pm} =
   \delta\mathcal R_{\pi}^0 \pm
        \left(
        \beta\delta\mathcal R_{\pi}^1-\delta\mathcal R_{\pi}^0
        \right)
   \frac{(\sigma_{\pi^+}-\sigma_{\pi^-})}{2\sigma_{\pi^\pm}},
\end{equation}
where the sign $+(-)$ corresponds to $\pi^+(\pi^-)$ scattering.

The nuclear corrections to the structure function $\widetilde F_L$,
which include the vector-current contribution and non-PCAC terms of the axial current, can be
treated similarly to the transverse structure function $F_T$. In
particular, we assume that the relative correction to $\widetilde F_L$ is
determined by the effective cross section in the longitudinal
state and $\delta\widetilde{\mathcal R}_L =
\delta\widetilde{F}_L^A/\widetilde{F}_L=\delta\sigmabar_L^A/\sigmabar_L$.
Taking into account both contributions, we have for the relative nuclear
correction to $F_L$
\begin{equation}
\delta \mathcal R_L(A/N) =
   r^{\rm PCAC} \delta \mathcal R_\pi(A/N)  +
   (1-r^{\rm PCAC}) \delta \widetilde{\mathcal R}_L(A/N),
\label{delR:L}
\end{equation}
where $r^{\rm PCAC}=\gamma^3 F_L^{\rm PCAC}/F_L$
should be computed for the isoscalar nucleon.

Nuclear correction to $F_2$ can then be calculated in terms of the
corresponding corrections for $F_L$ and $F_T$ using relation
(\ref{hel:123}). We have for the relative correction
\begin{equation}
\delta\mathcal R_2(A/N) =
   \frac{\delta\mathcal R_T^A + R\,\delta\mathcal R_L^A}{1+R},
\label{delR:2}
\end{equation}
where $R=F_L/F_T$, and $F_L$ and $F_T$ are the structure functions
of the isoscalar nucleon.


\subsubsection{Effective scattering amplitude and cross section}
\label{sec:aeff}

We now discuss the model of the effective scattering amplitude for
different isospin and helicity states $a_h^I$ which is used in our
calculations. This amplitude determines the rate of nuclear multiple
scattering corrections as discussed above. In Ref.\cite{KP04} the $C$-even
amplitude $a_T^0$ was studied phenomenologically using charged-lepton
data on the ratios $F_2^A/F_2^{A'}$,
and it was shown that a good description of data for a wide region
of nuclear targets can be achieved using a constant $\alpha_T^0$
and the effective cross section parametrized as
\begin{equation}
\sigmabar_T=\sigma_0/(1+Q^2/Q_0^2).
\label{xsec:ph}
\end{equation}
The values of the parameters were fixed to $\sigma_0=27\:$mb and
$\alpha_T=-0.2$ in order to reproduce the low-$Q^2$ limit
\cite{Bauer:iq}. The scale parameter $Q_0^2$, which describes transition to
high $Q^2$, was determined to be
$Q_0^2=1.43\pm0.06({\rm stat})\pm 0.20({\rm syst})\gevsq$.

In this paper we test the hypothesis that the $C$-even effective amplitude
$a_T^0$ also applies to the (anti)neutrino scattering and determine
$F_T^{\nu+\bar\nu}$.
The other amplitudes ($a_\Delta^0,\ a_T^1,\ a_\Delta^1$)
are not directly constrained by the CL DIS data.
These quantities determine nuclear corrections to different
combinations of neutrino and antineutrino structure functions
as discussed in Sec.~\ref{sec:nuke:coh}.
It is important to note that only Re/Im ratios of these amplitudes
are relevant for the calculation of nuclear corrections to
the (anti)neutrino structure functions.
In order
to evaluate Re/Im ratios we apply the Regge pole approach, which proved
to be very useful in analyses of high-energy hadron scattering (see,
\eg, \cite{Collins:1977jy}), and approximate the
scattering amplitudes by a single Regge pole
with proper quantum numbers. We recall that Re/Im of the Regge
pole amplitude in forward direction is determined by its
intercept $\alpha_R(0)$ and the signature (the latter corresponds to $C$
parity of the scattering amplitude)
$\mathrm{Re/Im}=-(C+\cos\pi\alpha_R(0))/\sin\pi\alpha_R(0)$. The Regge
poles of $\omega$, $\rho$, and $A_2$ mesons have appropriate quantum
numbers (see Table~\ref{tab:regge};
as known from hadron scattering phenomenology \cite{Collins:1977jy}
these Regge trajectories have an intercept close to 0.5).
In our calculations of (anti)neutrino nuclear
scattering discussed below, we use the amplitude $a_T^0$ derived from
the analysis of charged-lepton nuclear data \cite{KP04}.
The Re/Im ratio for the amplitudes $a_\Delta^0$
and $a_T^1$ are fixed from the studies of the Adler
and the Gross--Llewellyn-Smith sum rules for nuclear structure functions
(see Sec.~\ref{sec:ASR} and \ref{sec:GLS}).


\begin{table}[htb]
\shrinkvspace
\begin{tabular}{l|clll}
Amplitude &&          $a_\Delta^0$  & $a_T^1$     & $a_\Delta^1$ \\
\hline\hline
Regge pole $(I^C)$ && $\omega(0^-)$ & $\rho(1^-)$ & $A_2(1^+)$  \\
\hline
Re/Im &&              $1.15\;(1)$   & $1.35\;(1)$ & $(-1)$ \\
\hline
\end{tabular}
\caption{Regge poles contributing to the effective amplitude in
different isospin and $C$ parity states.
Shown are the values of the Re/Im ratio extracted from analysis of nuclear
corrections to the Adler and the GLS sum rules (see
Sec.~\ref{sec:ASR} and \ref{sec:GLS}) and those
calculated with the intercept $\alpha_R(0)=0.5$ common for
all Regge poles (in parenthesis).
\label{tab:regge}
}
\end{table}

The effective cross section $\sigmabar_L$ for longitudinally
polarized intermediate sates, which describes nuclear corrections to
$\widetilde F_L$ (see Sec.~\ref{sec:MS:L}), includes
contributions from the vector and the axial currents. We separate the PCAC
contribution to $F_L$, \eq{FL:PCAC:2}, and treat it explicitly as described
in Sec.~\ref{sec:sf:low-q} and \ref{sec:MS:L}. The remaining part
$\widetilde F_L$ is described by the vector and non-PCAC terms of the axial
current. We assume that $\widetilde F_L$ is similar to $F_L$ in CL
scattering.
In order to quantitatively evaluate $\sigmabar_L$, we assume the relation
$\sigmabar_L/\sigmabar_T=\widetilde{F}_L/F_T=\widetilde R$, which allows
to calculate $\sigmabar_L$ in terms of $\sigmabar_T$ and $\widetilde R$.
This relation follows from (\ref{sum-h-aprx}) if the normalization factors
are independent of helicity. This can be justified in the VMD model at low
$Q^2$ \cite{KP04} and we also apply this relation to the region of high
$Q^2$. For $Q^2>1\gevsq$ for $\widetilde R$ we use
the ratio $F_L/F_T$ calculated using phenomenological PDFs and HT terms of
\cite{a02}. In order to obtain $\widetilde R$ at $Q^2<1\gevsq$ we
extrapolate from large $Q^2$, assuming $\widetilde R\propto Q^2$ at $Q^2\to 0$.

\begin{figure}[htb]
\begin{center}
\hspace{-1.5em}
\includegraphics[width=0.52\textwidth]{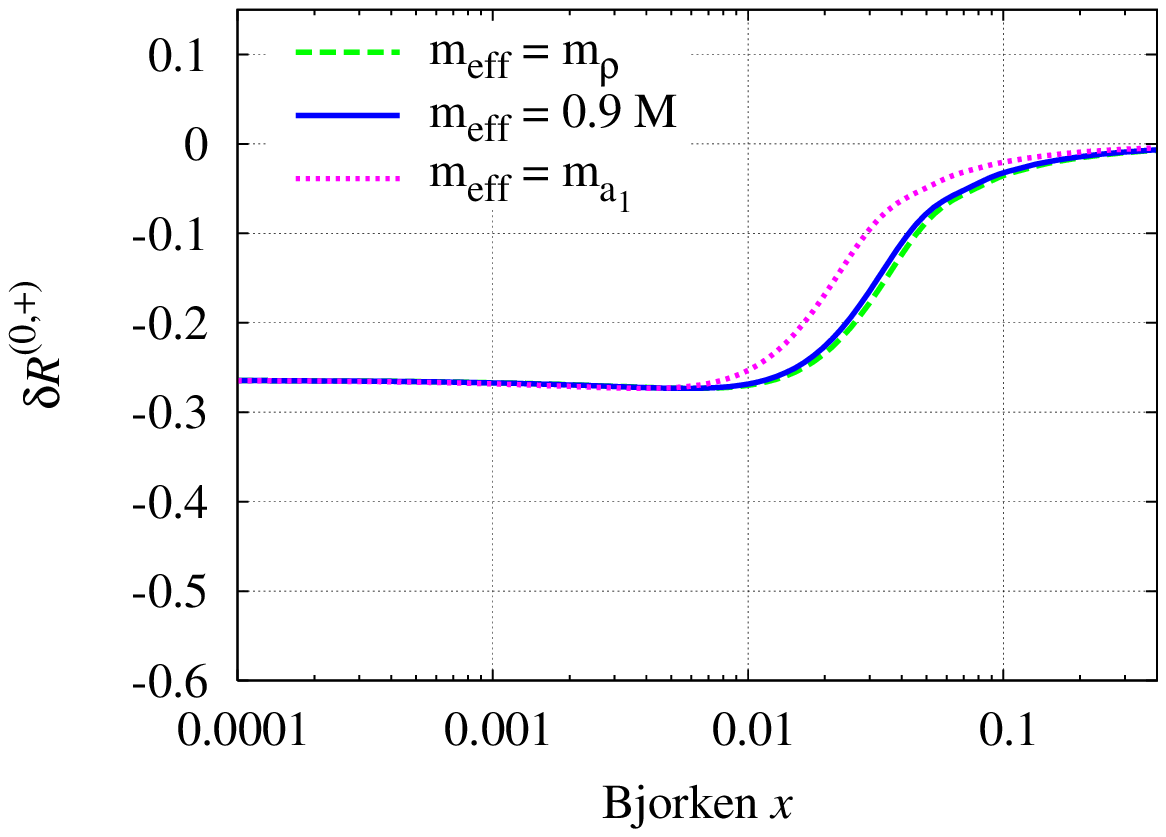}%
\hspace{-0.9em}\includegraphics[width=0.52\textwidth]{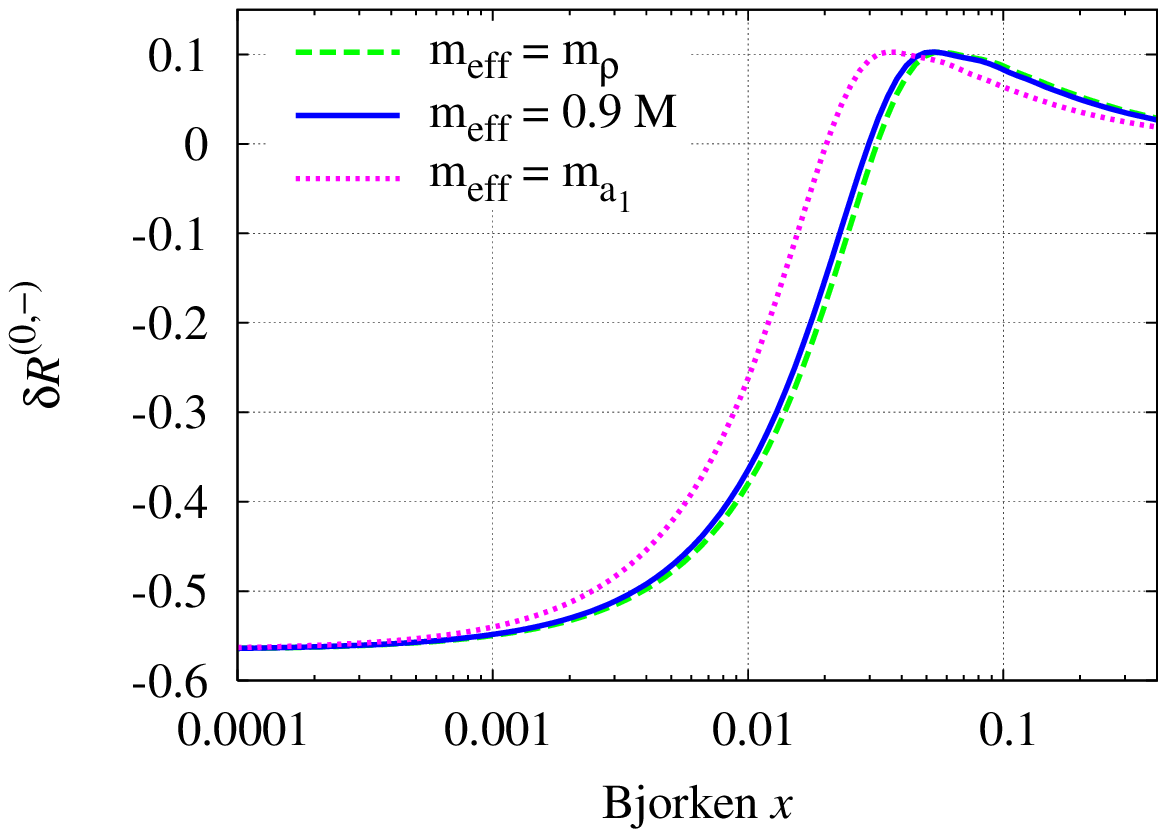}
\caption{%
The ratios $\delta\mathcal{R}^{(0,+)}$ and $\delta\mathcal{R}^{(0,-)}$
calculated for ${}^{207}$Pb at $Q^2=1\gevsq$ using the scattering
amplitudes from Sec.~\ref{sec:aeff} and different values of the parameter
$m_{\rm eff}$.
}
\label{fig:deltaR:mef}
\shrinkvspace
\end{center}
\end{figure}

It should be also commented that the nuclear MS effect in the transition region
of $x$ from 0.01 to 0.1 and $Q^2$ below $1\gevsq$ is affected by the value
of the mass of effective intermediate state $m_{\rm eff}$ through $k_L$
[see \eq{mult:A}]. In our studies of the charged-lepton data we used the
value $m_{\rm eff}=0.9M$ which nicely fits the shadowing data. We also apply
this parameter in the calculation of the vector-current contribution
in neutrino scattering. Apparently, for the axial current this parameter
could be different because of different set of intermediate states. In
order to account for this effect we use a simple
ansatz of rescaling $m_{\rm eff}$ with the ratio of $a_1$- and
$\rho$-meson masses, $m_{\rm eff}m_{a_1}/m_\rho$.
Figure~\ref{fig:deltaR:mef} illustrates the dependence of the ratios
$\delta\mathcal{R}^{(0,+)}$ and $\delta\mathcal{R}^{(0,-)}$ on the
parameter $m_{\rm eff}$. Note that this dependence is noticeable for the
transition region of $0.01<x<0.1$ and for low $Q^2$.


\shrinkvspace
\subsection{The Adler sum rule for nuclei}
\label{sec:ASR}

The Adler sum rule relates the integrated difference of the
isovector combination $F_2^{\nu - \bar\nu}$ to the isospin of the target:
\begin{equation}
\label{ASR}
\sadler=\int_0^{M_A/M}\hspace{-2em}\ud x\ F_2^{\bar\nu -\nu}(x,Q^2)/(2x)
= 2\,I_z,
\end{equation}
where the upper integration limit, the ratio of the nuclear and the
nucleon masses $M_A/M$, is the kinematical maximum of the Bjorken variable
$x=Q^2/(2Mq_0)$ for the nuclear target, and $I_z$ is the
projection of
the target isospin vector on the quantization axis ($z$ axis).
For the proton $\sadler^{p}=1$, and for the neutron $\sadler^{n}=-1$.
In the quark parton model the Adler sum is the difference between the
number of valence $u$ and $d$ quarks of the target. The Adler sum rule
survives the strong interaction effects because of CVC.
However, in the derivation of the Adler sum rule the effects of
nonconservation of the axial current as well as the Cabibbo mixing angle are
neglected (for more detail see, \eg, \cite{IoKhLi84}).

For an isoscalar nucleus the Adler sum rule is trivial since the integrand
in \eq{ASR} vanishes.
For a generic nucleus of $Z$ protons and $N$ neutrons the Adler
sum rule reads (we consider the nuclear structure functions per one
nucleon)
\begin{equation}
\label{ASR:A}
\sadler^A=(Z-N)/A=\beta.
\end{equation}
We now discuss in turn the contributions to $\sadler$ from different
nuclear effects. Let us first consider the Fermi motion and binding (FMB) corrections to
$F_2^{\nu-\bar\nu}$ from \eq{nuke:C-odd}.
We will consider the kinematics of high $Q^2$ and
neglect power $Q^{-2}$ terms in nuclear convolution. By performing the
direct integration by $x$ of \eq{nuke:C-odd} in the case of $F_2$
and using the isospin relations for the proton and neutron structure
functions, we obtain
\begin{subequations}
\begin{align}
\label{ASR:FMB}
\sadler &=
 \beta \int \frac{\ud x}{2x}\average{F_2^{(\bar\nu-\nu)p}}_1 =
       \beta + \delta S_{\rm A}^{\rm OS},
\\
\label{dASR:OS}
 \delta S_{\rm A}^{\rm OS} &=
\beta \int_0^1\frac{\ud x}{2x}\,F_2^{(\bar\nu-\nu)p}(x,Q^2)
                              \delta f_2(x)\average{v}_1,
\end{align}
\end{subequations}
where $\average{v}_1=\average{p^2-M^2}_1/M^2$ is the nucleon virtuality
averaged with the isovector spectral function $\mathcal P_1$. Note that
the FMB correction cancels out in (\ref{ASR:FMB}) in the impulse approximation.
The integral in \eq{dASR:OS} describes the variation of the Adler sum for
the off-shell proton [the structure functions in (\ref{dASR:OS}) should be
taken for the on-shell proton]. In the derivation of the OS correction to
the Adler sum rule (\ref{dASR:OS}) we assume a universal off-shell
function $\delta f_2$, i.e.
common for proton and neutron and independent from the probe (neutrino or antineutrino).

The nuclear pion correction to the Adler sum rule is computed
using \eq{picor:C-odd}. We have
\begin{equation}
\label{dASR:pi}
\delta S_{\rm A}^\pi =
2\left(
       n_{\pi^+/A} - n_{\pi^-/A}
\right),
\end{equation}
where $n_{\pi/A}=\int\ud y\, f_{\pi/A}(y)$ is the average nuclear pion
excess of the given pion type. These quantities also determine the
pion excess correction to the total nuclear charge. From the
charge conservation we have $n_{\pi^+/A} = n_{\pi^-/A}$ and, therefore,
$\delta S_{\rm A}^\pi=0$.

The nuclear shadowing effect in the Adler sum can be computed using the
results of Sec.~\ref{sec:nuke:coh}.
We integrate the shadowing correction $\delta F_2^{\bar\nu-\nu}$ [see
\eqs{delFA}] and obtain
\begin{equation}
\delta S_{\rm A}^{\rm NS} =
\beta \int_0^1\frac{\ud x}{2x}\,F_2^{(\bar\nu-\nu)p}(x,Q^2)
                              \delta\mathcal R_2^{(1,-)},
\label{dASR:NS}
\end{equation}
where $\delta\mathcal R_2^{(1,-)}$ is given in terms of
$\delta\mathcal R_T^{(1,-)}$ and $\delta\mathcal R_L^{(1,-)}$
by \eq{delR:2}.

\begin{figure}[htb]
\begin{center}
\hspace{-2.0em}
\includegraphics[width=0.52\textwidth]{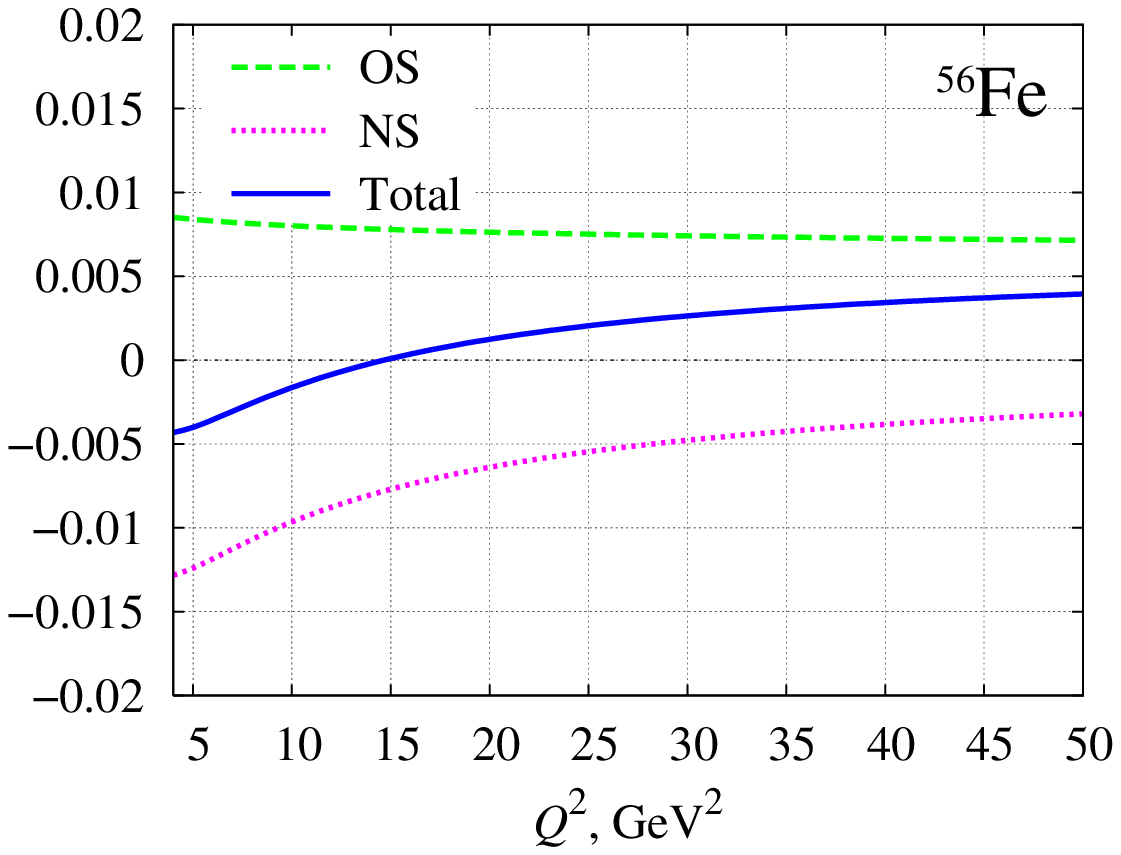}%
\hspace{-0.2em}\includegraphics[width=0.52\textwidth]{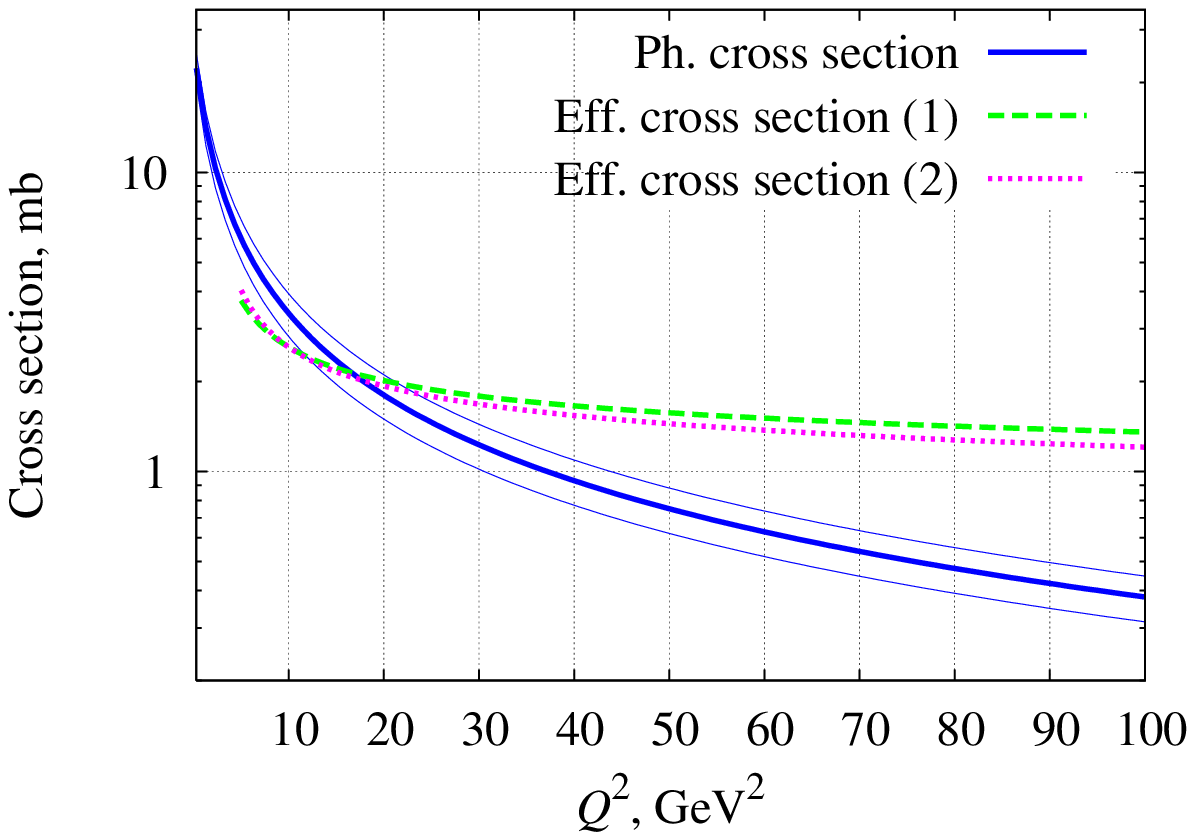}
\caption{%
Relative off-shell (OS), nuclear shadowing (NS) and total nuclear
correction to the Adler sum rule calculated for iron (left panel).
The right panel shows phenomenological cross section by
\eq{xsec:ph} with the error band and effective cross section computed by
requiring exact cancellation between OS and NS effects in nuclear Adler
sum rule (1) and in the normalization of nuclear valence quark
distribution (2)
(see text for more details). }
\label{fig:ASR}
\end{center}
\shrinkvspace
\end{figure}

It follows from  \Eqs{ASR:A}{ASR:FMB} that the total nuclear correction to
the Adler sum rule should vanish. This requirement provides an important
constraint on the isovector part of nuclear structure functions. We
verify that our approach is consistent with the Adler sum rule and
explicitly calculate the OS and NS corrections as a function of $Q^2$
using the off-shell function $\delta f_2$ by \eq{delf:fit} and effective
cross section (\ref{xsec:ph}) derived from analysis of nuclear DIS with
charged leptons \cite{KP04}.
It should be remarked that the quantity $\delta\mathcal R_T^{(1,-)}$ is
also determined by $\alpha_T^1$, the $\Re/\Im$ ratio of the isospin 1
transverse effective amplitude [see \eq{delR:mn}].
This quantity is not constrained by charged-lepton data.
We assume $\alpha_T^1$ to be energy- and $Q^2$-independent.
In order to constrain this parameter we apply the requirement of
minimization of the total nuclear correction to the Adler sum rule
$\delta S_{\rm A}=\delta S_{\rm A}^{\rm OS} + \delta S_{\rm A}^{\rm NS}$.
In particular, we start from $\alpha_T^1=1$, as suggested by simple Regge
arguments in Sec.~\ref{sec:aeff}, and try to adjust this parameter in
order to minimize $\delta S_{\rm A}$ for all $Q^2$. We found
that $\alpha_T^1=1.35$ provides a good cancellation between the OS and NS
corrections (see Fig.~\ref{fig:ASR}).


We also remark that phenomenological cross section (\ref{xsec:ph}) should
be applied in the limited region of $Q^2$. This is because this quantity
was constrained by nuclear shadowing data in DIS which are limited to the
region $Q^2<20\gevsq$ \cite{KP04}.
In order to evaluate the effective cross section at higher values of $Q^2$
we take the requirement $\delta S_{\rm A}=0$ as equation on the cross
section and solve it numerically. The result of calculation for iron is
shown in Fig.~\ref{fig:ASR}. Also shown is the effective cross section
obtained from a similar calculation for the normalization of the nuclear
quark valence number (see Sec.~\ref{sec:GLS}). Both approaches give
consistent results. We also examined this approach for different
nuclei and found similar results for effective cross section at high $Q^2$.

For convenience, we provide a parametrization which reasonably fits
the numerical solution to the effective cross section for $Q^2>4\gevsq$
\begin{equation}
\label{xsec:as}
\sigma^{\rm eff}_T(Q^2)=0.59+18.48 \ln(Q^2/0.37)^{-1.85},
\end{equation}
where the cross section is in mb and $Q^2$ in $\gevsq$.
In applications discussed in Sec.~\ref{sec:res}
for the cross section, we use \eq{xsec:ph} for
$Q^2<15\gevsq$ and \eq{xsec:as} for higher values of $Q^2$.

\subsection{The Gross--Llewellyn-Smith sum rule for nuclei}
\label{sec:GLS}

The Gross--Llewellyn-Smith (GLS) sum is the integrated
structure function $F_3^{\nu+\bar\nu}$,
\begin{equation}
\label{GLS}
\sgls^A = \frac12\int_0^{M_A/M}\hspace{-2em}\ud x\,
                         F_3^{\nu+\bar\nu}(x,Q^2),
\end{equation}
where we write the GLS sum per one nucleon for a generic nuclear
target.  In the quark parton model the GLS sum gives the number of
valence quarks (baryon number) of the target $\sgls^A=3$
\cite{Gross:1969jf}.
However, in QCD the direct relation between the baryon current and $\sgls$
only holds in the leading twist and leading order in $\as$.
In contrast to the Adler sum rule,
$\sgls$ depends on $Q^2$ and is affected by QCD radiative
corrections, target mass, and the higher-twist effects.

In this section we discuss the GLS
integral for nuclear targets. We explicitly separate nuclear corrections to
the GLS integral as $\sgls^A=\sgls^N + \delta \sgls$, where
$\sgls^N$ refers to the GLS integral for the nucleon (proton).
We first consider nuclear corrections to the GLS sum in the LT
approximation.
The correction due to nuclear binding and Fermi motion calcel out in
this approximation as follows from the direct integration of
\eq{FA-3} (see also \cite{Ku98}). The nuclear pion correction to
$F_3^{\nu+\bar\nu}$ also vanishes.
However, both, the nuclear shadowing (NS) and the off-shell (OS)
corrections to GLS, are generally present. We compute the OS correction
to $F_3$
using \Eqs{SF:OS}{delf} and assuming common
off-shell function $\delta f_2(x)=\delta f_3(x)$
(see discussion in Sec.~\ref{sec:OS}).
The NS correction is given by \eq{delFA:0-}. We write these corrections
for high-$Q^2$ kinematics and neglect power terms in the structure
functions
\begin{subequations}\label{GLS:cor}
\begin{align}
\label{GLS:OS}
\delta \sgls^{\rm OS} &= \frac12\int_0^1\hspace{-1ex}\ud x\,
        F_{3,{\rm LT}}^{(\nu+\bar\nu)p}(x,Q^2)
        \delta f(x) \average{v}_0,
\\
\label{GLS:NS}
\delta \sgls^{\rm NS} &= \frac12\int_0^1\hspace{-1ex}\ud x\,
        F_{3,{\rm LT}}^{(\nu+\bar\nu)p}(x,Q^2)
        \delta\mathcal{R}^{(0,-)}_T(x,Q^2),
\end{align}
\end{subequations}
where $\average{v}_0=\average{p^2-M^2}_0/M^2$ is the bound nucleon
virtuality averaged with the isoscalar nuclear spectral function
$\mathcal{P}_0$.

The subscript LT indicates that the structure functions
are calculated in the leading twist approximation in QCD. As mentioned
above, in LO approximation in $\alpha_S$, $\tfrac12 F_{3,{\rm
LT}}^{(\nu+\bar\nu)}$ reduces to the valence quark distribution $u_{\rm
val}+d_{\rm val}$ and the GLS integral gives the baryon number in the
target. For this reason the total nuclear correction to the GLS integral
should cancel out.
The phenomenological OS and NS corrections driven by
\Eqs{delf:fit}{xsec:ph} indeed largely cancel in the GLS integral, as was
shown in Ref.\cite{KP04} (see Fig.~5 in Ref.\cite{KP04}).
It should be remarked that phenomenological cross section effectively
incorporates contributions from  all twists since it is extracted from
data.
Futhermore, at practice the application of \eq{xsec:ph} should be
limited to $Q^2<20\gevsq$ (see also Sec.~\ref{sec:ASR}).
Higher twists are known to be important at low $Q^2$ and
for this reason we should not expect the exact cancellation between
(\ref{GLS:OS}) and (\ref{GLS:NS}) at low $Q^2$.

At high $Q^2$ the LO QCD approximation to structure functions becomes more
accurate and the total nuclear correction to the GLS sum rule should
cancel out for normalization reason. We evaluate the
effective cross section, which provides this cancellation,
from the equation $\delta S_{\rm GLS}=0$.
This equation is solved
numerically using the LO approximation for the structure function $xF_3$
and phenomenological off-shell function by \eq{delf:fit}.

It should be remarked that the quantity $\delta\mathcal R_T^{(0,-)}$,
which determines the nuclear shadowing correction to the GSL sum rule,
also depends on $\alpha_\Delta^0$, the $\Re/\Im$ ratio of the
transverse C-odd and isospin 0 effective amplitude [see
\eq{delFA:0-}], which is not well known.
We assume $\alpha_\Delta^0$ to be independent of energy and $Q^2$.
In order to fix its value we follow an iterative procedure. In
particular, we start from $\alpha_\Delta^0=1$, as suggested by simple
Regge arguments in Sec.~\ref{sec:aeff}, and calculate effective cross
section $\sigma^{\rm eff}_T(Q^2)$ from equation $\delta S_{\rm GLS}=0$.
Then the value of $\alpha_\Delta^0$ is ajusted in order to match
$\sigma^{\rm eff}_T(Q^2)$ at high $Q^2$ with the cross section
calculated from the Adler sum rule in Sec.~\ref{sec:ASR}. We found that
for $\alpha_\Delta^0=1.15$ the two solutions are in a reasonable
agreement. The results are shown in Fig.~\ref{fig:ASR} (right panel)
together with the results of similar calculation using the Adler sum rule.

\section{Numerical applications}
\label{sec:res}

In this section we apply the approach developed in this paper to
calculate SFs and cross sections for the targets relevant to recent
neutrino scattering experiments. In Sec.~\ref{sec:res:pcac} we
present results for the PCAC contribution to the neutrino SF,
in Sec.~\ref{sec:res:sf} we discuss the results for heavy
nucleus/nucleon ratios of different structure functions, in
Sec.~\ref{sec:res:sr} we deal with the numerical analysis of DIS sum rules for
nuclear targets, and in Sec.~\ref{sec:res:xsec} we present our results for
nuclear differential cross sections.

\subsection{Low-$Q^2$ limit}
\label{sec:res:pcac}

\begin{figure}[p]
\begin{center}
\shrinkvspace
\epsfig{file=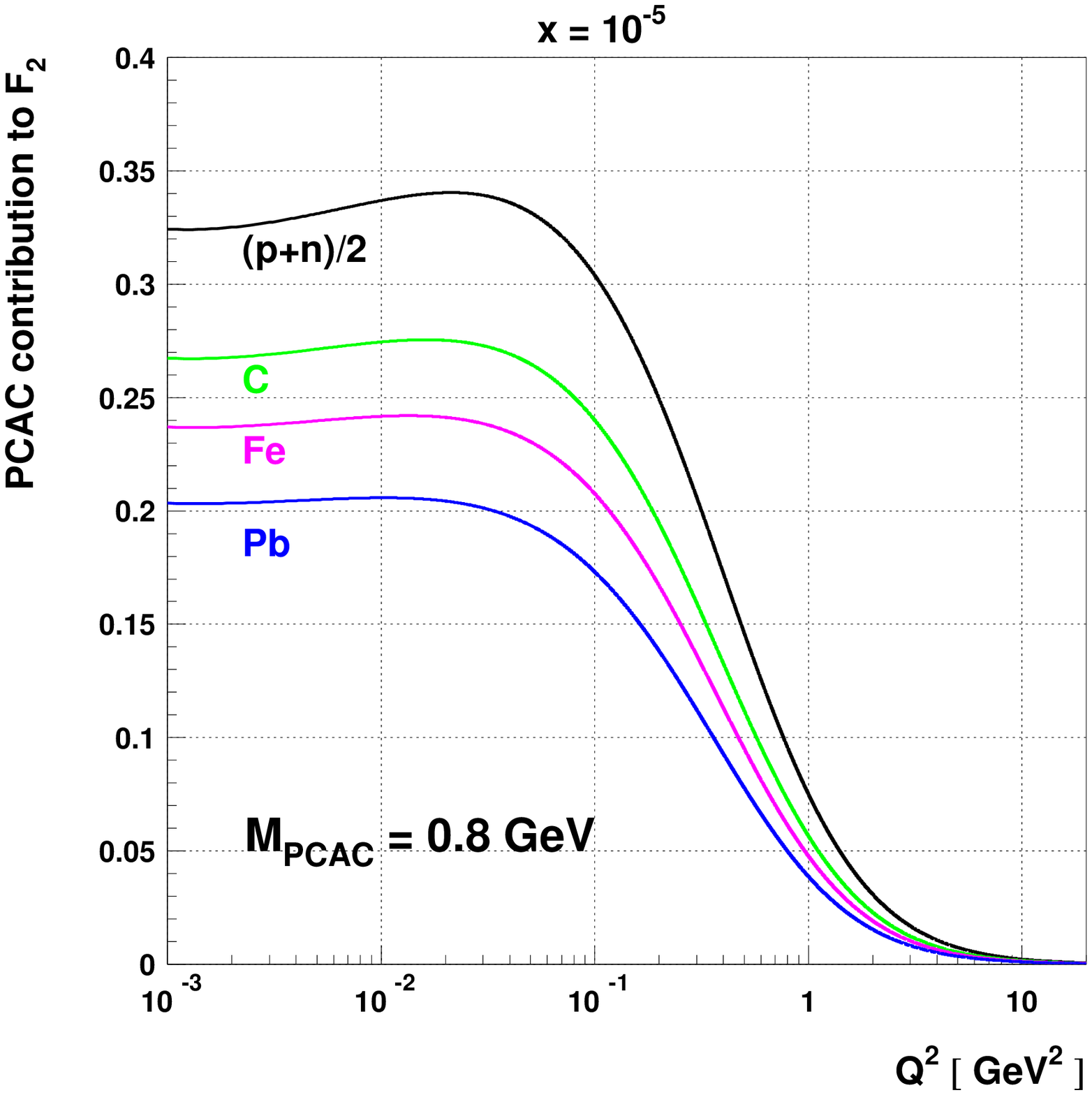,width=0.51\textwidth}%
\vspace{-1ex}\epsfig{file=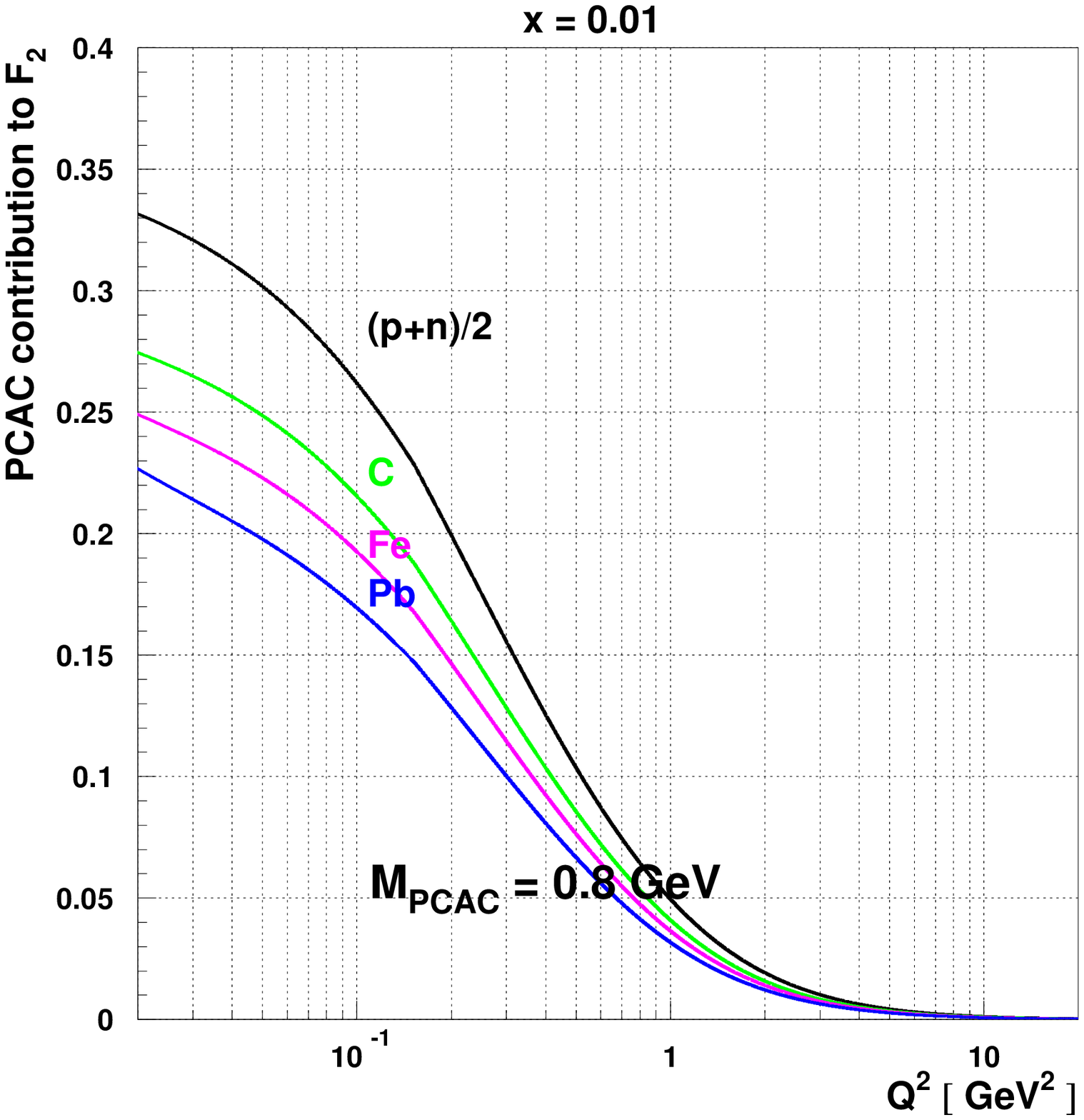,width=0.51\textwidth}
\caption{%
The PCAC term of the neutrino structure function $F_2$ ($\gamma
F_{L}^{\textsc{pcac}}$) calculated for $x=10^{-5}$ (left plot) and
$x=10^{-2}$ (right plot) as a function of $Q^2$ for a few different targets
(labels on the curves). A value $M_{\textsc{pcac}}=0.8$ GeV is assumed
(see Section~\ref{sec:res}).
}
\label{fig:pcacF2}
\epsfig{file=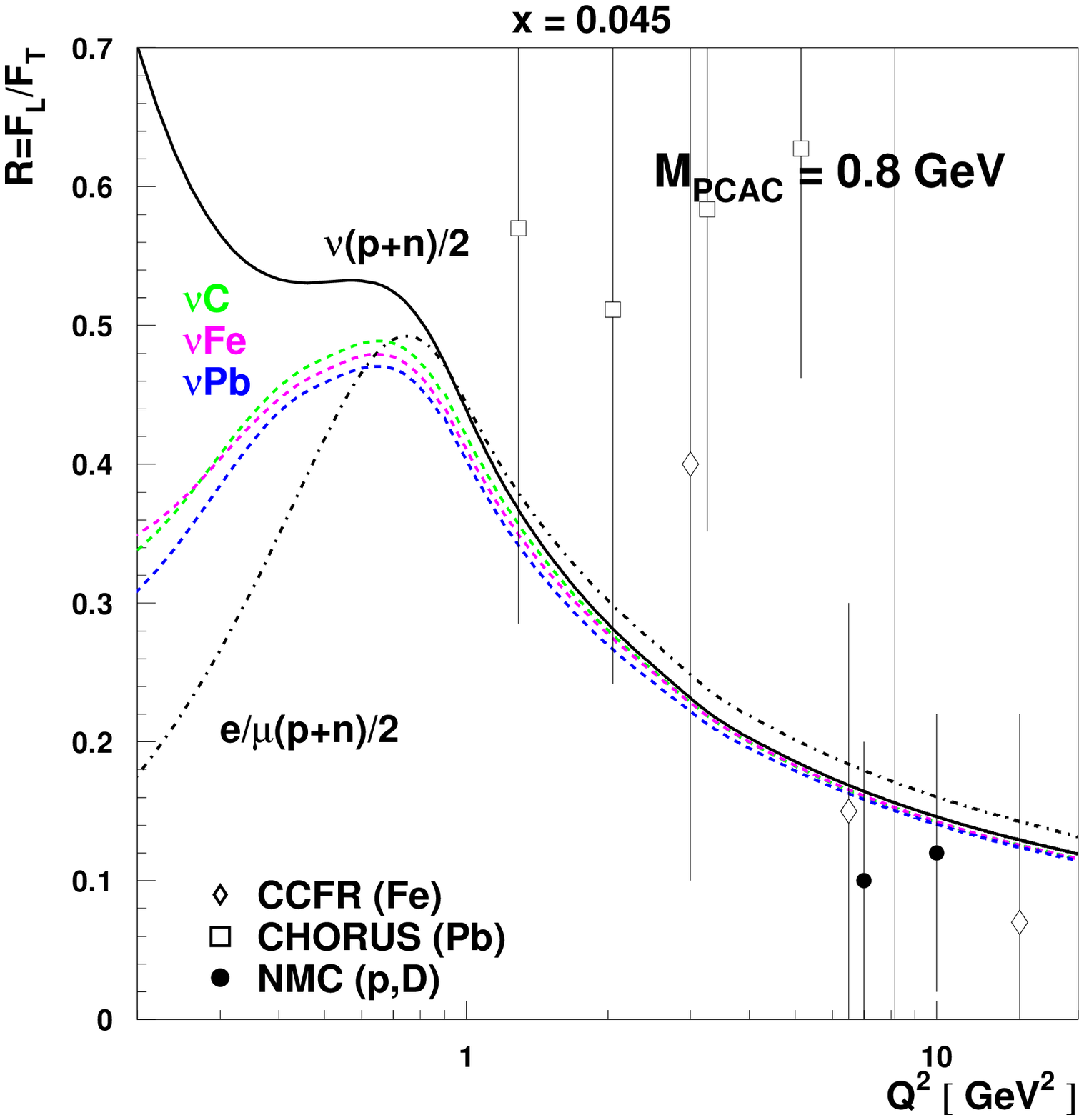,width=0.51\textwidth}%
\epsfig{file=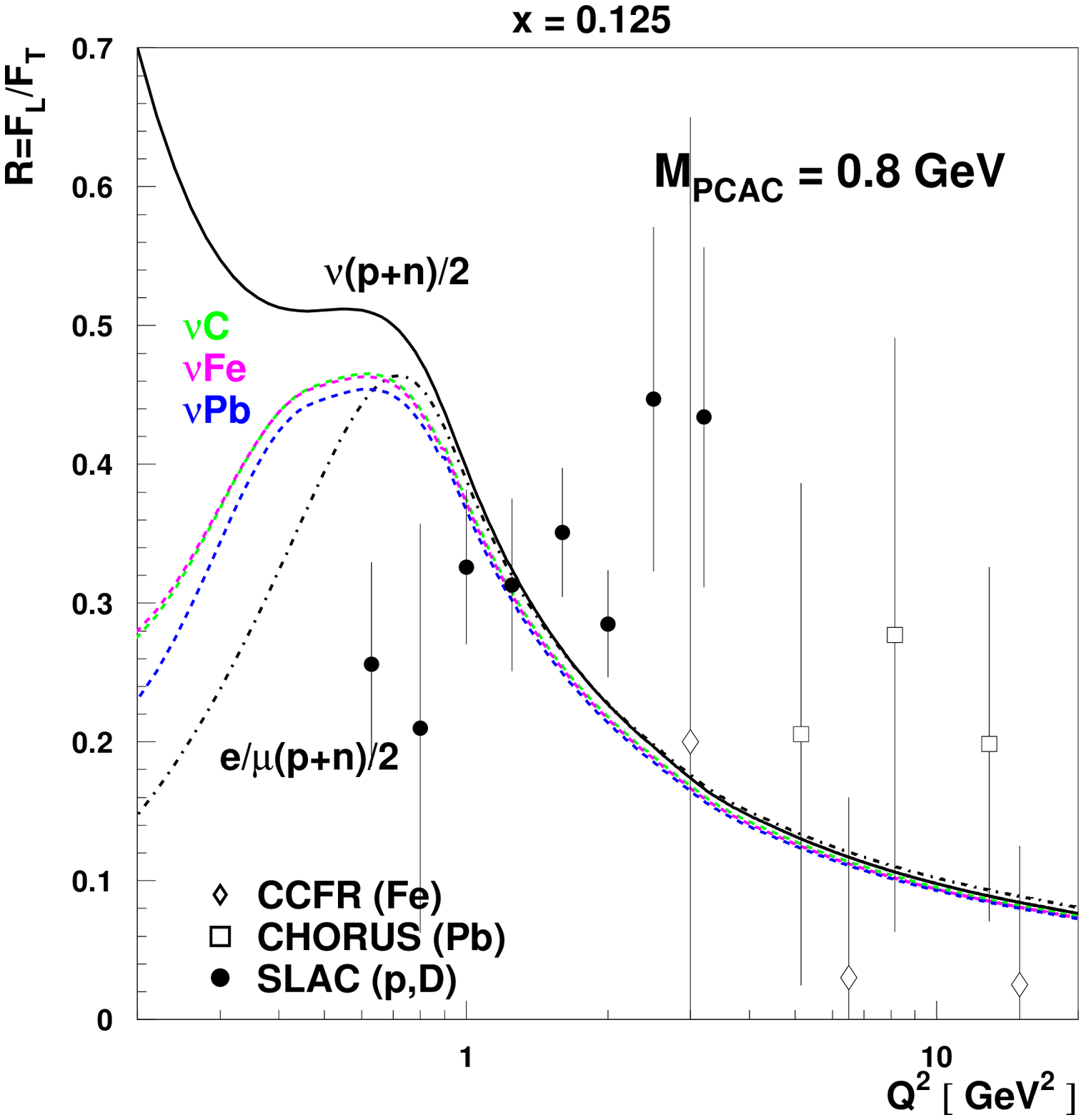,width=0.51\textwidth}
\caption{%
Comparison of the ratio $R=F_L/F_T$ calculated for the isoscalar
nucleon for the charged lepton (dashed-dotted line) and neutrino (solid
line) cases at fixed values of $x$. The left panel corresponds to
$x=0.045$, and the right panel to $x=0.125$. Also shown are the results for
different nuclear targets (${}^{12}$C, ${}^{56}$Fe and ${}^{207}$Pb from
top to bottom). A value $M_{\textsc{pcac}}=0.8$~GeV is assumed (see
Section~\ref{sec:res} for more details). Determinations from
SLAC~\cite{Whitlow:1990gk}, CCFR~\cite{Yang:2001xc} and CHORUS~\cite{chorus-xsec}
are given for comparison.
}
\label{fig:pcacR}
\shrinkvspace
\end{center}
\end{figure}

Figure~\ref{fig:pcacF2} illustrates the magnitude of the PCAC
contribution to $F_2$ for the nucleon and a few nuclear targets
calculated by \eq{FL:PCAC:2}.
In this calculation we use the Regge parameterization of $\pi N$
forward scattering amplitude and the total pion-nucleon cross section
of Ref.\cite{piN} (see Appendix~\ref{sec:piN}).
Note that the values of $F_2$ for heavy nuclear targets are
systematically smaller because of the nuclear shadowing effect for the
pion cross section (see Sec.~\ref{sec:nuke:coh}).
In Table~\ref{tab:F2Q0} we list the values of $F_2$ corresponding
to the limit $Q^2\to0$ and $x\to0$ for the same targets as in
Fig.~\ref{fig:pcacF2}. Our results are consistent with the value extracted
by the CCFR experiment on an iron target~\cite{CCFR-PCAC}.


\begin{table}[htb]
\begin{center}
\begin{tabular}{l|llll|l}
\hline
Target &\ $\tfrac12$(p+n) &\ ${}^{12}$C &\ ${}^{56}$Fe &\ ${}^{207}$Pb  &\ CCFR (${}^{56}$Fe) \\ \hline
$F_2$  &\ 0.325   &\ 0.268      &\ 0.235       &\ 0.204    &\   $0.210 \pm 0.02$\\ \hline
\end{tabular}
\caption{%
The value of $F_2$ in the limit of vanishing $Q^2$ 
for neutrino interactions on different targets. The numbers are extracted
at $x=10^{-5}$. The determination from CCFR~\cite{CCFR-PCAC} is also given for comparison.
\label{tab:F2Q0}
}
\shrinkvspace
\end{center}
\end{table}

The PCAC term determines the low-$Q^2$ limit of $F_L$, and $F_T$ should
vanish as $Q^2$ in the limit of $Q^2\to 0$.
In order to describe the structure functions in intermediate region, we
apply a smooth interpolation between the
high $Q^2$ regime, which is described in QCD in terms of LT and HT
contributions, as discussed in Sec.~\ref{sec:sf:high-q}, and the $Q^2\to0$
predictions derived from CVC and PCAC arguments.
We use $Q^2_{m} = 1\gevsq$ as the matching point between high- and
low-$Q^2$ regimes.

It is instructive to compare the low-$Q^2$ behavior of
$R=F_L/F_T$ for charged-lepton and neutrino scattering. In both cases
$F_T\propto Q^2$ as $Q^2\to0$. However, if $F_L\propto Q^4$ for the
electromagnetic current, for the weak current $F_L\to \FLpcac$, and thus
$F_L$ does not vanish in the low-$Q^2$ limit.
Then the behavior of $R$ at $Q^2\ll 1\gevsq$ is very different for
charged-lepton and neutrino scattering. In order to illustrate this effect
we calculate $R$ as a function of $Q^2$ for two different $x$ (the
$x$-bins of neutrino data \cite{nomad-xsec,nutev-xsec,chorus-xsec}) for
the isoscalar nucleon and a number of nuclei.
The results are shown in Fig.~\ref{fig:pcacR}.

\subsection{Nuclear structure functions}
\label{sec:res:sf}

\begin{figure}[p]
\begin{center}
\shrinkvspace
\epsfig{file=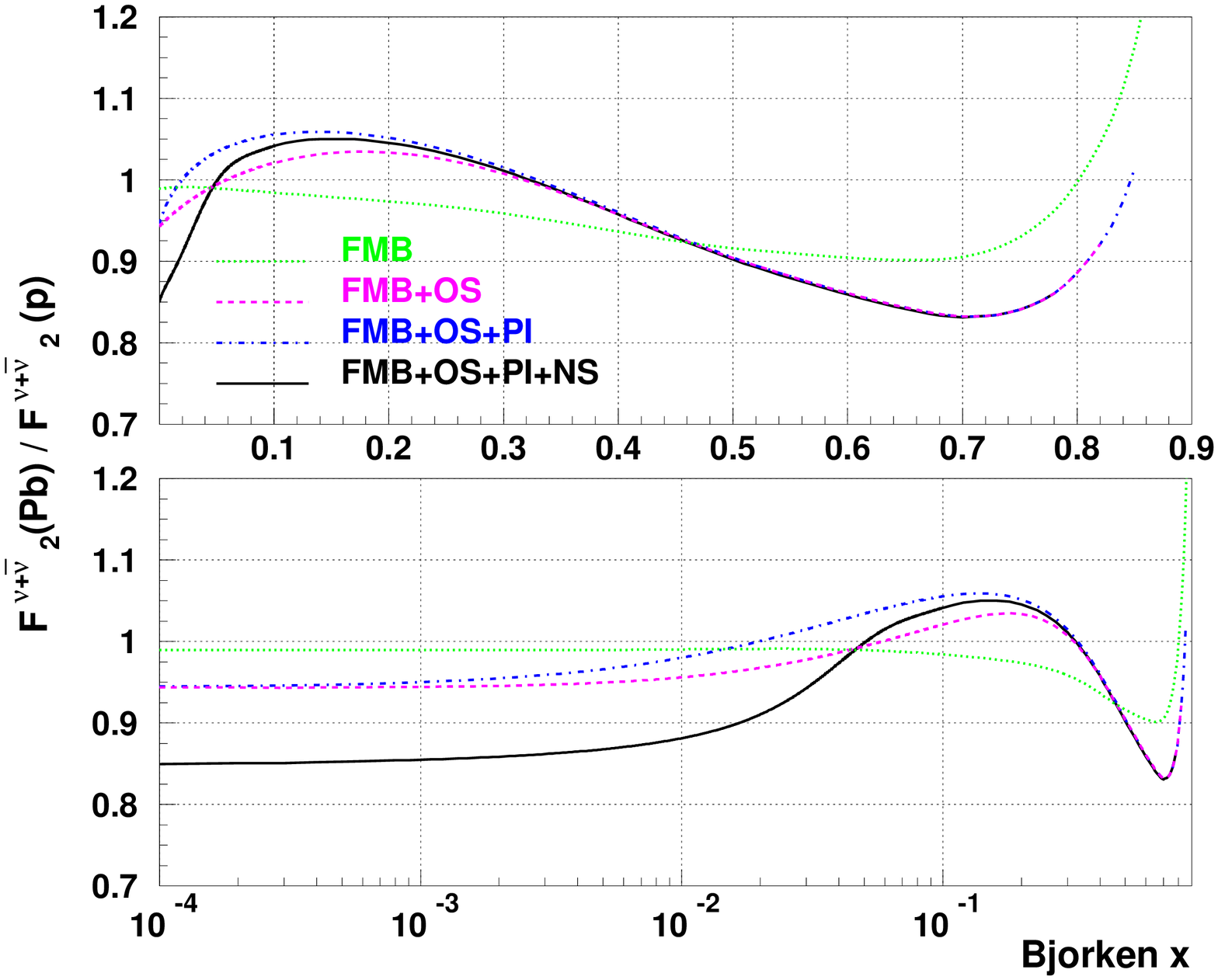,width=12.0cm}
\caption{%
Different nuclear effects on the ratio
$\tfrac{1}{A}F_2^{(\nu+\bar{\nu})A}/F_2^{(\nu+\bar{\nu})p}$ calculated for
${}^{207}$Pb at $Q^2=5\gevsq$. The labels on the curves
correspond to effects due to Fermi motion and nuclear binding (FMB),
off-shell correction (OS), nuclear pion excess (PI) and coherent nuclear
processes (NS).
}
\label{fig:F2sum}
%
%
\epsfig{file=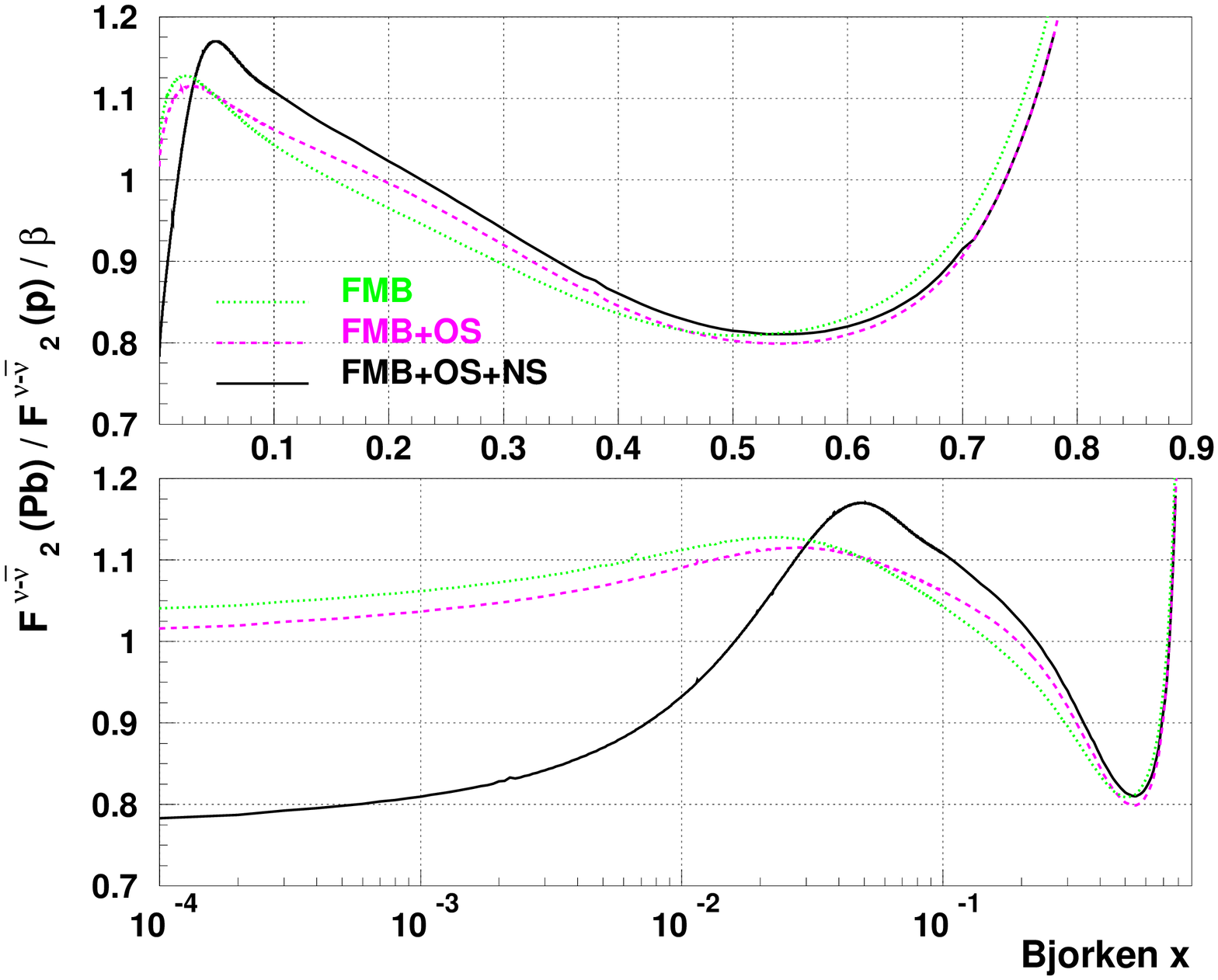,width=12.0cm}
\caption{%
Different nuclear effects on the ratio
$\tfrac{1}{A}F_2^{(\nu-\bar{\nu})A}/(\beta F_2^{(\nu-\bar{\nu})p})$ calculated for
${}^{207}$Pb at $Q^2=5\gevsq$. The labels on the curves
correspond to effects due to Fermi motion and nuclear binding (FMB),
off-shell correction (OS) and coherent nuclear processes (NS).
}
\label{fig:F2diff}
\shrinkvspace
\end{center}
\end{figure}

Figure~\ref{fig:F2sum} shows the result of calculation of 
the $C$-even structure function $F_2^{\nu + \bar{\nu}}$ of ${}^{207}$Pb.
The resulting EMC effect resembles that of CL $F_2$ (see Ref.\cite{KP04} for more details).
At large $x$ the nuclear correction is driven by FMB and OS effects. We
recall that the off-shell effect, which modifies the structure
functions of bound nucleon, is important to change the slope of the
EMC ratio, bringing it close to data.
At intermediate $x$ values we observe some
cancellation between the nuclear pion excess and the nuclear shadowing
correction. The latter is the dominant effect for $x<0.05$, providing a
suppression of the nuclear structure function.

The corresponding nuclear
corrections for the difference $F_2^{\nu - \bar{\nu}}$ are shown in
Fig.~\ref{fig:F2diff} after normalizing them to $\beta = (Z-N)/A$.
There are two distinct
contributions to the difference $F_2^{\nu - \bar{\nu}}$ for nuclear
targets [see Eq.(\ref{nuke:C-odd})]. The pure isovector contribution
[second term in Eq.(\ref{nuke:C-odd})] is shown in
Fig.~\ref{fig:non0}. A comparison with Fig.~\ref{fig:F2diff} indicates that
the dominant contribution to the difference $F_2^{\nu - \bar{\nu}}$
for nuclei is actually coming from the isoscalar part [first term in
Eq.(\ref{nuke:C-odd})], which is due to the Cabibbo mixing
and the heavy quark production effects.

It must be also noted that coherent nuclear effects
are particularly pronounced in the $C$-odd channel, as can be seen from
Fig.~\ref{fig:F2diff} (the shadowing effect at small $x$). The
antishadowing effect at $x\sim 0.05$ is a combined effect due to nuclear
smearing (FMB) and a constructive interference in the coherent nuclear
correction due to the real part of the effective amplitude.


\begin{sidewaysfigure}[p]
\begin{center}
\epsfig{file=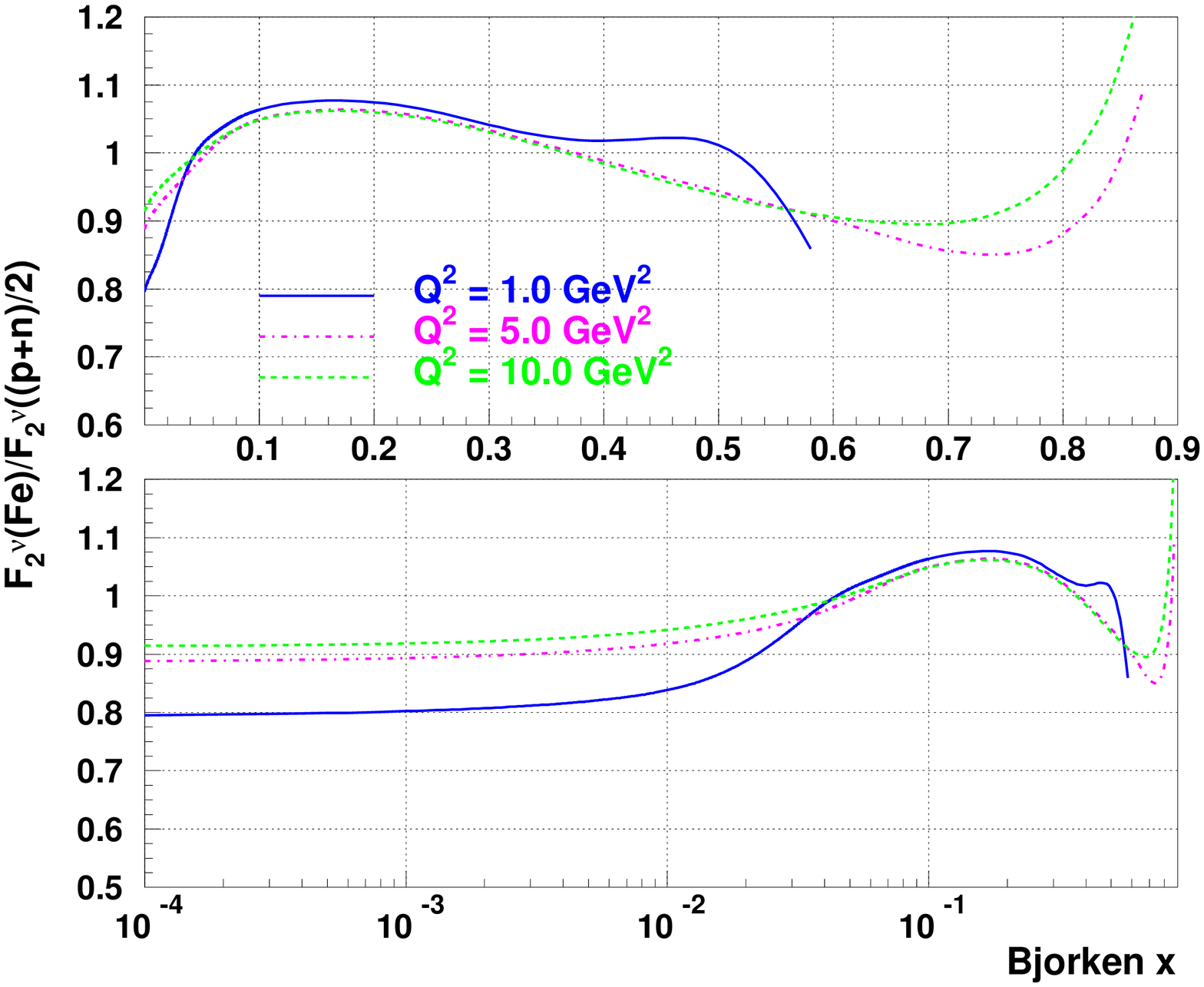,width=10.0cm}%
\epsfig{file=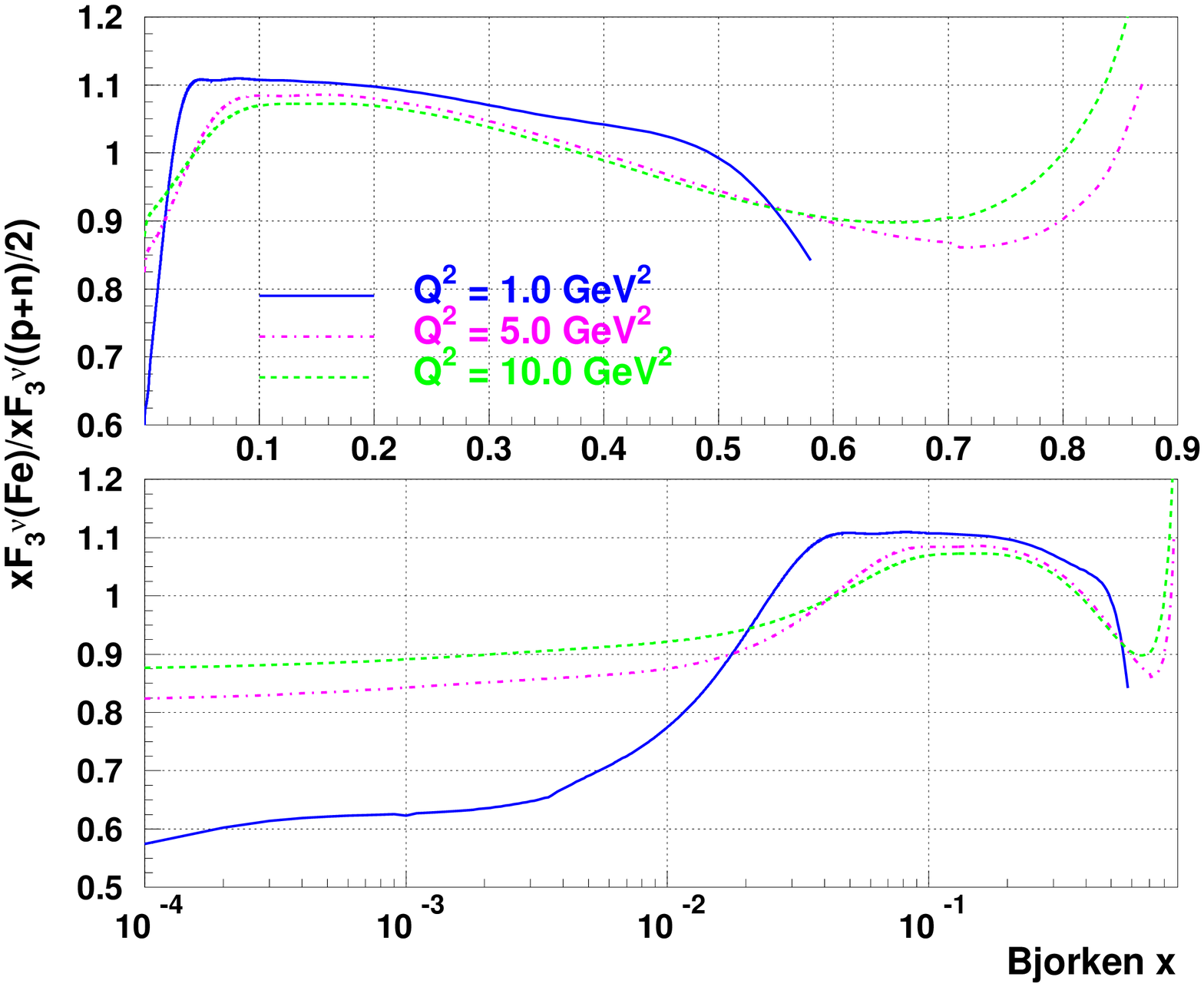,width=10.0cm}
\epsfig{file=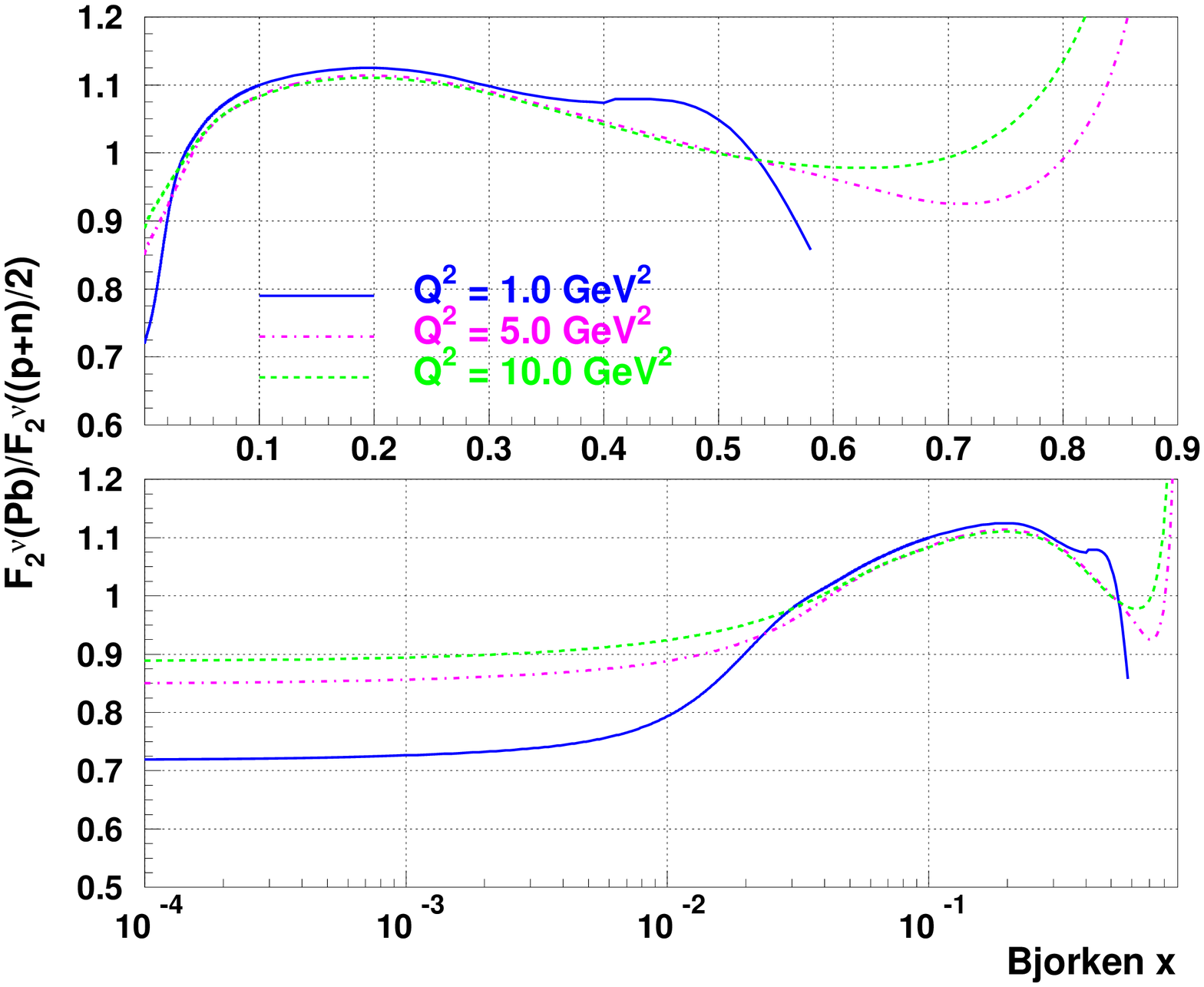,width=10.0cm}%
\epsfig{file=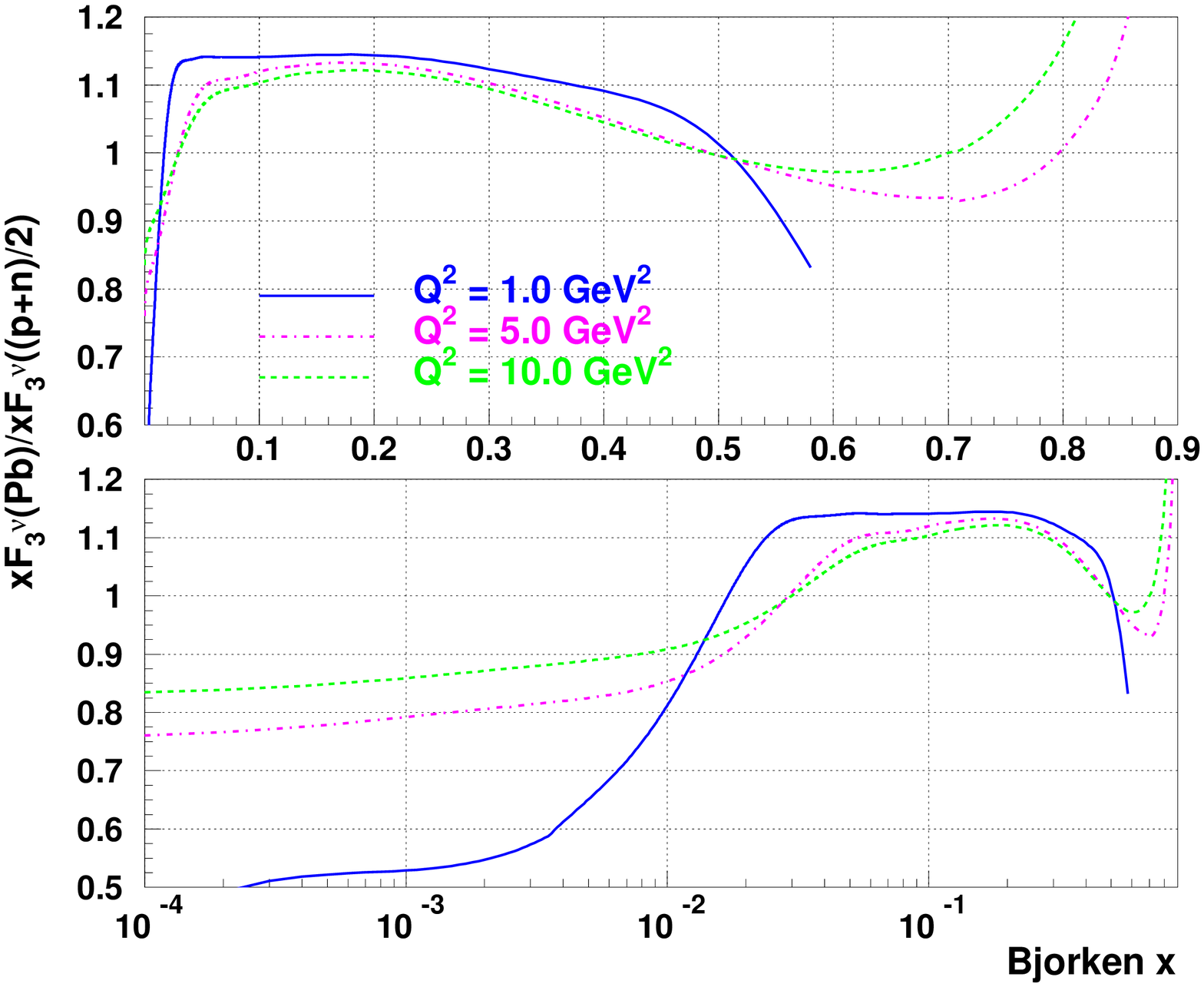,width=10.0cm}
\caption{%
Our predictions for the ratios of CC neutrino nuclear structure
functions normalized to one nucleon and those of the isoscalar nucleon
$(p+n)/2$ (left panel for $F_{2}$ and right panel for $xF_{3}$). The ratios
were calculated for ${}^{56}$Fe (upper panels) and ${}^{207}$Pb (lower
panels) as a function of $x$ at $Q^2=1,\ 5$, and $10\gevsq$.
}
\label{fig:nuratios2}
\end{center}
\end{sidewaysfigure}

Using the approach described in previous sections we calculate the ratios
$\mathcal{R}_2^\nu=F_2^{\nu A}/F_2^{\nu (p+n)/2}$ and
$\mathcal{R}_3^\nu=F_3^{\nu A}/F_3^{\nu (p+n)/2}$ for CC neutrino interactions for
the Fe and Pb targets. The results are shown in Fig.~\ref{fig:nuratios2}.
The bulk behavior of $\mathcal R_2^\nu$ is similar to that of $\mathcal
R_2^\mu$ in CL scattering (see \cite{KP04}) for both the $x$ and
$Q^2$ dependencies.
Note that for heavy nuclei with an
excess of neutrons over protons $\mathcal R_2^\nu > \mathcal R_2^\mu$ for
$x>0.1$. This is because the neutron excess correction is positive for
the neutrino case, while this correction is negative for CL scattering. Indeed,
this correction is given by the second term in \eq{nuke:FA}. Because
$F_2^{\mu(p-n)} > 0$ and $F_2^{\nu(p-n)} < 0$, we obtain the different sign
of the neutron excess correction in these cases. It should be also
remarked that this correction is negative for antineutrino SFs, similar to
the CL case. As a result there is a cancellation for the neutron excess
correction in the sum $F_2^{\nu + \bar{\nu}}$, as can be seen in
Fig.~\ref{fig:F2sum}. It is interesting to note that in the antineutrino
case different nuclear effects cancel to a high degree in the region $x\sim
0.1$ (see also Fig.~\ref{fig:xsec1}).

It is also instructive to compare nuclear effects for neutrino SFs $F_2$
and $xF_3$.
One observes from Fig.~\ref{fig:nuratios2} that at large $Q^2$
($=10\gevsq$) the ratios $\mathcal R_2^\nu$ and $\mathcal R_3^\nu$ turn
out to be similar (although not identical) in the entire kinematical region of $x$.
We recall in this context that the effect of coherent nuclear interactions
at small $x$ is quite different for $F_2$ and $xF_3$ (see
Sec.~\ref{sec:nuke:coh}). Futhermore, the nuclear pion correction to
$xF_3$ cancels out, and the antishadowing effect for $F_2$ and $xF_3$ is
generated by different mechanisms (the combined effect of the real part of
the $C$-odd effective amplitude and the off-shell correction for $xF_3$
and the interplay between pion and off-shell effect for $F_2$).
Nevertheless, at high $Q^2$ the total nuclear correction to $xF_3$ is
very similar to that for $F_2$.

The differences between $\mathcal R_2$ and $\mathcal R_3$ are more
pronounced at lower $Q^2$, indicating different $Q^2$ dependence of nuclear
effects for these structure functions. The shadowing and antishadowing
effects for $F_3$ become stronger in this region as illustrated in
Fig.~\ref{fig:nuratios2}.

Also one should remark that the target mass correction to the structure
functions strongly affects the EMC ratio in the region of large $x$ and
low $Q^2\sim 1\gevsq$. This effect causes a noticeable $Q^2$ dependence
of the EMC ratio at $x>0.6$.
We also note that in Fig.~\ref{fig:nuratios2} the curves for $Q^2=1\gevsq$
are given for the region $W > 1.25\gev$ and for this reason the region
$x>0.6$ is not shown.

\subsection{Neutrino DIS sum rules}
\label{sec:res:sr}

As discussed in Sections~\ref{sec:ASR} and
\ref{sec:GLS}, we observe a remarkable cancellation between the off-shell
and the nuclear shadowing corrections. This cancellation, which becomes
more accurate at high $Q^2$ where the leading twist dominates, can be
attributed to underlying symmetry through the conservation of the isospin
(Adler) and the nuclear valence quark number (GLS) in nuclear targets.
Assuming exact cancellation we evaluate the effective
cross-section describing the nuclear shadowing at high-$Q^2$.
Both the Adler and GLS sum rules can be used for this purpose, providing
consistent results. In this paper we choose the Adler integral since the
corresponding isospin relation is accurate to all orders. Furthermore, the
evaluation of the Adler sum rule for nuclear targets involves the
off-shell function $\delta f_2$, which was determined from charged-lepton DIS data.

\begin{figure}[htb]
\begin{center}
\shrinkvspace
\epsfig{file=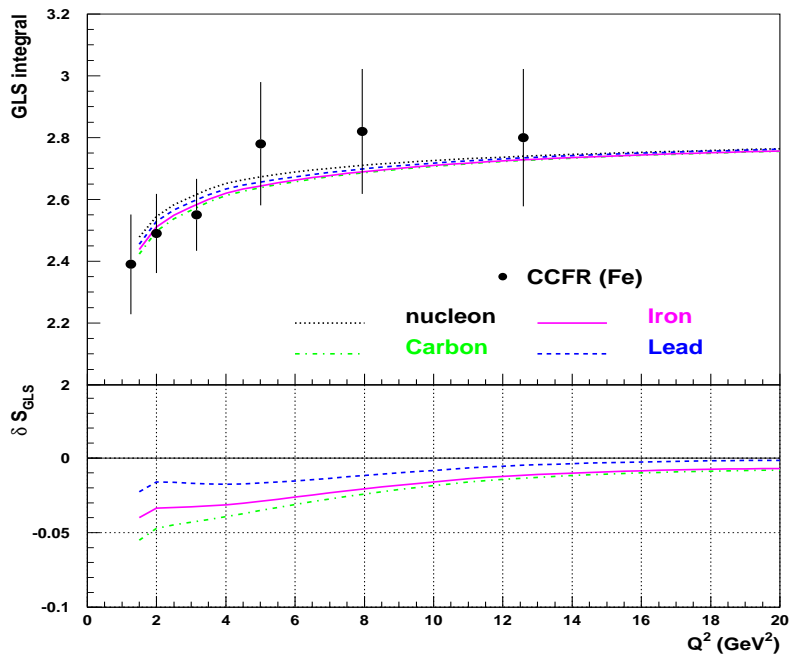,width=0.7\textwidth, height=0.6\textwidth}
\caption{%
The GLS integral for different nuclear targets as a
function of $Q^2$. The nucleon $xF_3$ is calculated in LT NNLO
approximation with the target mass correction as described in
Sec.~\ref{sec:sf:high-q}. The bottom panel shows variations with respect to
the result obtained for average isoscalar nucleon. Data points are the
CCFR extraction of the GLS integral for iron \cite{CCFR:GLS}.
}
\label{fig:gls}
\shrinkvspace
\end{center}
\end{figure}

We can then evaluate the GLS integral $\sgls(Q^2)$ for different nuclear targets.
Results are shown in Fig.~\ref{fig:gls}. The nucleon
integral $\sgls^N(Q^2)$ is calculated using the NNLO coefficient functions
and the NNLO PDFs of~\cite{a02}. We observe that the nuclear correction
$\delta \sgls$ decreases progressively
by increasing $Q^2$ and ranges from about 2\% at $Q^2=2\gevsq$ to less
than 0.3\% at $Q^2=20\gevsq$. The results for different targets are
similar and noticeable differences are present only at the lowest $Q^2$
values. We also note the general $Q^2$ dependence for ${}^{56}$Fe is in
agreement with the CCFR measurement~\cite{CCFR:GLS}. The values of
$\sgls^N$ and $\delta \sgls$ are listed in Table~\ref{tab:gls} for a
number of nuclear targets and fixed value of $Q^2=10\gevsq$.


\begin{table}[hbt]
\begin{center}
\begin{tabular}{lc|clc|ccc|cc}
Target &&& Comment &&& $\sgls$ &&& $\delta \sgls$  \\ \hline\hline
p &&& free proton &&& $2.726$ &&& 0 \\ \hline
D &&& full calculation &&& $2.717$ &&& $-0.009$ \\ \hline
${}^{12}$C &&& full calculation &&& $2.707$ &&& $-0.019$ \\ \hline
${}^{56}$Fe &&& full calculation &&& $2.710$ &&& $-0.016$ \\ \hline
${}^{207}$Pb &&& full calculation &&& $2.717$ &&& $-0.009$ \\ \hline
\end{tabular}
\caption{
The GLS sum rule calculated for different nuclei at
$Q^{2}=10\gevsq$. The nucleon structure functions were calculated in LT
NNLO approximation using the PDFs of Ref.\cite{a02}. We use the lower
cutoff $x_{\rm min}=10^{-5}$ in the calculation of the GLS integral.
\label{tab:gls}
}
\shrinkvspace\shrinkvspace
\end{center}
\end{table}

Power corrections due to target mass effect are included in the
calculation. Apart from the target mass effect the dynamical power
corrections should generally be present. The contribution of such terms to
the GLS integral was evaluated in \cite{BraunKolesnichenko:1987} using QCD
sum rule approach, predicting a negative correction of $-0.1$ at
$Q^2=3\gevsq$.
We also comment that phenomelogical studies of HT terms in
$F_3$ using CCFR data were reported in \cite{kataev} with, however,
high statistical uncertainties. In this context we note the current
analysis of combined (anti)neutrino cross-section data \cite{AKP07} that
would allow to greatly reduce statistical uncertainty in the neutrino HT
terms.

The nuclear effects somewhat modify the $Q^2$ evolution of the
GLS integral. This is mainly due to the explicit $Q^2$ dependence
of effective cross section describing the
nuclear shadowing effect.
As can be seen from Fig.~\ref{fig:gls} the correction
is negative and of the order of 2\% at low $Q^2$. This must be
compared with the variation related to the running $\alpha_s$, which
is about 10\% between 2 and 20 $\gevsq$. Nuclear corrections should be then
taken into account for precise extractions of $\alpha_s$
from the GLS sum rule (see also \cite{hep-ph/0009150}). The effect of nuclear corrections
would indeed reduce the measured value of the strong coupling constant, by
reducing the $Q^2$ slope in the GLS integral.

\subsection{Comparison with cross section data}
\label{sec:res:xsec}

\begin{sidewaysfigure}[p]
\begin{center}
\epsfig{file=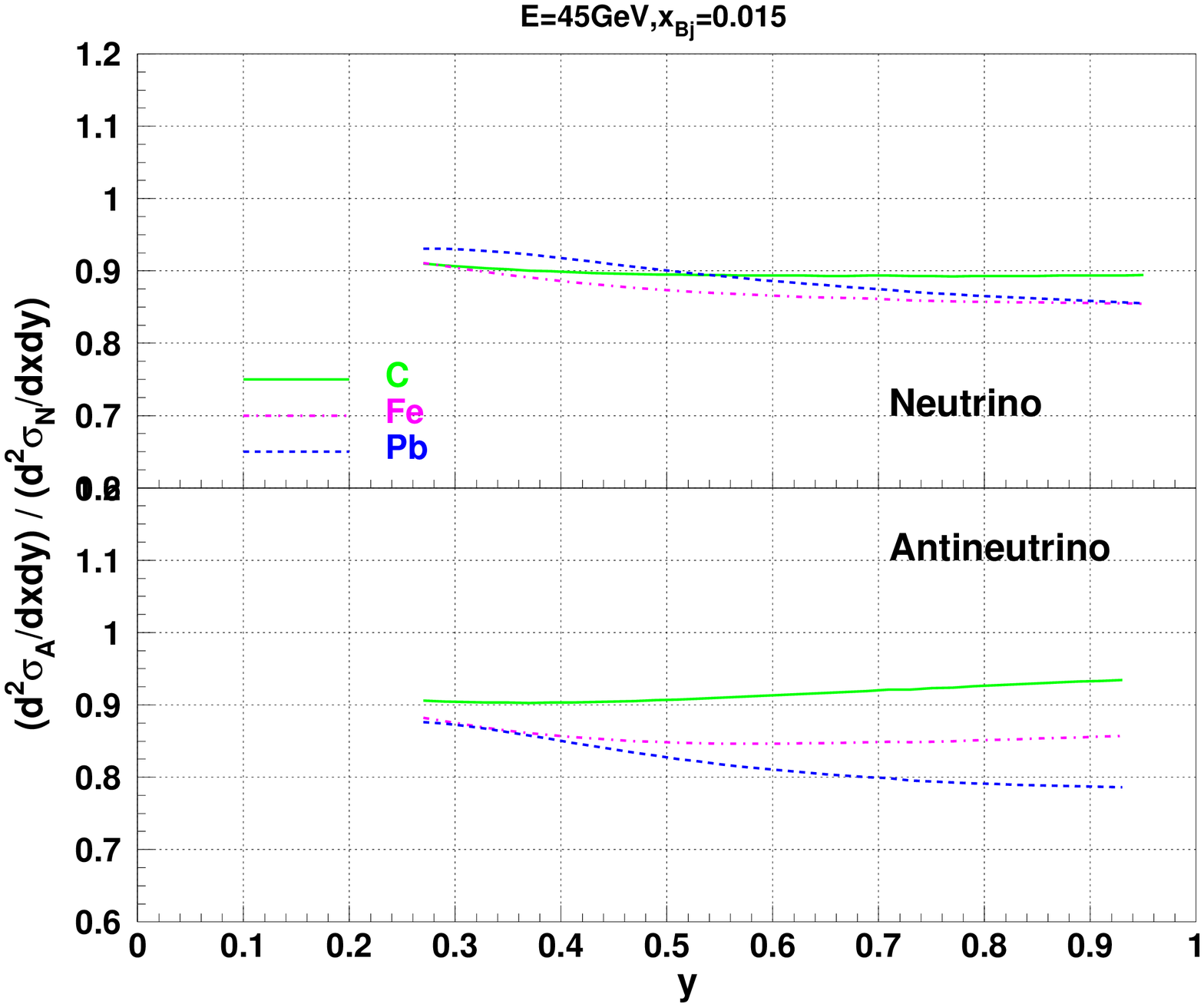,width=10.0cm}%
\epsfig{file=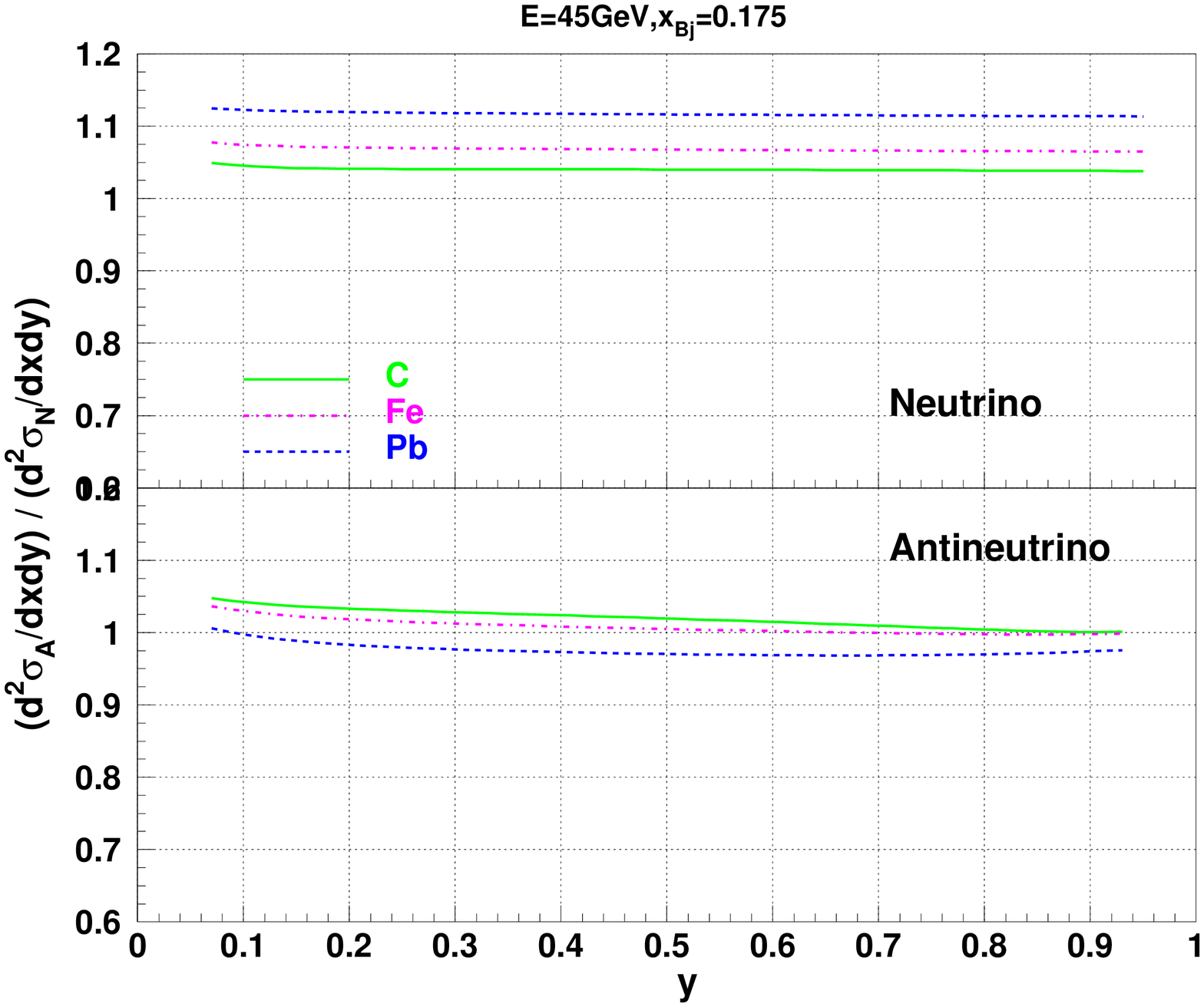,width=10.0cm}
\epsfig{file=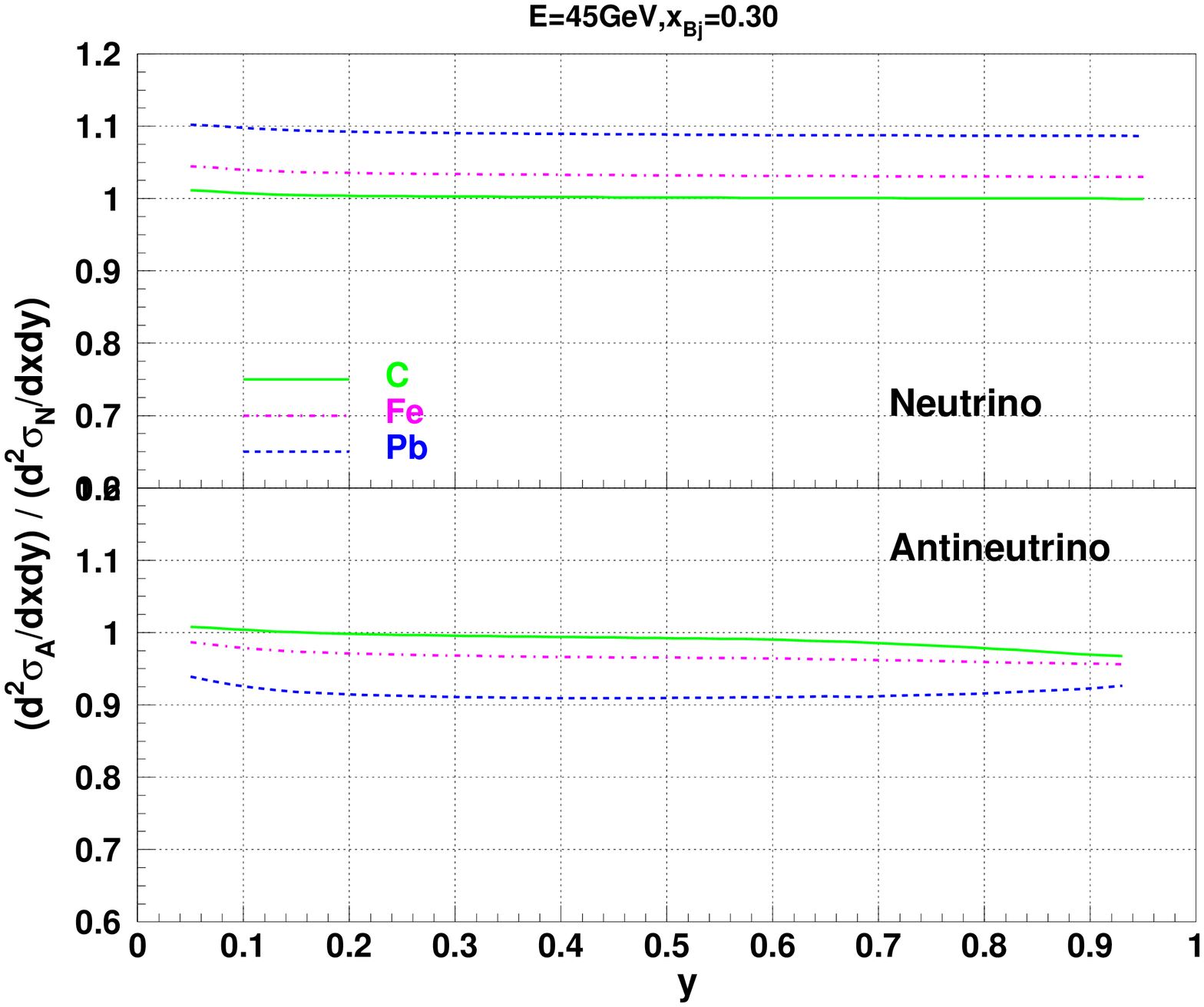,width=10.0cm}%
\epsfig{file=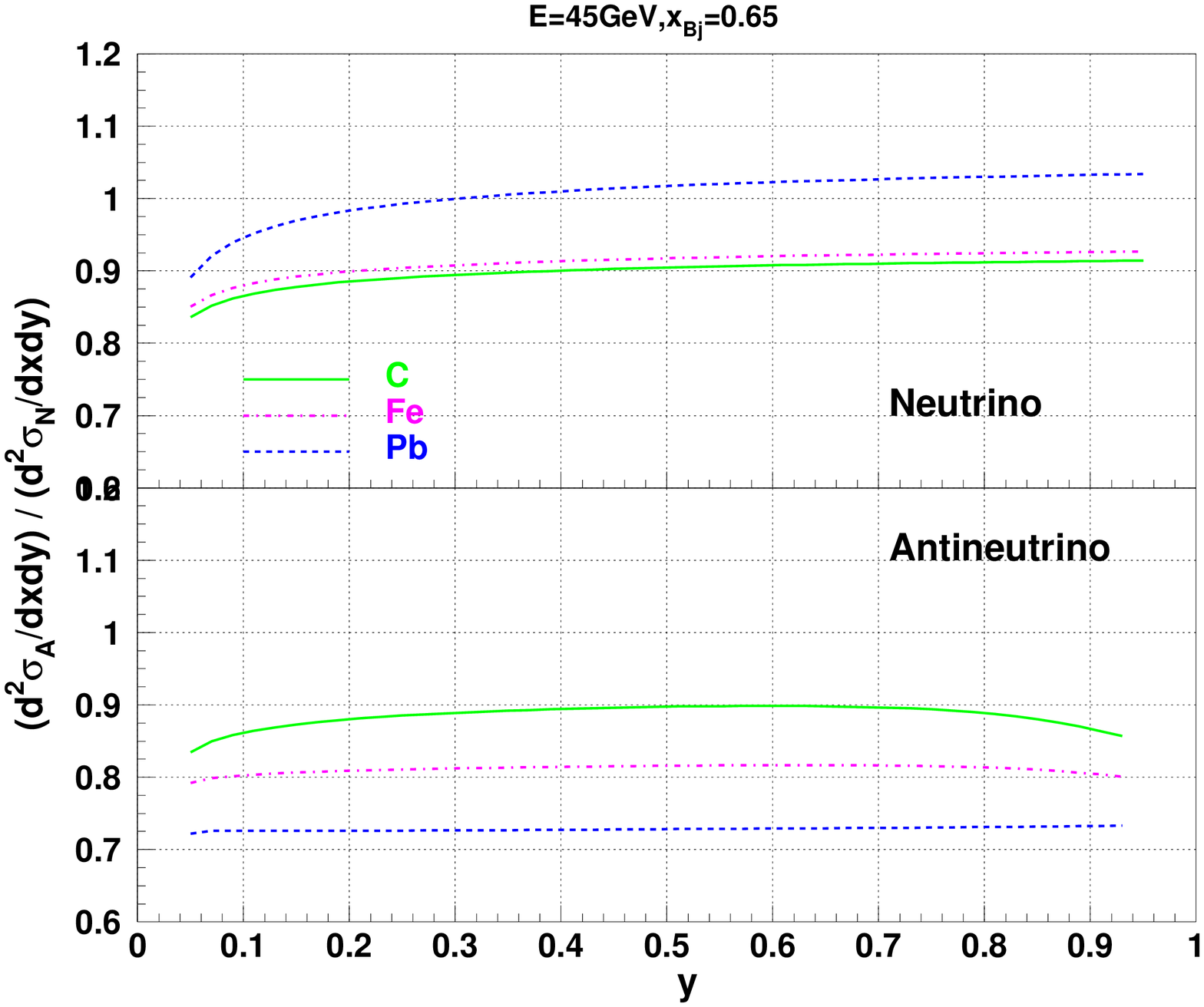,width=10.0cm}
\caption{%
The ratio of (anti)neutrino nuclear and the isoscalar nucleon differential
cross sections calculated for ${}^{12}$C, ${}^{56}$Fe and ${}^{207}$Pb
targets at $E=45\gev$.
The electroweak correction is taken into account
using  Ref.\cite{Arbuzov-Bardin}.
}
\label{fig:xsec1}
\end{center}
\end{sidewaysfigure}

\begin{sidewaysfigure}[p]
\begin{center}
\epsfig{file=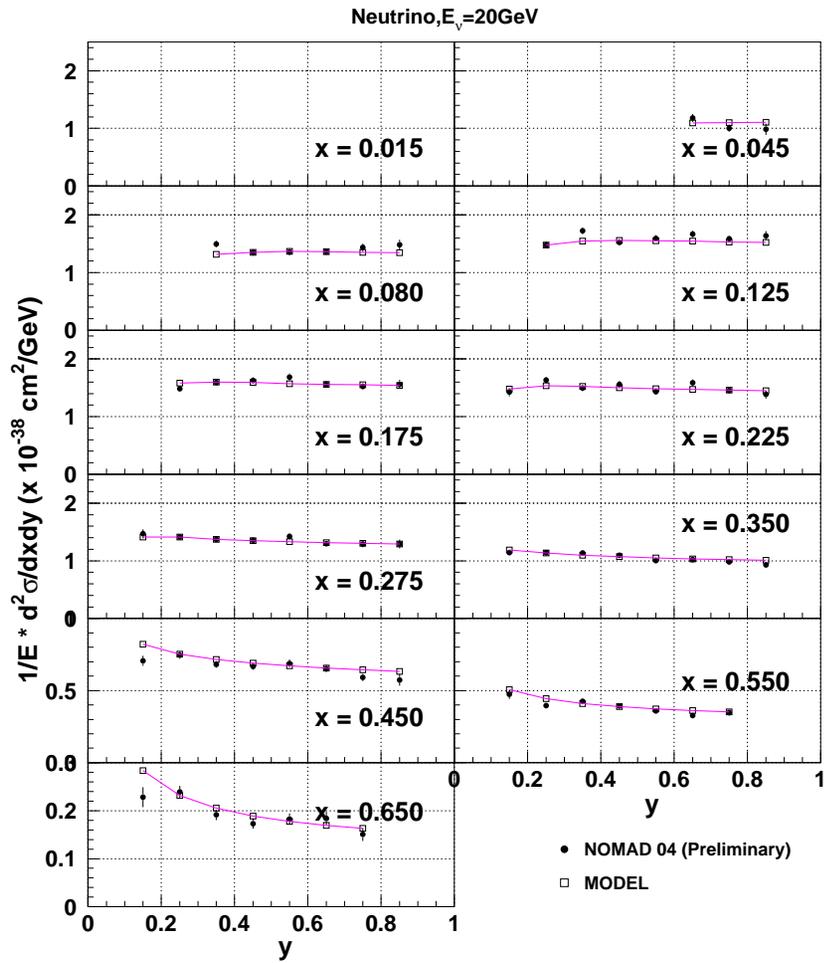,width=12.0cm}%
\hspace*{-0.30cm}\epsfig{file=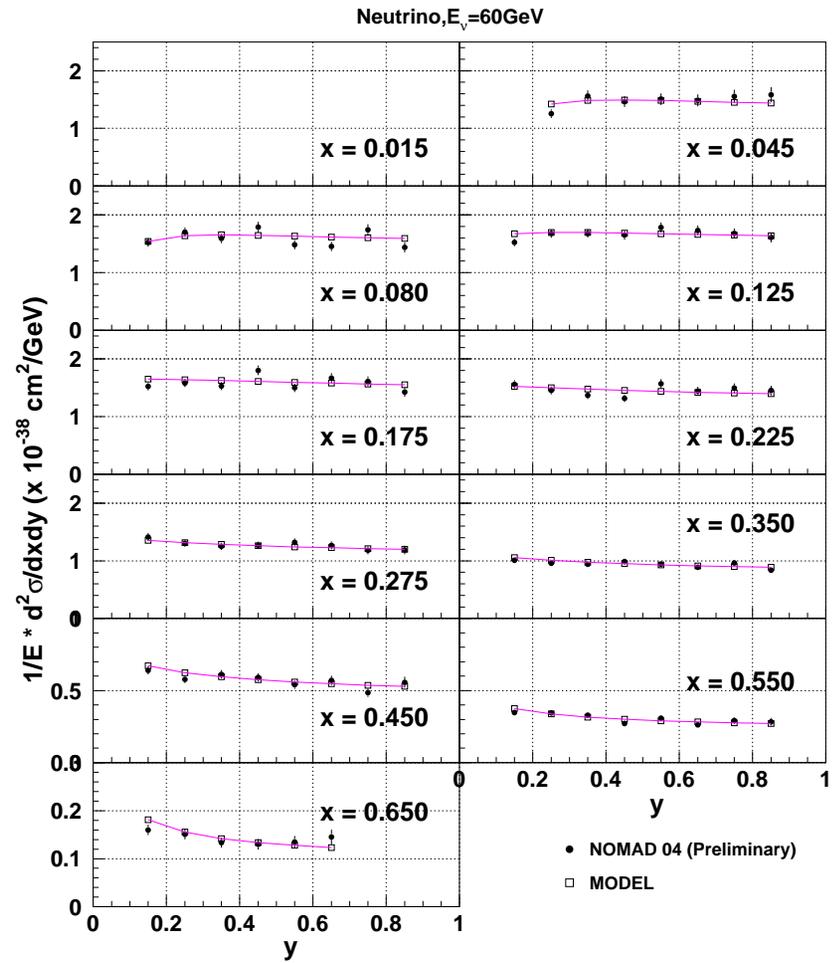,width=12.0cm}
\caption{Comparison of our predictions (open symbols) with NOMAD data
(full symbols) for neutrino differential cross-sections on ${}^{12}$C at
$E=20\gev$ (left plot) and $E=60\gev$ (right plot).
See text for details.
}
\label{fig:NOMAD}
\end{center}
\end{sidewaysfigure}

\begin{sidewaysfigure}[p]
\begin{center}
\epsfig{file=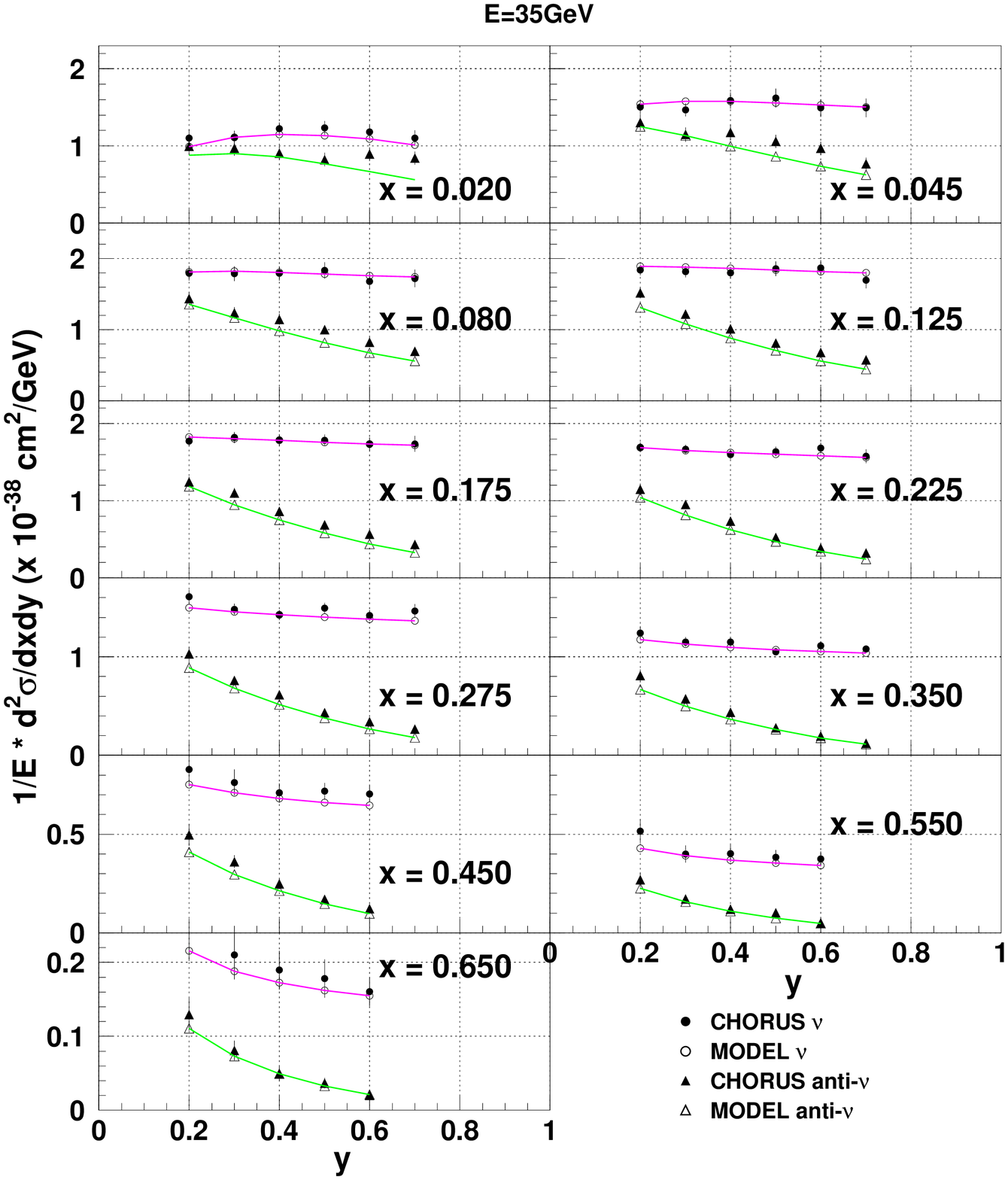,width=12.0cm}%
\hspace*{-0.30cm}\epsfig{file=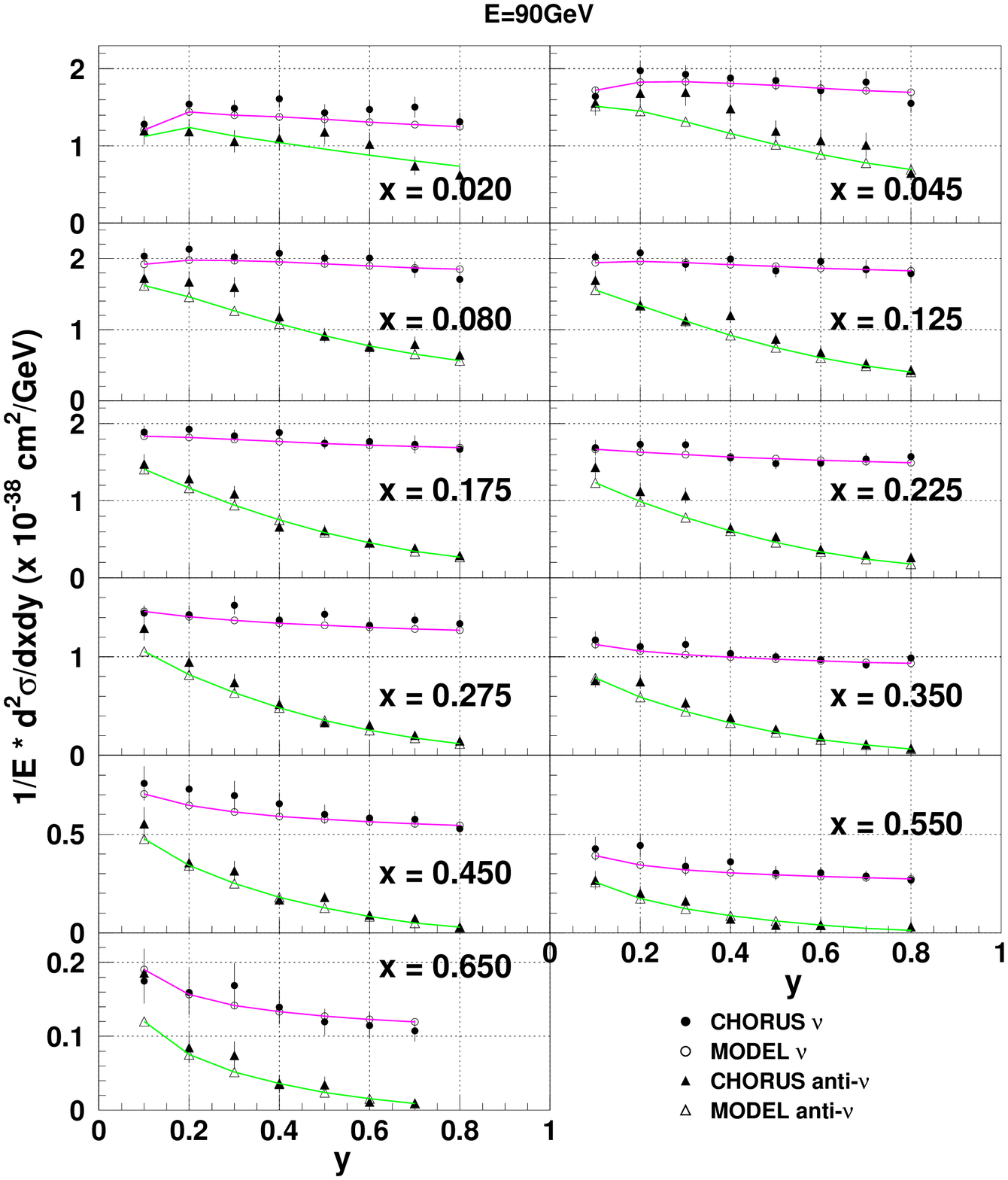,width=12.0cm}
\caption{Comparison of our predictions (open symbols) with CHORUS data
(full symbols) for neutrino (circles) and antineutrino
(triangles) differential cross-sections on ${}^{207}$Pb at
$E=35\gev$ (left plot) and $E=90\gev$ (right plot).
See text for details.
}
\label{fig:CHORUS}
\end{center}
\end{sidewaysfigure}

\begin{sidewaysfigure}[p]
\begin{center}
\epsfig{file=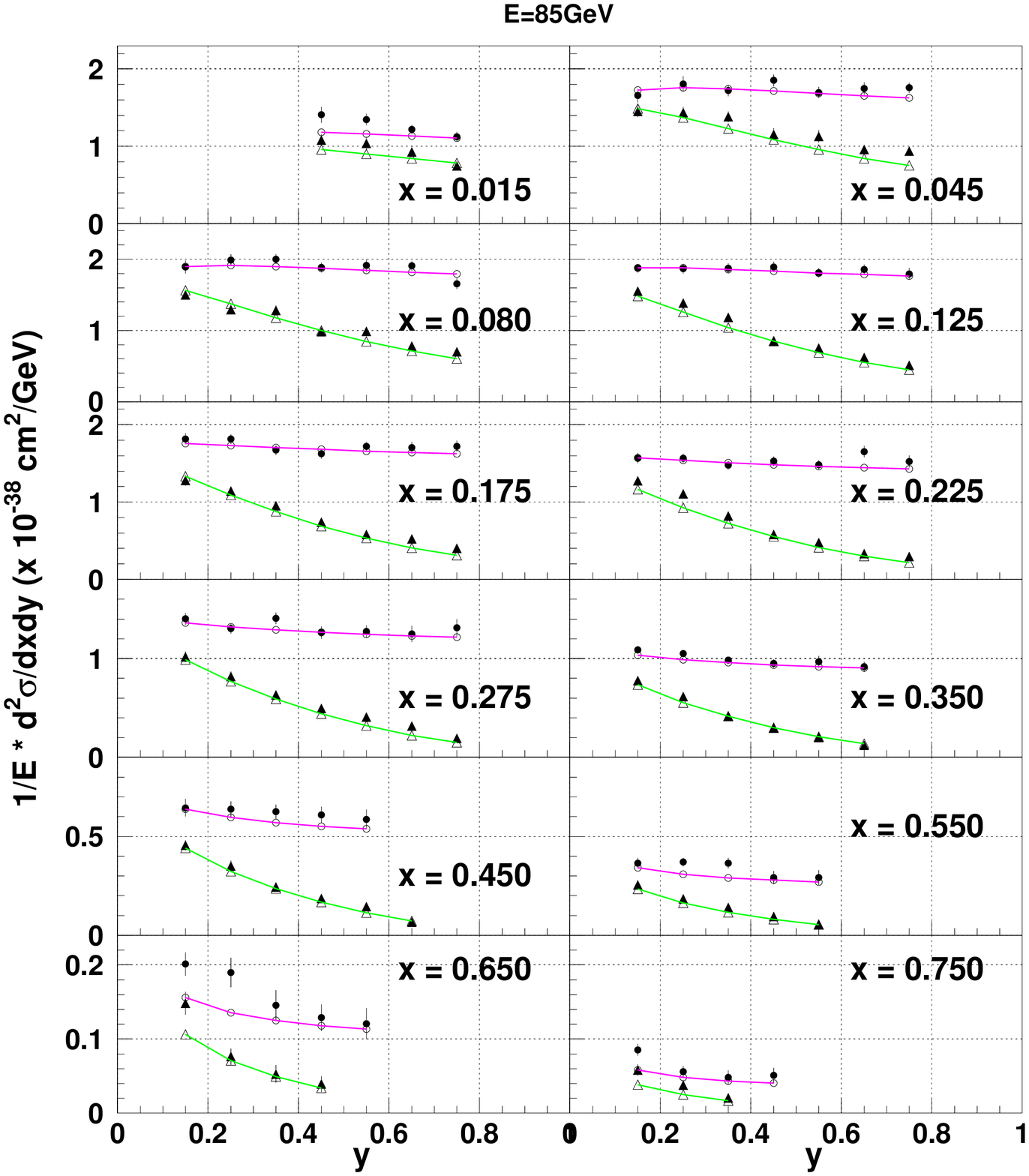,width=12.0cm}%
\hspace*{-0.30cm}\epsfig{file=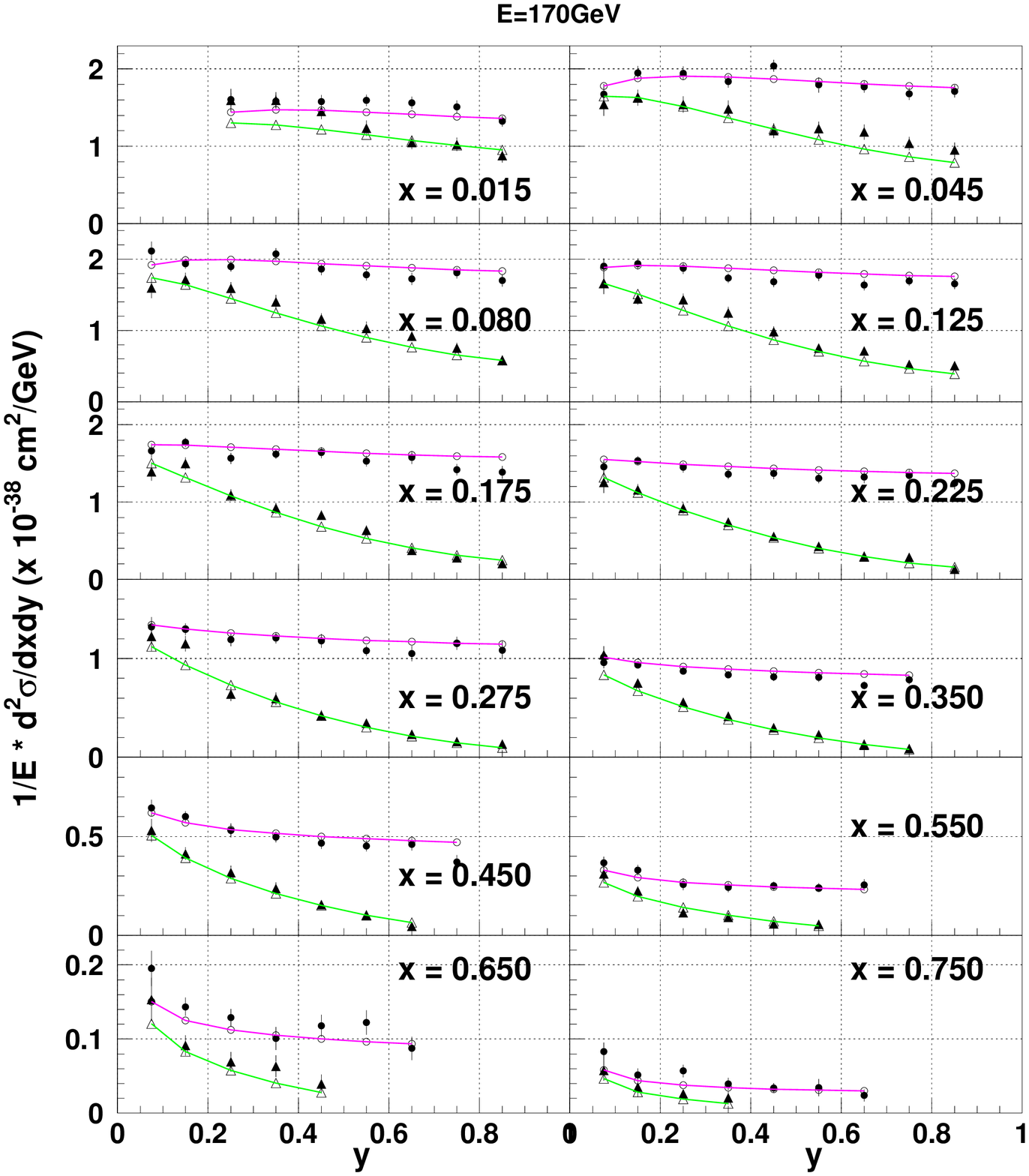,width=12.0cm}
\caption{Comparison of our predictions (open symbols) with NuTeV data
(full symbols) for neutrino (circles) and antineutrino
(triangles) differential cross-sections on ${}^{56}$Fe at
$E=85\gev$ (left plot) and $E=170\gev$ (right plot).
See text for details.
}
\label{fig:NuTeV}
\end{center}
\end{sidewaysfigure}

Using the results on nuclear structure functions we calculate
the neutrino and antineutrino inelastic inclusive differential
cross-sections. Figure~\ref{fig:xsec1} summarizes our results for the
corresponding nuclear dependence.
We note a cancellation between different nuclear effects for antineutrino
at $x\sim 0.1\div 0.2$. As a result, the antineutrino cross section
is not very sensitive to the target material in this region.
Note also the cancellation of nuclear effects for $x\sim 0.3$ for the
isoscalar target ($^{12}$C). Nuclear effect in this region is driven by
the neutron excess.

We can then compare our predictions with the available data on
differential cross sections for different targets. Table~\ref{tab:nudata}
summarizes the most recent and precise measurements on ${}^{12}$C (NOMAD),
${}^{56}$Fe (NuTeV) and ${}^{207}$Pb (CHORUS).\footnote{We use the
cross-section data for all our studies and not structure function data
provided by experiments. This procedure minimizes potential biases related
to different model assumptions used in extractions of structure
functions by individual experiments.}
It must be noted that when comparing
with data the predictions have to include electroweak (EW) corrections
related to virtual loop diagrams, soft photon emission and hard photon
emission. The effect of such processes is to shift the measured kinematic
variables and hence to modify the differential cross sections. To this end,
in this paper we use the recent one-loop calculations of
Ref.~\cite{Arbuzov-Bardin} (see also Ref.~\cite{Dittmaier}). We also
implicitly assume that EW and nuclear corrections factorize so that they
can be applied to the final nuclear structure functions.\footnote{We note
that the magnitude of EW corrections in some specific kinematic regions
can be comparable with nuclear corrections.}

\begin{table}
\begin{center}
\begin{tabular}{lc|ccc|ccc|ccc|ccc|ccc|ccc|ccc}
Experiment &&& Beam &&& Target &&& Statistics &&& $E$ values &&& $x$ values &&& $y$ values &&& Number of  \\
 &&&  &&&  &&&  &&& $[{\rm GeV}]$ &&&  &&&  &&& points  \\ \hline\hline
NOMAD\cite{nomad-xsec} &&& $\nu$ &&& ${}^{12}$C &&& 750k &&& 20$\div$200 &&& 0.015$\div$0.65 &&& 0.15$\div$0.85 &&& 563  \\
NuTeV\cite{nutev-xsec} &&& $\nu$ &&& ${}^{56}$Fe &&& 860k &&& 35$\div$340 &&& 0.015$\div$0.75 &&& 0.05$\div$0.95 &&& 1423  \\
 &&& $\bar{\nu}$ &&& ${}^{56}$Fe &&& 240k &&& 35$\div$340 &&& 0.015$\div$0.75 &&& 0.05$\div$0.85 &&& 1195  \\
CHORUS\cite{chorus-xsec} &&& $\nu$ &&& ${}^{207}$Pb &&& 930k &&& 25$\div$170 &&& 0.020$\div$0.65 &&& 0.10$\div$0.80 &&& 607  \\
  &&& $\bar{\nu}$ &&& ${}^{207}$Pb &&& 160k &&& 25$\div$170 &&& 0.020$\div$0.65 &&& 0.10$\div$0.80 &&& 607  \\ \hline
\end{tabular}
\caption{The list of neutrino and antineutrino data samples (after analysis
cuts) corresponding to the cross-section measurements used in this paper.
\label{tab:nudata}
}
\shrinkvspace
\end{center}
\end{table}

In Figs.~\ref{fig:NOMAD} 
to \ref{fig:NuTeV} we report the
comparisons between mesurements and our calculations for several values of the
(anti)neutrino energy ranging from 20 GeV to 170 GeV. In this calculations
we use both the nucleon parton distributions and the HT terms obtained from
the fits to charged-lepton DIS data \cite{a02}. In the presented results the
LT structure functions were evaluated in the NNLO approximation. In order
to evaluate the HT contribution to neutrino structure functions $F_2$ and $F_T$ we use the
phenomenological charged-lepton HT terms rescaled according to the
corresponding ratio of neutrino to the CL structure function in the LT
approximation. No HT contribution is included for $xF_3$. A more detailed
study of (anti)neutrino data in the context of QCD fits and of the
extraction of HT terms will be published elsewhere~\cite{AKP07}.
Note that the approach discussed was successfully applied in the analysis of
charged-lepton nuclear data (the nuclear EMC effect) \cite{KP04}. The
comparison with neutrino data provides an independent verification of our
model and also allows a compatibility check between (anti)neutrinos and
charged leptons.

In general, we observe good agreement between data and our predictions
for all nuclei. We note that this also includes the lowest $Q^2$ data
points, which in the case of CHORUS are at $Q^2\sim 0.25 \gevsq$. The
$Q^2$ dependence in the low $x$ bins is also reproduced, as can be seen
from the $y$ distributions and by comparing different (anti)neutrino
energies. The existing cross-section data in the low $x$ and low $Q^2$
region support the presence of the PCAC term
in the longitudinal structure function. A value of the scale
$M_{\textsc{pcac}}=0.8$ GeV seems to provide the best agreement with data.

After verifying the consistency of our cross-section model with (anti)neutrino data,
it is interesting to use the model to examine the compatibility of data
from different experiments and nuclear targets. At small and intermediate
values of $x$ the three experiments are in agreement for both the neutrino
and antineutrino samples. At large values of $x>0.45$ the NuTeV
data seem to be systematically above calculations for most
neutrino energies. On the other hand, the corresponding predictions for
${}^{12}$C and ${}^{207}$Pb at large $x$ are in agreement with NOMAD and
CHORUS data, respectively.

The low-$Q^2$ and low-$x$ behaviour of the (anti)neutrino cross section is
dominated by the divergence of the axial current and PCAC relation.
However, as discussed in Section~\ref{sec:sf:low-q}, the scale
controlling the PCAC contribution to the longitudinal structure function
is not well known.
The use of heavy targets introduces additional uncertainties.
The MINER$\nu$A
experiment~\cite{minerva}, recently proposed at Fermilab and currently
under construction, has the possibility to address
PCAC effect (as well as other nuclear effects) in heavy
nuclei. To this end, the possibility to have cryogenic hydrogen and
deuterium targets would greatly enhance the corresponding physics
potential.

\section{Summary}
\label{sec:sum}

We discussed  inelastic inclusive scattering of high-energy (anti)neutrino
off nuclei and developed a quantitative model for nuclear structure
functions using an
approach which takes into account the QCD treatment of the nucleon
structure functions and addresses the basic nuclear effects including
nuclear shadowing, Fermi motion and nuclear binding, nuclear pions and
off-shell corrections to bound nucleon structure functions.

The presence of both the axial and the vector currents in neutrino
interactions results in contributions with different $C$-parity to
neutrino cross sections.
The interference between the vector and the axial current is described by
the structure function $F_3$ which is not present in charged-lepton
scattering.
We discussed in detail how the nuclear effects depend on the structure
function type ($F_2$ vs $xF_3$) and on the $C$-parity of structure
functions ($C$-even $\nu+\bar\nu$ and $C$-odd $\nu-\bar\nu$
combinations).

The axial-vector current plays a special role in neutrino scattering in the region
of low $Q^2$ and low $x$. In this region the cross sections are dominated
by contributions from the divergence of the axial current which is linked
to the virtual pion cross section via the Adler relation.
Using PCAC we examined the derivation of low $Q^2$ and low $x$ limit for
neutrino structure functions and discussed
the scale in $Q^2$ at which the Adler relation
can be applied. We studied this problem phenomenologically using low-$Q^2$
and low-$x$ neutrino cross-section data.

We examined the Adler and the
Gross--Llewellyn-Smith sum rules for nuclear structure functions.
A remarkable cancellation between nuclear shadowing and
off-shell corrections was found, underlying the conservation of isospin and
valence quark number in nuclei. This fact was
used to constrain the effective scattering amplitude
controlling the nuclear shadowing for different $C$-parity and isospin
states.

We applied our results to calculate nuclear structure functions
for the targets and kinematical conditions typical for recent neutrino
experiments.
Our predictions for the (anti)neutrino inelastic differential
cross-sections agree well with the recent data on ${}^{12}$C \cite{nomad-xsec},
${}^{56}$Fe \cite{nutev-xsec}, and ${}^{207}$Pb \cite{chorus-xsec}.

We conclude by commenting that the nuclear corrections prove to be
particularly important for QCD phenomenology of $\nu(\bar{\nu})$ data on
heavy targets, where they significantly affect the fit results~\cite{AKP07}.

\section*{Acknowledgments}
We thank S. Alekhin, A. Butkevich, A. Kataev, and S. Mishra for
useful discussions and comments, J. Panman for
providing the tables of CHORUS cross-section data, M.
Tzanov for useful information about the NuTeV cross sections, and the NOMAD
Collaboration for the use of their preliminary data.
S.K. was partially supported by Russian Foundation for Basic Research
Project No. 06-02-16659 and 06-02-16353 and by INTAS Project No. 03-51-4007.
R.P. thanks USC for supporting this research.

\appendix
\section{Parameterization of the pion--nucleon scattering amplitude}
\label{sec:piN}

The parametrization of the pion--nucleon forward scattering amplitude based
on a Regge fit to $\pi N$ scattering data is given in Ref.\cite{piN}:
\begin{align}
\label{amp:pi}
a_{\pi^\pm p} &= a_\pi^0 \pm \frac12 a_\pi^1,
\\
a_\pi^0(s)   &= \frac{X}{2} s^\epsilon
           \left[i-\cot\left(\frac{\pi}{2}(1+\epsilon)\right)\right]
+ \frac{Y_1}{2} s^{-\eta_1}
    \left[i-\cot\left(\frac{\pi}{2}({1-\eta_1})\right)\right],
\\
a_\pi^1(s)   &= -Y_2 s^{-\eta_2}
         \left[i+ \tan\left(\frac{\pi}{2}({1-\eta_2})\right)\right].
\end{align}
In \eq{amp:pi} the amplitude $a_\pi^{0(1)}$ corresponds to the pion coupling
to the isoscalar (isovector) nucleon configuration and
the sign $+(-)$ corresponds to the $\pi^+(\pi^-)$ meson.
The pion-neutron scattering amplitude is derived from isospin relations
$a_{\pi^\pm n} = a_{\pi^\mp p}$.
The parameters are
\begin{align}
X &= 12.08 \pm 0.29, &\epsilon &=0.0933\pm 0.0024,\\
Y_1 &= 26.2\pm 0.74, &\eta_1   &=0.357\pm 0.015,\\
Y_2 &= 0.560\pm 0.017, &\eta_2 &=0.560\pm 0.017.
\end{align}
We also note that the forward scattering amplitude is normalized as
$\Im a_\pi=\sigma_\pi/2$ with $\sigma_\pi$ the total cross section, and
the units of the parameters $X$ and $Y_{1,2}$ are such that $a_\pi$ is
measured in mb if $s$ is in GeV$^2$.


\end{document}